\magnification=\magstep2  
\input amstex
\documentstyle{amsppt}
\hsize=16.5 truecm
\vsize=24 truecm
\parindent=1truecm  
\NoBlackBoxes
\tolerance=2000
\font\BF=cmbx12 
\font\cmbsyfnt=cmbsy10
\font\BF=cmbx12 scaled\magstep 2
\def\Par{\leavevmode\hbox{\cmbsyfnt\char120}} 
\font\rsfs=rsfs10
\def\a{\alpha}  \def\b{\beta}  \def\g{\gamma}  \def\G{\Gamma}  \def\t{\theta}
\def\s{\sigma} \def\u{\tau}
\def\mod{\mathop{\hbox{\rm mod}}}
\def\reg{\mathop{\hbox{\rm reg}}}
\def\Re{\mathop{\hbox{\rm Re}}}
\def\ord{\mathop{\hbox{\rm ord}}}

\def\det{\mathop{\hbox{\rm det}}}
\def\sgn{\mathop{\hbox{\rm sgn}}}
\def\vraisup{\mathop{\hbox{\rm vraisup}}}
\def\loc{\mathop{\hbox{\rm loc}}}
\def\tg{\mathop{\hbox{\rm tg}}}
\def\arctg{\mathop{\hbox{\rm arctg}}}
\font\msbm=msbm10
\def\RR{\hbox{\msbm R}}  
\def\CC{\hbox{\msbm C}}  
\def\QQ{\hbox{\msbm Q}}  
\def\KK{\hbox{\msbm K}}  

\font\cyr=cmcyr10
\def\No{\hbox{\cyr\char'031\hskip2pt}}  
\def\spt{\mathop{\hbox{\rm spt}}}
\def\AC{\mathop{\hbox{\rm AC}}}
\def\script#1{\hbox{\rsfs #1}}
\TagsOnRight
\let\leq\leqslant
\let\geq\geqslant

\document

\vskip 3cm
\centerline {\BF V.~S.~Vladimirov}
\medskip
\centerline {\bf Steklov Mathematical institute, 117966 GSP-1,}
\centerline {\bf Moscow, Gubkin str., 8, Russia. E-mail: vsv$\@$vsv.mian.su}
\vskip 2cm
\centerline {\BF Tables of Integrals}
\centerline {\BF of Complex-valued Functions}
\centerline {\BF of p-Adic Arguments}
\vskip 1cm
\centerline {\bf Steklov, 05-1997}
\vfill\eject
\centerline {\bf Content}
\bigskip
\centerline {\bf Part I. Some Facts From p-Adic Analysis}
\medskip

{\Par 1. The Field of $p$-Adic Numbers $\QQ_p$.}

{\Par 2. Some Functions on $\QQ_p$.}

{\Par 3. Analytic Functions on $\QQ_p$.}

{\Par 4. The Haar Mesure on $\QQ_p$.}

{\Par 5. $n$-Dimensional Space $\QQ_p^n$.}

{\Par 6. Generalized Functions on $\QQ_p^n$.}

{\Par 7. The Fourier-transform.}

{\Par 8. Homogeneous Generalized Functions.}

{\Par 9. Quadratic Extentions of the Field $\QQ_p$.}

{\Par 10. Operator $D^\a$.}
\bigskip
\centerline {\bf Part II. Tables of Integrals}
\medskip

{\Par 11. The Simplest Integrals. One Variable.}

{\Par 12. The Fourier Integrals. One Variable.}

{\Par 13. The Gaussian Integrals. One Variable.}

{\Par 14. Two Variables.}

{\Par 15. Many Variables.}

{\Par 16. Integrals of Generalized Functions.}

{\Par 17. The Fourier-transform of Generalized Functions.}

\vskip 1cm	  

This work was supported in part by the RFFI, Grants \No 96-01-01008 and 
96-15-96131. 

\vfill\eject	  	      	       

\centerline {\BF Part I}
\bigskip
\centerline {\BF Some Facts from p-Adic Analysis}
\bigskip

Everywhere henceforth we shall assume, unless otherwise stipulated, that $p$ 
takes all prime numbers, $p=2,3,5,\ldots,137,\ldots$, and $\g$ takes all 
integer (rational) numbers, $\g=0,\pm 1,\pm 2,\ldots$, $\g\in Z$; By $Z_+$ we
shall denote the set of natural numbers $\g=1,2,\ldots.$ If $\KK$ is some field
(or ring), by $\KK^\times$ we shall denote its multiplicative group. 
\bigskip
\centerline {\bf \Par 1. The Field of p-Adic Numbers $\QQ_p$}
\medskip

{\bf Denote:} by $Q$ the field of rational numbers, by $\RR$ the field of real
numbers, by $\CC$ the field of complex numbers.

Let $p$ be a prime number. Any rational number $x\neq 0$ uniquely represented
in the form
$$x=\pm p^\g{a/b}$$
where $\g\in Z$ and $a,b$ are  natural numbers not divisible by $p$ and without
common divisors. $p$-{\it Adic norm} $|x|_p$ of the number $x\in\QQ$ is defined
by the formulas
$$|x|_p=p^{-\g}, x\neq 0, \quad |0|_p=0.$$
The completion of the field $\QQ$ with respect to the norm $|\cdot |_p$ is the
{\it field of $p$-adic numbers} $\QQ_p$.

{\it The canonical form} of a $p$-adic number $x\neq 0$ is
$$x=p^\g (x_0+x_1p+x_2p^2+\ldots) \eqno (1.1)$$
where $\g=\g (x)\in Z$, $x_j=0,1,\ldots,p-1, x_0\neq 0, j=0,1,\ldots$, besides
$|x|_p=p^{-\g}$. The number $-\g$ is called {\it the order} of number $x$ 
and it is denoted by $\ord x, \quad \ord x=-\g(x), \quad \ord 0=-\infty.$ 

The norm $|\cdot |_p$ possesses the following characteristic properties:
$$1) |x|_p\geq 0, \quad |x|_p=0 \leftrightarrow x=0,$$
$$2) |xy|_p=|x|_p|y|_p,$$
$$3) |x+y|_p\leq\max (|x|_p,|y|_p). \eqno (1.2)$$ 
Besides,
$$3') |x+y|_p=\max (|x|_p,|y|_p), \quad |x|_p\neq |y|_p,$$
$$3'') |x+y|_p\leq |2x|_p, \quad |x|_p=|y|_p.$$

Thus, owing to (1.2), the norm $|\cdot |_p$ is {\it non-Archimedean} and the 
space $\QQ_p$ is {\it ultrametric.}

{\rm Denote:} by
$$B_\g(a)=[x\in\QQ_p: |x-a|_p\leq p^\g]$$
a {\it disk} with a center at the point $a\in\QQ_p$ of radius $p^\g$,
$B_\g=B_\g(0)$; by
$$S_\g(a)=[x\in\QQ_p: |x-a|_p=p^\g]$$
a {\it circumference} with the same center and radius, $S_\g=S_\g(0).$

Obvious relations are valid:
$$B_\g (a)=\cup_{\g'\leq\g}S_{\g'}(a), \quad S_\g (a)=B_\g (a)\backslash B_{\g-1}(a),$$
$$\QQ_p= \cup_{\g\in Z}B_\g (a), \quad \QQ_p^\times= \cup_{\g\in Z}S_\g(a).$$  

The geometry of the space $\QQ_p$ is very unusual: all triangles in it are
isosceles; every point of a disk is its center; a disk has no boundary; a disk
is a finite union of disjoint disks of smaller radii; if two disks have a
common point, so one of them is contained in another; a disk is open compact.

A set of $\QQ_p$ which is closed and open is called {\it clopen} set.

{\rm Denote:} by $Z_p=B_0$ the maximal compact subring of the field $\QQ_p$ 
(the ring of integer $p$-adic numbers); by $Z_p^\times =S_0$ multiplicative 
group of the ring $Z_p$ (it is the group of unities of the field $\QQ_p$); by
$I_p=pZ_p=B_{-1}$ maximal ideal of the ring $Z_p$.

The residue classes $Z_p/I_p$ form the finite field which is isomorphic to the
residue classes module $p: \{0,1,\ldots,p-1\}.$

Introduce special sets:
$$G_p=[x\in\QQ_p: |x|_p\leq |2p|_p];$$
$$J_p=[x\in Z_p^\times : 1-x\in G_p],$$
$J_p$ is a multiplicative group;
$$S_{\g,k_0k_1\ldots k_n}=[x\in S_\g:x_0=k_0,x_1=k_1,\ldots,x_n=k_n];$$
$$S_\g^{k_0k_1\ldots k_n}=[x\in S_\g:x_0\neq k_0,x_1\neq k_1,\ldots,x_n\neq k_n],$$
where $k_j=0,1,\ldots,p-1, k_0\neq 0, j=1,2,\ldots,n.$ 

The sets just introduced are open compacts in $\QQ_p.$

{\it Rational part} $\{x\}_p$ of a number $x\in\QQ_p$ is $\{x\}_p=0$ if 
$\g (x)\geq 0$, and it is 
$$\{x\}_p=p^\g (x_0+x_1p+\ldots+x_{-\g-1}p^{-\g-1}) \hbox{ if } \g (x)\leq{-1}. \eqno (1.3)$$
Denote by $\QQ_p^{\times 2}$ the multiplicative {\it group of squares} of 
$p$-adic numbers.

{\it In order a number $x\in\QQ_p^\times$ belongs to $\QQ_p^{\times 2}$, it is
necessary and safficient that $\g (x)$ is even and}
$$\Bigl ({x_0\over p}\Bigr )=1, p\neq 2; \quad x_1=x_2=0, p=2.$$
Here
$$\Bigl ({a\over p}\Bigr ), \quad a\in Z, a\not\equiv 0(\mod p)$$
is the {\it Legendre symbol} which equal to 1 or -1 subject to if the number 
$x$ is quadratic residue or non-residue module $p.$

Thus the group ${\QQ_p^\times}$/${\QQ_p^{\times 2}}$ consists of four elements
$(1,\epsilon,\-p,\epsilon p)$ where $\epsilon$ is any unit of the field 
$\QQ_p$ which is not a square in $\QQ_p$ if $p\neq 2$, and it consists of eight
elements $\{1, 2, 3, 5, 6, 7, 10, 14\}$ if $p=2$. 
\bigskip
\centerline {\bf \Par 2. Some Functions on $\QQ_p$}
\medskip

{\bf Characters of the field $\QQ_p$.} Let $\chi (x)$ be an {\it additive 
character} of the field $\QQ_p$,
$$\chi (x+y)=\chi (x)\chi (y), \quad |\chi (x)|=1, x,y\in\QQ_p. \eqno (2.1)$$
{\it Standard additive character} of the field $\QQ_p$ has the form
$$\chi_p (x)=\exp (2\pi i\{x\}_p) \eqno (2.2)$$
where $\{x\}_p$ is the rational part of $x\in\QQ_p$ which is defined by the 
formula (1.3).

{\it The general form of an additive character} of the field $\QQ_p$ is 
$$\chi (x)=\chi_p (\xi x)=\exp (2\pi i\{\xi x\}_p) \eqno (2.3)$$
for some $\xi\in\QQ_p.$

Let $\pi (x)$ be {\it multiplicative character} of the field $\QQ_p$,
$$\pi (xy)=\pi (x)\pi (y), \quad |\pi (x)|=1, x,y\in\QQ_p^\times. \eqno (2.4)$$
{\it The general form of a multiplicative character} of the field $\QQ_p$ is
$$\pi (x)=\pi_{i\a,\t}(x)=|x|_p^{i\a}\t (x), \quad x\in\QQ_p^\times \eqno (2.5)$$
where $\a\in\RR$ is defined by the equality $\pi (p)=p^{-i\a}$ and 
$\t (t), t\in Z_p^\times$ is a character of the compact group $Z_p^\times$ 
normalized by the condition $\t (p)=1$. (The set of the latters is countable
and discrete.)

If the unitarity condition $|\pi (x)|=1$ in (2.4) is not fulfilled then the
function $\pi (x)$ is a {\it representation} of the group $\QQ_p^\times$ in 
$\CC$, and its general form is given by the formula (2.5) in which $i\a$ is any
complex number, so that
$$\pi_{\a,\t}(x)=|x|_p^{\a-1}\t(x), \quad x\in\QQ_p^\times, \a\in\CC. \eqno (2.5')$$
Such functions are called {\it quasi-characters}. A quasi-character $\pi (x)$
$=|x|_p^{\a-1}$ for which $\t=1$ is called {\it principal quasi-character}.

Let $d\not\in\QQ_p^{\times 2}$. Without loss of generality it is possible to
suppose that $d$ is square free of $p$-adic numbers, that is it is one of the
listed in \Par 1 forms,
$p,\epsilon,p\epsilon, |\epsilon |_p=1, \epsilon\not\in\QQ_p^{\times 2}$ for
$p\neq 2$, and $2,3,5,6,7,10$,$14$ for $p=2$.

Denote by $\QQ_p^\times (d)$ the set of $p$-adic numbers in $\QQ_p^\times$ 
which are representable in the form $\a^2-d\b^2$, $\a,\b \in\QQ_p$; \quad
$\QQ_p^\times (d)$ is a multiplicative group.

The {\it Hilbert symbol} $\Bigl ({{x,y}\over p}\Bigr ), x,y \in\QQ_p^\times$ by
definition is equal to 1 or -1 subject to the form $x\a^2+y\b^2-\g^2$ 
represents nontrivially zero in $\QQ_p$ or not.

The Hilbert symbol has the following obvious properties [5]:
$$\Bigl ({{x,y}\over p}\Bigr )=\Bigl ({{y,x}\over p}\Bigr ), \quad \Bigl ({{x,-x}\over p}\Bigl )=1, \quad \Bigr ({{x,yz}\over p}\Bigr )=\Bigl ({{x,y}\over p}\Bigr )\Bigl ({{x,z}\over p}\Bigr ),$$
and besides
$$\Bigl ({{p,\epsilon}\over p}\Bigr )=\Bigl ({{\epsilon_0}\over p}\Bigr ), \quad \Bigl ({{\epsilon,\eta}\over p}\Bigr )=1, \quad p\neq 2;$$
$$\Bigl ({{2,\epsilon}\over 2}\Bigr )=(-1)^{{(\epsilon^2-1)}/2}, \quad \Bigl ({{\epsilon,\eta}\over 2}\Bigr )=(-1)^{{(\epsilon-1)(\eta-1)}/4}, \quad p=2.$$
Here $\epsilon$ and $\eta$ are any units of the field $\QQ_p$.  
  
From here it follows a criterion in order that a $p$-adic number $x$ belongs to
$\QQ_p^\times (d)$ for $p\neq 2$. {\it In order that $x\in\QQ_p^\times (d)$ it
is necessary and sufficient: for $d=\epsilon$ $\g (x)$ is even; for $d=p$
$\g (x)$ is even and $\Bigl ({{x_0}\over p}\Bigr )=1$ or $\g (x)$ is odd and
$\Bigl ({{-x_0}\over p}\Bigr )=1$; for $d=p\epsilon$ $\g (x)$ is even and
$\Bigl ({{x_0}\over p}\Bigr )=1$ or $\g (x)$ is odd and 
$\Bigl ({{-x_0}\over p}\Bigr )=-1$.} (Similar criterion takes place and for 
$p=2$.) 
  
Hence, {\it The group ${\QQ_p^\times}/{\QQ_p^\times (d)}$ is isomorphic to the
group $(1,-1)$, and the function
$${\textstyle\sgn_{p,d}} x=\cases{1, \quad x\in\QQ_p^\times (d),} \\ 
{-1, \quad x\not\in\QQ_p^\times (d) } \endcases \tag 2.6$$
is a multiplicative character of the group $\QQ_p^\times$.}

Directly from the definitions it follows
$${\textstyle\sgn_{p,d}} x=\Bigr ({{x,-dx}\over p}\Bigl ), x\in\QQ_p^\times, \quad d\not\in\QQ_p^{\times 2}.$$
(Note that always $\Bigl ({{x,-dx}\over p}\Bigr )=1$ if 
$d\in\QQ_p^{\times 2}$.)
  
$\lambda_p$-{\it function of field} $\QQ_p$ is defined by the following way 
[1a)],[6a)]
$$\lambda_p (x)=\cases{1, \quad \g (x)=2k, \quad p\neq 2}, \\ 
{\sqrt{\Bigl ({{-1}\over p}\Bigr )}\Bigl ({{x_0}\over p}\Bigr ), \quad \g (x)=2k+1, \quad p\neq 2}, \\ 
{\exp [\pi i({1/4}+x_1)], \quad \g (x)=2k, \quad p=2}, \\
{\exp [\pi i({1/4}+{x_1/2}+x_2)], \quad \g (x)=2k+1, \quad p=2.} \endcases $$ 

Properties of $\lambda_p$-function $\QQ_p^\times\rightarrow\CC$.
$$|\lambda_p (x)|=1, \quad \lambda_p (x)\lambda_p (-x)=1;$$
$$\lambda_p (x)=\lambda_p (y), \quad xy\in\QQ_p^{\times 2};$$
$${{\lambda_p (x)\lambda_p (y)}\over\lambda_p (x+y)}=\lambda_p\Bigl ({{xy}\over{x+y}}\Bigr );$$
$$\lambda_p (x)\lambda_p (y)=\Bigl ({{x,y}\over p}\Bigr )\lambda_p (xy)\lambda_p (1). \eqno (2.7)$$
Puting in (2.7) $y=-dx$ and using the formula (2.6) we obtain relation [6a)]   
$${\textstyle\sgn_{p,d}} x=\lambda_p (x)\lambda_p (-dx)\lambda_p (d)\lambda_p (-1), x\in\QQ_p^\times, \quad d\not\in\QQ_p^{\times 2}. \eqno (2.8)$$

Note the following formulae [6a)]
$${\textstyle\sgn_{p,d}}x=\cases{\Bigl ({{x_0}\over p}\Bigr )^{\g (d)}\Bigl ({{d_0}\over p}\Bigr )^{\g (x)}\Bigl ({{-1}\over p}\Bigr )^{\g (x)\g (d)}, \quad p\neq 2, } \\
{(-1)^{d_1x_1+(d_1+d_2)\g (x)+(x_1+x_2)\g (d)}, \quad p=2. } \endcases \tag 2.9$$
In particular, for $d\equiv 3(\mod 4)$ we have [6b)]
$${\textstyle\sgn_{p,d}}x=\cases{1, \quad \Bigl ({d\over p}\Bigr )=1, } \\
{(-1)^{\g(x)}, \quad \Bigl ({d\over p}\Bigr )=-1, p\neq 2, p\neq d,} \\
{\Bigl ({d\over p}\Bigr )(-1)^\g(x), \quad p=d,} \\
{(-1)^{x_1}, \quad p=2, d\equiv 7(\mod 8),}\\
{(-1)^{x_1+\g(x)}, \quad p=2, d\equiv 3(\mod 8).} \endcases $$
 Note the following infinite products, which are valid for $x, y\in\QQ^\times$
$$|x|_\infty\prod_{p=2}^\infty |x|_p=1, \quad |x|_\infty =|x|; \eqno (2.10)$$
$$\chi_\infty (x)\prod_{p=2}^\infty\chi_p (x)=1, \quad \chi_\infty (x)=\exp (-2\pi ix); \eqno (2.11)$$    
$$\lambda_\infty (x)\prod_{p=2}^\infty\lambda_p (x)=1, \quad \lambda_\infty (x)=\exp (-{{i\pi}/4}\sgn x); \eqno (2.12)$$
$$\Bigl ({{x,y}\over\infty}\Bigr )\prod_{p=2}^\infty\Bigl ({{x,y}\over p}\Bigr )=1 \eqno (2.13)$$
where $x,y\in\QQ_p^\times$ and
$$\Bigl ({{x,y}\over\infty}\Bigr )=\cases{-1, \quad x<0, y<0}, \\ 
{1, \hbox{ otherwise };} \endcases$$
$${\textstyle\sgn_{\infty,d}} x\prod_{p=2}^\infty{\textstyle\sgn_{p,d}} x=1 \eqno (2.14)$$
where
$${\textstyle\sgn_{\infty,d}}x=\cases{\sgn x, \quad d<0,} \\ 
{1, \quad d>0.} \endcases$$ 
Infinite products in formulas (2.10)--(2.14) converge for all rational $x$ and 
$y$ as only finite number of factors in them are different from 1. 
Formulas of such kind are called {\it adelic}. 

{\rm Denote:} $\Omega (|x|_p)$ is the characteristic function of disk $B_0$, so
$\Omega (t)=1$, if $0\leq t\leq 1$ and $\Omega (t)=0$, if $t>1$;
$\delta (|x|_p-p^\g)$ is the characteristic function of circumference $S_\g$; 
$\delta (x_\ell-k)$ is the characteristic function of the set 
$[x\in\QQ_p: x_\ell=k],$ $k=1,2,\ldots,p-1$ for $\ell=0$ and $k=0,1,\ldots,p-1$
for $\ell=1,2,\ldots.$
\bigskip
\centerline{\bf \Par 3. Analytic Functions}
\medskip

Let $\script{O}$ be an open set in $\QQ_p$. Function 
$f:\script{O}\rightarrow\QQ_p$ is called {\it analytic} in $\script{O}$ if
for any point $a\in\script{O}$ there exists a $\g\in Z$ such that in the disk
$B_\g(a)\subset\script{O}$ it is represented by a convergent power series
$$f(x)=\sum_{k=o}^\infty c_k(x-a)^k. \eqno(3.1)$$

{\it Radius of convergence} $r=r(f)$ of the series (3.1) is
$$r=p^\s, \quad \s=-{1\over{\ln p}}\overline{\lim}_{k\to\infty}{1\over k}\ln |f_k|_p.$$
The series (3.1) converges if, and only if, the series
$$\sum_{k=o}^\infty |c_k|p^{\g k}$$
converges, and it is possible to differentiate it term by term in $B_\g (a)$ 
infinite numbers of times,
$$f^{(n)}(x)=\sum_{k=n}^\infty k(k-1)\ldots (k-n+1)c_k(x-a)^{k-n}, \quad n=1,2,\ldots, \eqno(3.2)$$
and also
$$c_k={{f^{(k)}(a)}\over{k!}}, \quad k=0,1,\ldots. \eqno (3.3)$$
By every differentiation of series (3.1) the radius of convergence of the
differentiated series (3.2) may only increase.

The functions $e^x$, $\ln x$, $\sin x$, $\cos x$, $\tg x$, $\arcsin x$, 
$\arctg x$ are analytic, they are defined by the following series
$$e^x=\sum_{k=0}^\infty{{x^k}\over{k!}}, \quad x\in G_p, \eqno(3.4)$$
$$\ln x=\ln [1-(1-x)], \quad x\in J_p; \quad \ln x=-\sum_{k=1}^\infty{{x^k}\over k}, \quad x\in G_p, \eqno(3.5)$$
$$\sin x=\sum_{k=0}^\infty{{(-1)^k}\over{(2k+1)!}}x^{2k+1}, \quad x\in G_p, \eqno(3.6)$$
$$\cos x=\sum_{k=0}{{(-1)^k}\over{(2k)!}}x^{2k}, \quad x\in G_p, \eqno(3.7)$$
$$\tg x={{\sin x}\over{\cos x}}, \quad x\in G_p, \eqno(3.8)$$
$$\arcsin x=\sum_{k=o}^\infty {{(2k)!}\over{2^{2k}(k!)^2(2k+1)}}x^{2k+1}, \quad x\in G_p, \eqno(3.9)$$
$$\arctg x=\sum_{k=0}^\infty{{(-1)^k}\over{2k+1}}x^{2k+1}, \quad x\in G_p. \eqno(3.10)$$    
The following relations are valid
$$(e^x)'=e^x, \quad e^xe^y=e^{x+y}, \quad x, y\in G_p, \eqno (3.11)$$
$$|e^x|_p=1, \quad |e^x-1|_p=|x|_p, \quad x\in G_p, \eqno (3.12)$$
$$\ln (xy)=\ln x+\ln y, \quad x, y\in G_p, \eqno (3.13)$$
$$|\ln (1+x)|_p=|x|_p, \quad x\in G_p, \eqno (3.14)$$
$$\ln e^x=x, \quad x\in G_p; \quad e^{\ln x}=x, \quad x\in J_p. \eqno (3.15)$$

The function $e^x$ realizes the analytic diffeomorphism of additive group $G_p$
onto multiplicative group $J_p$. The invers map is realized by the function 
$\ln x$.

All formulas of classical trigonometry are valid. Their proofs easily follow 
from the formal relation
$$e^{ix}=\cos x+i\sin x, \quad x\in G_p \eqno (3.16)$$
where the symbol $e^{ix}$ is defined by series (3.3) provided that $i^2=-1.$
In particular,
$$\sin^2x+\cos^2x=1, \quad x\in G_p. \eqno (3.17)$$
$$e^{\t x}=\cos x+\t\sin x, \quad x\in G_p, \quad \t^2=-1, \t\in\QQ_p \eqno (3.18)$$ 
(the last is possible only for $p\equiv 1(\mod 4)$).

Functions $\sin x$ and $\tg x$ realize the analytic isomorphysm of group $G_p$
onto $G_p$; invers maps are given by functions $\arcsin x$ and $\arctg x$ 
respect.
\bigskip
\centerline{\bf \Par 4. The Haar Measure on $\QQ_p$.}
\medskip

As $\QQ_p$ is a commutative group on addition so on it there exists an ivariant
measure (uniqe up to a factor), the Haar measure, which we denote by $d_px$,
$$d_p(x+a)=d_px, a\in\QQ_p; \quad d_p(ax)=|a|_pd_px, a\in\QQ_p^\times.$$
Normalize the measure $d_px$ by the condition
$$\int_{Z_p}d_px=1. \eqno (4.1)$$
The normed Haar measure $d_p^\times x$ on $\QQ_p^\times$ is
$$d_p^\times x=(1-p^{-1})^{-1}{{d_px}\over{|x|_p}}, \quad d_p^\times (ax)=d_p^\times x, a,x\in\QQ_p^\times \eqno(4.2)$$
so
$$\int_{Z_p^\times }d_p^\times x=1.$$

Let $M\subset\QQ_p$ be a measurable set (on the Haar measure). Integral of a
function $f:M\rightarrow\CC$ on the set $M$ we will write in the form
$$\int_M f(x)d_px, \quad \int f(x)d_px=\int_{\QQ_p}f(x)d_px.$$
Let $1\leq q\leq\infty$ be. The set of functions $f:\QQ_p\rightarrow\CC$ for
which $f(x)=0, x\not\in M$ and
$$\|f\|_q=\Bigl [\int_M |f(x)|^qd_px\Bigr ]^{1/q}<\infty, \hbox{ if } q<\infty,$$
$$\|f\|_\infty =\vraisup_{x\in M}|f(x)|<\infty, \hbox{ if } q=\infty,$$
we denote by $\script{L}^q(M), \quad \script{L}^q=\script{L}^q(\QQ_p)$. If
$\script{O}$ is an open set in $\QQ_p$ then the set of functions 
$f:\script{O}\rightarrow\CC$ for which for any compact 
$K\subset\script{O} \quad f\in \script{L}^q(K)$ we denote by
$\script{L}_{\loc}^q(\script{O})$, 
$\script{L}_{\loc}^q=\script{L}_{\loc}^q(\QQ_p)$. 

Functions of the set $\script{L}_{\loc}^1(\script{O})$ are called
{\it locally-integrable} in $\script{O}$.

Let a function $f$ be in $\script{L}_{\loc}^1(\QQ_p^\times)$. {\it (Improper)
integral} of a function $f$ on $\QQ_p$, 
$$\int f(x)d_px=\sum_{\g=-\infty}^{\infty}\int_{S_\g}f(x)d_px,$$ 
is called the limit (if it exists)
$$\lim_{N,M\to\infty}\int_{B_N\backslash B_{-M-1}}=\lim_{N,M\to\infty}\sum_{\g=-M}^N\int_{S_\g}f(x)dx.$$

{\bf Example.} Integral
$$\int_{Z_p}|x|^{\a-1}d_px={{1-p^{-1}}\over{1-p^{-\a}}} \eqno (4.3)$$
exists for $\Re\a>0.$

{\bf The formula of change of variables in integral:} {\it if $x(y)$ is 
an analytic diffeomorphism of a clopen set $D'\subset\QQ_p$ onto 
$D\subset\QQ_p$, and also $x'(y)\neq 0, y\in D',$ then for any 
$f\in\script{L}^1(D)$ the formula is valid}
$$\int_D f(x)d_px=\int_{D'} f(x(y))|x'(y)|_pd_py. \eqno (4.4)$$ 

{\bf Example}. Let $x=(py)^{-1}, d_p x=p|y|_p^{-2}d_py$ be. Then owing to (4.3)
we have
$$\int_{|x|_p>1}|x|_p^{\a-1}d_px=p^\a\int_{Z_p}|y|_p^{-\a-1}d_py={{1-p^{-1}}\over{p^{-\a}-1}}, \Re\a<0.$$

{\bf Example}. The linear-fractional transformation is
$$x={{ay+b}\over{cy+d}}, \quad \pmatrix
a & b \cr
c & d \cr
\endpmatrix \in GL(\QQ_p,2),
$$
$$d_p x={{|ad-bc|_p}\over{|cx+d|_p^2}}d_p y.$$

\bigskip
\centerline{\bf \Par 5. $n$-Dimensional Space $\QQ_p^n$}
\medskip

Space $\QQ_p^n=\QQ_p\times\QQ_p\times\ldots\times\QQ_p$ ($n$ times) 
consists of points $x=(x_1,x_2,\ldots,x_n), x_j\in\QQ_p, j=1,2,\ldots,n$ 
supplied with the norm
$$|x|_p=\max_{1\leq j\leq n} |x_j|_p. \eqno (5.1)$$
This norm possesses properties 1)--3) \Par 1 so the space $\QQ_p^n$ is
ultrametric (non-Archimedian).

{\it Scalar product}
$$(x,y)=x_1y_1+x_2y_2+\ldots+x_ny_n, \quad x,y\in\QQ_p^n$$
satisfies the inequality
$$|(x,y)|_p\leq |x|_p|y|_p, \quad x,y\in\QQ_p^n.$$

We denote the Haar measure on $\QQ_p^n$ by $d_p^nx=d_px_1d_px_2\ldots d_px_n$,
$d_px_1=d_px,$
$$d_p^n(x+a)=d_p^nx, a\in\QQ_p^n, \quad d_p^n(Ax)=|\det A|_pd_p^nx$$
where $x\to Ax$ is a linear isomorphism of $\QQ_p^n$ onto 
$\QQ_p^n$ ($\det A\neq 0$).

Henceforth we agree in integrals on whole space $\QQ_p^n$ to omit a domain of
inegration,
$$\int_{\QQ_p^n}f(x)d_p^nx=\int f(x)d_p^nx.$$
Spaces of functions $\script{L}^q(M)$ and $\script{L}_{loc}^q(\script{O})$, 
$M,\script{O}\in\QQ_p^n$ are defined analogously to the case $n=1$ (see 
\Par 4).

As in the case $n=1$ with the help of the notions introduced we define: 
$B_{\g}^n(a)$ is the ball of radius $p^\g$ with the center at point 
$a=(a_1,a_2,\ldots,a_n)\in\QQ_p^n$ and $S_{\g}^n(a)$ is the sphere of radius 
$p^{\g}$ with the center at point $a; B_{\g}^n(0)=B_{\g}^n$, 
$B_{\g}^1(a)=B_{\g}(a)$, $S_{\g}^n(o)=S_{\g}^n$, $S_{\g}^1(a)=S_{\g}(a),$
$$B_{\g}^n(a)=B_{\g}(a_1)\times B_{\g}(a_2)\times\ldots\times B_{\g}(a_n).$$

{\bf The Fubini theorem.} {\it If a function $f:\QQ_p^{n+m}\to\CC$ is such that
the repeated integral
$$\int\Bigr [\int |f(x,y)|d_p^my\Bigl ]d_p^nx$$
exists then $f$ is in $\script{L}^1(\QQ_p^{n+m})$ and the aqualities are 
valid}
$$\int\Bigr [\int f(x,y)d_p^my\Bigl ]d_p^nx=\int f(x,y)d_p^nxd_p^my=\int\Bigr [\int f(x,y)d_p^nx\Bigl ]d_p^my. \eqno (5.2)$$

{\bf Change of variables.} {\it It $x=x(y)$ is an analytic diffeomorphism of a
clopen set $D'\subset\QQ_p^n$ onto set $D\subset\QQ_p^n$ and also
$$\det{{\partial x(y)}\over{\partial y}}=\det\biggl ({{\partial x_k}\over {\partial y_j}}\biggr )\neq 0, y\in D'$$
then for any $f\in\script{L}^1(D)$ the equality is valid}
$$\int_Df(x)d_p^nx=\int_{D'}f(x(y))|\det{{\partial x(y)}\over{\partial y}}|_pd_p^ny. \eqno (5.3)$$ 

{\bf The Lebesgue theorem on passage to the limit under the sign of integral.}
{\it If a sequence ${f_k, k\to \infty}$ of functions $f_k\in\script{L}^1$ 
converges almost everywhere to a function $f(x)$ and there exists a function
$\psi\in\script{L}^1$ such that
$$|f_k(x)|\leq\psi (x) \hbox { for almost every } x\in\QQ_p^n$$
then the equality is valid}
$$\lim_{k\to\infty}\int f_k(x)d_p^nx=\int f(x)d_p^nx.$$
\bigskip
\centerline{\bf \Par 6. Generalized Functions on $\QQ_p^n$}
\medskip

Let $\script{O}$ be an open set in $\QQ_p^n$. A function 
$\varphi:\script{O}\to\CC$ is called {\it locally-constant in} $\script{O}$
if for any point $x\in\script{O}$ there exists $\g\in Z$ such that
$$\varphi (x+x')=\varphi (x), x'\in B_\g^n, \quad x\in\script{O}.$$
The set of all locally-constant functions in $\script{O}$ we denote by 
$\script{E}(\script{O})$; $\script{E}=\script{E}(\QQ_p^n).$ Every functions in
$\varphi\in\script{E}(\script{O})$ is continuous on $\script{O}.$ Its support,
which is the closure of points $x\in\script{O}$ for which $\varphi (x)\neq 0$,
we will denote by $\spt\varphi.$

{\bf Examples.} 
$$|x|_p\in\script{E}(\QQ_p^n\backslash\{0\}),$$
$$\chi_p ((\xi,x))\in\script{E}, \quad \xi\in\QQ_p^n.$$
A function $\varphi\in\script{E}(\script{O})$ is called {\it test function in}
$\script{O}$ (the Bruhat-Schwartz function) if its support is compact in 
$\script{O}.$ The set of test functions in $\script{O}$ we denote by
$\script{S}(\script{O});$ $\script{S}=\script{S}(\QQ_p^n).$ Every function
in $\script{S}(\script{O})$ is uniformly locally-constant in $\script{O}.$

{\bf Examples.}
$$\Omega_k(x)=\Omega (p^{-k}|x|_p)\in\script{S}, \quad k\in Z, \eqno (6.1)$$
$$\Delta_k(x)=p^k\Omega (p^k|x|_p)\in\script{S}, \quad k\in Z. \eqno (6.2)$$
$$|x|_p\Omega (|x|_p)\in\script{S}(\QQ_p^n\backslash\{0\}).$$ 
$$\chi_p((\xi,x))\Omega (|x|_p)\in\script{S}, \quad \xi\in\QQ_p^n.$$
$$\delta (|x|_p-p^\g)\in\script{S}(S_\g), \quad \g\in Z.$$
$$\delta (|x|_p-p^\g)\delta (x_0-k)\in\script{S}(S_\g), k=1,2,\ldots,p-1, \quad \g\in Z.$$
If $K$ is an open compact in $\QQ_p^n$ then $\theta_K$ is in $\script{S}(K).$ 
Here $\theta_M$ is the characteristic function of a set $M\subset\QQ_p^n:$
$\theta_M(x)=1, x\in M,$ $\theta_M(x)=0, x\not\in M.$

{\it Convergence in} $\script{S}(\script{O})$,
$$\varphi_k\to 0, k\to\infty \hbox{ ¢ } \script{S}(\script{O}),$$
means:

(i) there exists a compact $K\subset\script{O}$ not depending on $k$ such that 
$\spt\varphi_k\subset K$; 
  
(ii) there exists $\g\in Z$ depending neither $k$ nor $x$ such that 
$$\varphi_k(x+x')=\varphi_k(x), x'\in B_\g^n, \quad x\in K;$$

$(iii) \varphi_k(x)\Rightarrow 0, x\in K, k\to\infty.$

{\it Generalized function} on $\script{O}$ is called any linear continuous 
functional $f:\varphi\to (f,\varphi)$ on $\script{S}(\script{O}).$ The set of 
all generalized functions on $\script{O}$ we denote by 
$\script{S}'(\script{O}); \script{S}'=\script{S}'(\QQ_p^n).$

{\it Convergence in} $\script{S}'(\script{O}),$ 
$$f_k\to 0, k\to\infty \hbox{ in } \script{S}'(\script{O}),$$ 
is defined as the weak convergence of functionals in 
$\script{S}'(\script{O}),$ that is 
$$(f_k,\varphi )\to 0, k\to\infty, \quad \varphi\in\script{S}(\script{O}).$$

{\it Every linear on $\script{S}(\script{O})$ functional $f$ is continuous on
$\script{S}(\script{O})$}, that is $f\in\script{S}'(\script{O}).$

In an open set $\script{O}$ there exists {\it "decomposition of unity"} with
functions in $\script{S}(\script{O})$, namely if
$$\script{O}=\cup_{k=1}^\infty G_k, \quad G_k\cap G_j=\emptyset, k\neq j$$
where $G_k, k=1,2,\ldots$ are clopen compacts, so the equality holds 
$$\sum_{k=1}^\infty\theta_{G_k}(x)=1, \quad x\in\script{O}. \eqno (6.3)$$
A generalized function $f$ in $\script{S}'(\script{O})$ {\it vanishes in an
open set} $\script{O}'\subset\script{O}$ if 
$(f,\varphi )=0, \varphi\in\script{S}(\script{O}'),$ besides we write: 
$f(x)=0, x\in\script{O}'$. Generalized functions $f$ and $g$ in
$\script{S}'(\script{O})$ {\it coincide in} (equal in) 
$\script{O}'\subset\script{O}$, $f=g$ in $\script{O}'$, iff $f(x)-g(x)=0$ 
for $x\in\script{O}'$. The largest open set in which vanishes 
$f\in\script{S}'(\script{O})$ is called {\it null-set} of $f$, and it is 
denoted by $\script{O}_f\subset\script{O}.$ A closed in $\script{O}$ set 
$\script{O}\backslash\script{O}_f$ is called {\it support of $f$}, and it is
denoted by $\spt f$, $\spt f=\script{O}\backslash\script{O}_f$. 

We denote the set of generalized functions with compact support in 
$\script{O}$ by $\script{E}'(\script{O})$, $\script{E}'=\script{E}'(\QQ_p^n)$; 
$\script{E}'(\script{O})$ is the strongly conjugate space to 
$\script{E}(\script{O})$.

{\bf Example.} {\it $\delta$-Function}
$$(\delta,\varphi )=\varphi (0), \quad \spt\delta=\{0\}. \eqno (6.4)$$
Conversely, every $f\in\script{S}', \spt f=\{0\}$ has the form
$$f=C\delta\eqno (6.5)$$
where $C\neq 0$ is an arbitrary constant.

A sequence $\{\delta_k, k\to\infty\}$ of functions $\delta_k(x)$ in
$\script{S}$ is called {\it $\delta-$like} if it is bounded in
$\script{L}^1$ and for any $\g\in Z$ the limit relation holds
$$\int_{B_\g^n}\delta_k(x)d_p^nx\to 1, \quad \int_{\QQ_p^n\backslash B_\g^n}|\delta_k(x)|d_p^nx\to 0, \quad k\to\infty.$$
Thus,
$$\delta_k\to\delta, k\to\infty \hbox{ in } \script{S}'. \eqno (6.6)$$

A sequence $\{\omega_k, k\to\infty\}$ of functions $\omega_k(x)$ in 
$\script{S}$ is called {\it 1-like} if it is the Fourier-transform (see below
\Par 7) of some $\delta-$like sequence $\{\delta_k, k\to\infty\}$. 

1-Like sequence is bounded in $\script{L}^\infty$, and for any$\g\in Z$ 
$$\omega_k(x)\Rightarrow 1, x\in B_\g^n, k\to\infty.$$
Thus,
$$\omega_k\to 1, k\to\infty \hbox{ in } \script{S}'. \eqno (6.7)$$
If $f\in\script{L}_{\loc}^1(\script{O})$ so $f\in\script{S}'(\script{O}),$ 
besides 
$$(f,\varphi)=\int f(x)\varphi (x)d_p^nx, \quad \varphi\in\script{S}(\script{O}). \eqno (6.8)$$

Generalized functions of the form (6.8) are called {\it regular} in 
$\script{O};$ the others are called {\it singular.} $\delta$-Function is
singular in $\QQ_p^n$, and it is regular in $\QQ_p^n\backslash\{0\}.$

Let 0 be in $\script{O}.$ If $f\in\script{S}'(\script{O}\backslash\{0\})$ then
it admit an {\it extension (regularization)} $f_1\in\script{S}'(\script{O})$ on
$\script{O}$ and all its regularizations, $\reg f,$ are given by the formula
$$\reg f=f_1+C\delta, \eqno (6.9)$$
where $C$ is an arbitrary constant and $f_1$ can be choosen in the form
$$(f_1,\varphi )=(f,\varphi-\Omega_\g\varphi (0)), \quad \varphi\in\script{S}(\script{O}),$$
besides $\g\in Z$ is such that $B_\g^n\subset\script{O}.$ Note, that this fact
does not take place for generalized functions of real arguments! As an example
of such $f$ is function $f(x)=\exp x^{-1}.$ 

For $f=|x|_p^{-1}$ as a regularization it is possible to take the functional
$$(\reg |x|_p^{-1},\varphi )=\int |x|_p^{-1}[\varphi (x)-\Omega (|x|_p)\varphi (0)]d_px, \quad \varphi\in\script{S}.$$
The generalized function $\reg |x|_p^{-1}$ gives another example of singular
generalized function on $\QQ_p^n$.

{\bf The product} of a generalized function $f\in\script{S}'(\script{O})$ on a 
function $a\in\script{E}(\script{O})$ is defined by the formula
$$(af,\varphi)=(f,a\varphi), \varphi\in\script{S}(\script{O}), \quad af\in\script{S}'(\script{O}). \eqno (6.10)$$

{\bf Examples.}
$$a(x)\delta (x)=a(0)\delta (x).$$
If $f\in\script{L}_{loc}^1(\script{O})$, so $af$ coincides with ussual product
of functions $a(x)$ and $f(x).$

If $f\in\script{S}'(\script{O})$ and $\spt f$ is clopen set in $\script{O}$, so
$$f(x)=\t_{\spt f}(x)f(x). \eqno (6.11)$$
Finelly, if $f\in\script{S}'$, so
$${\textstyle\omega_k} f\to f, k\to\infty \hbox{ in } \script{S}' \eqno (6.12)$$
where $\{\omega_k, k\to\infty\}$ is any 1-like sequence.

In $\script{S}'(\script{O})$ {\it theorem on "piecewise sewing"} is valid. 
Let a collection of generalized functions 
$f_k\in\script{S}'(G_k), k=1,2,\ldots$ be given where $G_k, k=1,2,\ldots$ are
clopen compacts satisfying conditions $G_k\cap G_j=\emptyset, k\neq j.$ Then
there exists a (unique) generalized function $f\in\script{S}'(\script{O})$ such
that $f=f_k$ in $G_k, k=1,2,\ldots$ and $\script{O}=\cup_{k\leq 1}G_k.$
 
{\bf The®rem on "nucleus".} {\it Let $\varphi\to A(\varphi )$ be a linear map
of $\script{S}(\script{O}), \script{O}\in\QQ_p^n$ into
$\script{S}'(\script{O}'), \script{O}'\in\QQ_p^m.$ Then there exists a (unique)
generalized function $f\in\script{S}'(\script{O}\times\script{O}')$ such that}
$$\Bigl (A(\varphi),\psi\Bigr )=(f,\varphi (x)\psi (y)), \varphi\in\script{S}(\script{O}), \psi\in\script{S}(\script{O}').$$

The spaces $\script{S}(\script{O})$ and $\script{S}'(\script{O})$ are complete,
reflexive and nuclear; $\script{S}(\script{O})$ is dense in 
$\script{S}'(\script{O}).$
 
{\bf Linear change of variables} $y=Ax+b, \quad \det A\neq 0,$ maps a 
generalized function $f(y)$ in $\script{S}'(\script{O}')$ in the generalized
function $f(Ax+b)$ in $\script{S}'(\script{O})$ by the formula
$$\Bigl (f(Ax+b),\varphi\Bigr )={1\over{|\det A|_p}}\Bigl (f(y),\varphi (A^{-1}(y-b))\Bigr ), \quad \varphi\in\script{S}(\script{O}). \eqno (6.13)$$ 

{\bf Examples.} $\delta (x)=\delta (-x),$ 
\quad $(\delta (x-x_0),\varphi )=\varphi (x_0).$ 

{\bf The direct product} $f(x)\times g(y)$ of generalized functions 
$f\in\script{S}'(\script{O}_1)$, $\script{O}_1\subset\QQ_p^n$ and 
$g\in\script{S}'(\script{O}_2),$ $\script{O}_2\subset\QQ_p^m$ is defined by the
formula
$$(f(x)\times g(y),\varphi )=\bigl (f(x),(g(y),\varphi (x,y))\bigr ), \quad \varphi\in\script{S}(\script{O}_1\times\script{O}_2).$$
The direct product is {\it commutative}, so
$$f(x)\times g(y)=g(y)\times f(x)\in\script{S}'(\script{O}_1\times\script{O}_2). \eqno (6.14)$$
For $g=1$ the formula (6.14) takes the form
$$\bigl (f(x),\int_{\script{O}_2}\varphi (x,y)d^my\bigr )=\int_{\script{O}_2}\bigl (f(x),\varphi (x,y)\bigr )d^my, \quad f\in\script{S}'(\script{O}_1), $$
$$\quad \varphi\in\script{S}(\script{O}_1\times\script{O}_2) \eqno (6.15)$$
(generalization of the Fubini theorem, see \Par 5).

{\bf Convolution} $f\ast g$ of generalized functions
$f\in\script{E}, \spt f\in B_N^n$ and $g\in\script{S}'$ is defined by the 
equality
$$(f\ast g,\varphi )=(f(x)\times g(y),\Omega_N(x)\varphi (x+y)), \quad \varphi\in\script{S}. \eqno(6.16)$$
On the base of this definition the convolution of generalized functions $f$ and
$g$ in $\script{S}'$ is defined by
$$(f\ast g,\varphi )=\lim_{k\to\infty}(f(x)\times g(y),\Omega_k (x)\varphi (x+y)=\lim_{k\to\infty}((\Omega_k f)\ast g,\varphi )$$
if the limit exists for any $\varphi\in\script{S}$, so $f\ast g\in\script{S}'.$

If the convolution $f\ast g$ exists then the convolution $g\ast f$ also exists
and they both are equal ({\it commutativity of convolution}),
$$f\ast g=g\ast f. \eqno (6.17)$$

{\bf Examples.}
$$f\ast\delta=\delta\ast f=f, \quad f\in\script{S}'. \eqno (6.18)$$
If $f\in\script{S}'$ and $\psi\in\script{S}$  then the convolution $f\ast\psi$
is a locally-constant function in $\QQ_p^n$, besides 
$$(f\ast\psi )(x)=(f(y),\psi (x-y)), x\in\QQ_p^n. \eqno (6.19)$$
If $\{\delta_k,k\to\infty\}$ is a $\delta-$like sequence then
$$f\ast\delta_k\to f, k\to\infty \hbox{ in } \script{S}', \quad f\in\script{S}'. \eqno (6.20)$$
If $f,g$ in $\script{L}_{\loc}^1$ and there exists a function 
$q\in\script{L}_{\loc}^1$ such that 
$$\int_{B_k}f(x-y)g(y)d_p^ny\to q(x), k\to\infty \hbox{ in } \script{S}'$$
then
$$f\ast g=q(x). \eqno (6.21)$$
If $f\in\script{S}'$ and the convoluton $f\ast 1$ exists then it is a constant.
We call this constant {\it integral} of generalized function $f$ on the whole
space $\QQ_p^n$, and we denote it by
$$G\!\!\!\!\int f(x)d_p^nx=f\ast 1. \eqno (6.22)$$
This definition is equivalent to the following:
$$G\!\!\!\!\int f(x)d_p^nx=\lim_{k\to\infty}(f,\Omega_k) \eqno (6.23)$$
if the limit exists.

If $f\in\script{S}'$ and $\spt f\subset D$ where $D$ is a clopen set in 
$\QQ_p^n$, so $f=\t_Df$, and the integral (6.22) we denote by
$$G\!\!\!\!\int_Df(x)d_p^nx.$$
In particular, if $f\in\script{S}'$, $\varphi\in\script{S}$ and 
$\spt\varphi\subset B_\g$, so
$$G\!\!\!\!\int_{B_\g}f(x)\varphi (x)d_p^nx=(f,\varphi ). \eqno (6.24)$$
If $f\in\script{S}', \spt f\subset B_\g$, so
$$G\!\!\!\!\int_{B_\g}f(x)d_p^nx=(f,\Omega_\g). \eqno (6.25)$$  
The notion of integral of a generalized function introduced is in fact an 
extension of the notion of integral on the Haar measure (see \Par\Par 1,4).

{\bf Example.} 
$$G\!\!\!\!\int\delta (x)d_px=1.$$

{\bf Multipliation of generalized functions.} Let $f,g$ be in $\script{S}'$. 
We call {\it product} $f\cdot g$ the functional defined by the equality
$$f\cdot g=\lim_{k\to\infty}(f\ast\Delta_k)g$$
if the limit exists in $\script{S}'$, so $f\cdot g\in\script{S}'$. 

If the product $f\cdot g$ exists, so the product $g\cdot f$ also exists and they
are equal ({\it commutativity of product})
$$f\cdot g=g\cdot f. \eqno (6.26)$$

{\bf Examples.}
$$a\cdot f=af, \quad a\in\script{E}, f\in\script{S}'.$$
In particular,
$$f\cdot 1=1\cdot f=f, \quad f\in\script{S}',$$
$$a(x)\cdot\delta (x)=a(0)\delta (x)$$
if $a$ is a continuous function in a vicinity of 0,
$$|x|_p^\a\cdot\delta (x)=0, \a>0, \quad |x|_p\cdot\reg |x|_p^{-1}=1. \eqno(6.27)$$
As in the case of real field, a question arises: is it possible to define the
product of any generalized functions by such a way that it was associative and
commutative? The answer is negative. Well-known example by L.~Schwartz in 
$p$-adic case seems so. If such product would exist so owing to (6.27) we would 
have the following contradictory chain of equalities:
$$0=0\cdot\reg |x|_p^{-1}=(|x|_p\cdot\delta (x))\cdot\reg |x|_p^{-1}=\delta (x)\cdot (|x|_p\cdot\reg |x|_p^{-1})=\delta (x)\cdot 1=\delta (x).$$
\bigskip
\centerline {\bf \Par 7. The Fourier Transform}
\medskip
Let $\varphi$ be in $\script{S}.$ The {\it Fourier transform} 
$\tilde\varphi =F[\varphi]$ is defined by the formula
$$\tilde{\varphi}(\xi )=\int\varphi (x)\chi_p((\xi,x))d_p^nx, \quad x\in\QQ_p^n.$$
The Fourier transform is a linear isomorphism of $\script{S}$ onto $\script{S}$
and the {\it inversion formula} for the Fourier transform is valid
$$\varphi (x)=\int\tilde{\varphi}(\xi )\chi_p(-(x,\xi))d_p^n\xi, \quad \varphi\in\script{S}.$$
{\bf Examples.} 
$$\tilde{\Omega}_k=\Delta_k, \quad \tilde{\Delta}_k=\Omega_k, \quad k\in Z. \eqno (7.1)$$

The Fourier transform $\tilde f=F[f]$ of a generalized function
$f\in\script{S}'$ is defined by the formula
$$(\tilde f,\varphi )=(f,\tilde{\varphi} ), \quad \varphi\in\script{S},$$
so that $\tilde f\in\script{S}'$.

The Fourier transform $f\to\tilde f$ is a linear isomorphism of $\script{S}'$ 
onto $\script{S}'$ and the inversion formula is valid
$$f=F^{-1}[\tilde f]=F[\check{\tilde f}], \quad f\in\script{S}'$$
where $\check f(x)=f(-x).$

{\bf Examples.}
$$\tilde\delta=1, \quad \tilde 1=\delta; \eqno (7.2)$$
$$F[f(Ax+b)]=|\det A|_p^{-1}\chi_p\bigl (-(A^{-1}b,\xi )\bigr )F[f(A^{-1}\xi )], \quad \det A\neq 0. \eqno (7.3)$$
In particular,
$$F[f(x-b)]=\chi_p ((b,\xi ))F[f(\xi )]; \eqno (7.4)$$
$$\tilde{\check f}=\check{\tilde f}. \eqno (7.5)$$
If $f\in\script{L}^1$ then
$$\tilde f(\xi )=\int f(x)\chi_p((\xi,x))d^nx, \eqno (7.6)$$ 
and also $\tilde f$ is continuous in $\QQ_p^n$ and $\tilde f(\xi)\to 0,$ 
$|\xi |_p\to\infty$ (analogy of the {\it Riemann-Lebesgue theorem}).

If $f\in\script{L}_{\loc}^1$ and there exists a function 
$q\in\script{L}_{\loc}^1$ such that
$$\int_{B_k^n}f(x)\chi_p((\xi,x))d_p^nx\to q(\xi), k\to\infty \hbox{ in } \script{S}'$$
then
$$\tilde f=q. \eqno (7.7)$$
If $f\in\script{S}', \spt f\subset B_\g^n$ then
$$\tilde f(\xi)=\bigl (f(x),\Omega_\g(x)\chi_p((\xi,x))\bigr ). \eqno (7.8)$$
If $f\in\script{L}^2$ then
$$\int_{B_k^n}f(x)\chi_p ((\xi,x))d_p^nx\to\tilde {f}(\xi), \quad k\to\infty \hbox{ in } \script{L}^2. \eqno (7.9)$$
The operator $f\to\tilde f$ is unitary in $\script{L}^2$ so the 
{\it Parseval-Steklov equality} is valid
$$\|f\|=\|\tilde f\|, \quad f\in\script{L}^2 \eqno (7.10)$$
where the norm $\|f\|=\|f\|_2=(f,f)^{1/2}$ is defined in \Par 4 and the
{\it scalar product} $(f,g)$ in $\script{L}^2$ is equal to
$$(f,g)=\int f(x)\bar g(x)d_p^nx, \quad f, g\in\script{L}^2.$$
The {\it Cauchy-Buniakowski inequality} is valid
$$|(f,g)|\leq\|f\|\|g\|, \quad f, g\in\script{L}^2.$$
If $f\in\script{L}^2$ then
$$\lim_{k\to\infty}p^{-{k/2}}\int_{B_k}|f(x)|d_p^n=0. \eqno (7.11)$$

{\bf Theorem.} {\it Let $f,g$ be in $\script{S}'.$ The convolution $f\ast g$ 
exists if, and only if, there exists the product $\tilde f\cdot\tilde g$
and the equalities are valid}
$$\widetilde{f\ast g}=\tilde f\cdot\tilde g, \quad \widetilde{f\cdot g}=\tilde f\ast\tilde g. \eqno (7.12)$$

Note the following useful formula
$$\int_{S_\g^n}\chi_p((x,\xi ))d_p^nx=(1-p^{-n})p^{\g n}\Omega (p^\g |\xi |_p)-q^{(k-1)n}\delta (|\xi |_p-p^{1-\g}) \eqno(7.13)$$
whence
$$\int_{B_\g^n}\chi_p((x,\xi ))d_p^nx=p^{\g n}\Omega (p^\g |\xi |_p). \eqno(7.14)$$
The {\it Gaussian integral} $G_p(a;\xi )$ is called the Fourier transform of the
function $\chi_p(ax^2), a\in\QQ_p^\times, p=\infty,2,3,5,\ldots,$
$$G_p(a,\xi )=\int\chi_p(ax^2+\xi x)d_px=\lambda_p(a)|2a|_p^{-{1/2}}\chi_p\bigr (-{{\xi^2}/{4a}}\bigl ). \eqno (7.15)$$
The following adelic formula is valid
$$G_\infty (a;\xi )\prod_{p=2}^\infty G_p(a;\xi )=1, \quad a\in\QQ^\times, \xi\in \QQ \eqno (7.16)$$
which follows from the adelic formulae (2.10)--(2.12).

\bigskip
\centerline {\bf \Par 8. Homogeneous Generalized Functions}
\medskip
Let $\pi (x)=\pi_{\a,\t}(x)=|x|_p^{\a-1}\t(x)$ be a quasi-character of the 
field $\QQ_p$ (see $(2.5')$). A generalized function $f\in\script{S}'$ is 
called {\it homogeneous} with respect to a quasi-character $\pi_{\a,\t}$ if
$$f(tx)=\pi_{\a,\t}(t)f(x), t\in\QQ_p^\times, \quad x\in\QQ_p^\times. \eqno (8.1)$$
Homogeneous generalized functions with respect to a principal quasi-character
$$\pi_{\a,1}(x)=|x|_p^{\a-1}$$ 
are called homogeneous of degree $\a-1.$

A quasi-character $\pi_{\a,\t}(x)$ defines a homogeneous with respect to itself
generalized function $\pi_{\a,\t}$ by the formula
$$(\pi_{\a,\t},\varphi )=\int |x|_p^{\a-1}\t(x)\varphi (x)d_px, \quad \varphi\in\script{S}. \eqno (8.2)$$
{\it The generalized function $\pi_{\a,\t}$ for $\t\neq 1$ is entire on $\a$; 
for $\t=1$ it is holomorphic on $\a$ everywhere except simple poles
$$\a_k={{2k\pi i}/{\ln p}}, \quad k\in Z$$
with residue ${{1-p^{-1}}\over{\ln p}}\delta (x).$}

Note that the generalized function $|x|_p^{\a-1}$ defined in domain $\Re\a>0$ 
by the formula (8.2) is analytically continued from this domain to the domain
$\Re\a\leq 0, \quad \a\neq\a_k, k\in Z$ by the formula
$$(|x|_p^{\a-1},\varphi)=(1-p^{-\a})^{-1}\int |x|_p^{\a-1}[\varphi (x)-\varphi ({x/p})]d_px$$
$$=\int |x|_p^{\a-1}[\varphi (x)-\varphi (0)]d_px, \quad \varphi\in\script{S} \eqno (8.3)$$
as
$$\int |x|_p^{\a-1}=0, \quad \a\neq\a_k, k\in Z.$$
 For $\a=\a_k, k\in Z$ the quasi-character $\pi_{0,1}(x)=|x|_p^{-1}$ 
corresponds the generalized function $\delta (x)$ of degree $-1$; conversely, 
every homogeneous generalized function $f\in\script{S}'$ of degree $-1$ has the
form $f(x)=C\delta (x)$ where $C$ is some constant.

The Fourier transform of $\pi_{\a,\t}$ is a homogeneous generalized function 
$\tilde{\pi}_{\a,\t}$ with respect to the quasi-character
$$\pi_{\a,\t}^{-1}(\xi )|\xi |_p^{-1}=|\xi |_p^{-\a}\bar{\t}(\xi )=\pi_{1-\a,\bar{\t}}(\xi ), \eqno (8.4)$$
so
$$\tilde{\pi}_{\a,\t}=\G_p(\pi_{\a,\t})\pi_{1-\a,\bar{\t}}. \eqno (8.5)$$
Here $\G_p(\pi_{\a,\t})$ is {\it gamma-function} of field $\QQ_p$ for 
quasi-character $\pi_{\a,\t}(x)$,
$$\G_p(\pi_{\a,\t})=\tilde{\pi}_{\a,\t}(1)=\int |x|_p^{\a-1}\t (x)\chi_p(x)d_px. \eqno (8.6)$$
In particular, for $\t=1$, if we denote
$$\G_p(\a)=\G_p(|x|_p^{\a-1}),$$
we get for the gamma-function $\G_p(\a)$ of a principal quasi-character
$|x|_p^{\a-1}$ the representation 
$$\G_p(\a)=\int |x|_p^{\a-1}\chi_p(x)d_px={{1-p^{\a-1}}\over{1-p^{-\a}}}, \quad \a\neq\a_k, k\in Z. \eqno (8.7)$$
For $\epsilon\not\in\QQ_p^{\times 2}, |\epsilon|_p=1, p\neq 2$
$$\t (x)={\textstyle\sgn_{p,\epsilon}}x=|x|_p^{{\pi i}/{\ln p}}=(-1)^{\g ( x)}$$
we denote
$$\tilde{\G}_p(\a)=\G_p(|x|_p^{\a-1}{\textstyle\sgn_{p,\epsilon}}x).$$
For $\tilde{\G}_p$-function from (8.7) it follows the expression
$$\tilde{\G}_p(\a)=\G_p(\a+{{\pi i}/{\ln p}})={{1+p^{\a-1}}\over{1+p^{-\a}}}, \quad \a\neq\a_k-{{\pi i}/{\ln p}}, k\in Z. \eqno(8.8)$$

Note the particular formulae for gamma-function when d=-1 (cf. \Par2),
$$\G_p({\textstyle\sgn_{p,-1}}x|x|^{\a-1})=\cases{\G_p(\a)={{1-p^{\a-1}}\over{1-p^{-\a}}}, p\equiv 1 (\mod 4),} \\
{\tilde{\G}_p(\a)={{1+p^{\a-1}}\over{1+p^{-\a}}}, p\equiv 3 (\mod 4),} \\ {2i4^{\a-1}, p=2.} \endcases$$
 
The following equality is valid
$$\G_p(\pi_{\a,\t})\G_p(\pi_{1-\a,\bar{\t}})=\t(-1). \eqno (8.9)$$
In particular,
$$\G_p(\a)\G_p(1-\a)=1. \eqno (8.10)$$

Convolution of homogeneous generalized function $\pi_{\a,\t}$ and 
$\pi_{\b,\t'}$ exists and it is a homogeneous generalized function with respect
to quasi-character
$$\pi_{\a,\t}(x)\pi_{\b,\t'}(x)|x|_p^{-1}=\pi_{\a+\b,\t\t'}(x),$$
and thus
$$\pi_{\a,\t}\ast\pi_{\b,\t'}=B_p(\pi_{\a,\t},\pi_{\b,\t'})\pi_{\a+\b,\t\t'}. \eqno (8.11)$$
Here $B_p(\pi_{\a,\t},\pi_{\b,\t'})$ is {\it beta-function} of field $\QQ_p$
for quasi-characters $\pi_{\a,\t}$ and $\pi_{\b,\t'}$,
$$B_p(\pi_{\a,\t},\pi_{\b,\t'})=\bigl (\pi_{\a,\t}\ast\pi_{\b,\t'}\bigr )(1)={{\G_p(\pi_{\a,\t})\G_p(\pi_{\b,\t'})}\over{\G_p(\pi_{\a+\b,\t\t'})}}$$
$$=\G_p(\pi_{\a,\t})\G_p(\pi_{\b,\t'})\G_p(\pi_{\g,\t''})\t''(-1), \quad \a+\b+\g=1, \t\t'\t''=1. \eqno(8.12)$$
In particular, for principal quasi-characters ($\t=\t'=1$) formula (8.12) 
takes the form
$$B_p(\a,\b)=\G_p(\a)\G_p(\b)\G_p(\g), \quad \a+\b+\g=1 \eqno (8.13)$$
where is denoted
$$B_p(\a,\b)=B_p(|x|_p^{\a-1},|x|_p^{\b-1}).$$

Note another symmetric expression for the beta-function $B_p(\a,\b)$ [13]:
$$B_p(\a,\b)=(1-p^{-1})\bigl [(1-p^{-\a})^{-1}+(1-p^{-\b})^{-1}+(1-p^{-\g})^{-1}-1\bigr ]-1,$$ 
$$\quad \a+\b+\g=1. \eqno (8.14)$$
By introducing the analogy of the Euler gamma- and beta-functions,
$$\g_p(\a)=\int_{Z_p}|x|_p^{\a-1}\chi_p(x)d_px={{1-p^{-1}}\over{1-p^{-\a}}},$$
$$b_p(\a,\b)=\int_{Z_p}|x|_p^{\a-1}|1-x|_p^{\b-1}d_px=\g_p(\a)+\g_p(\b)-1,$$
we get the equality (8.14) in the form
$$B_p(\a,\b)={1\over 2}b_p(\a,\b)+{1\over 2}b_p(\a,\g)+{1\over 2}b_p(\b,\g)+{1\over p}-{1\over 2}, \quad \a+\b+\g=1. \eqno(8.15)$$

{\it Rank} $\rho (\t )$ of a quasi-character $\pi_{\a,\t}$ (and a chracter 
$\t$) is called such integer number $k\geq 0$ that $\t(t)=1$ for 
$|1-t|_p\leq p^{-k}, t\in Z_p^\times$ and $\t(t)\neq 1$ for 
$|1-t|_p=p^{1-k}, t\in Z_p^\times$. It is clear that zero rank has only the
principal quasi-character $|x|_p^\a$.                       

For quasi-characters of the rank $k\geq 1$ the following formulas are valid 
[4]:
$$\G_p(\pi_{\a,\t})=p^{\a k}a_{p,k}(\t), \eqno (8.16)$$
$$a_{p,\g}(\t)=\int_{S_0}\t(t)\chi_p(p^{-\g}t)d_pt, \quad \g\geq 1, \eqno (8.17)$$
$$a_{p,\g}(\t)=0, \g\neq k, \quad |a_{p,k}(\t)|=p^{-{k/2}},\eqno (8.18)$$
$$a_{p,k}(\t)a_{p,k}(\bar\t)=p^{-k}\t(-1), \eqno (8.19)$$
$$\int_{S_k}\t(p^kx)\chi_p(\xi x)d_px=p^ka_{p,k}(\t)\bar\t(\xi )\delta (|\xi|_p-1), \eqno (8.20)$$
$$\G_p(\pi_{\a,\t})\G_p(\pi_{\a,\t}^{-1})=p^k\t(-1), \eqno (8.21)$$
$$\G_p(\pi_{\a+1,\t})=p^k\G_p(\pi_{\a,\t}). \eqno (8.22)$$

{\bf Example.} The rank of a quasi-character
$$\pi_{\a,\t}(x)=|x|_p^{\a-1}{\textstyle\sgn_{p,d}}x, \quad |d|_p=1/p, p\neq 2 \eqno(8.23)$$
is equal 1. Therefore and owing to (8.16) and (8.19)
$$\G_p(\pi_{\a,\t})=\pm p^{\a-{1/2}}\sqrt{{\textstyle\sgn_{p,d}}(-1)}. \eqno (8.24)$$

The operator (8.2)
$$\varphi\to (\pi_{\a,\t},\varphi )\equiv M^\pi[\varphi ]$$
is called {\it the Mellin transform} of function $\varphi\in\script{S}$ with
respect to a quasi-character $\pi_{\a,\t}(x)$. For $\t=1$ the function
$M^{|x|_p^{\a-1}}[\varphi ]\equiv M^\a[\varphi ]$ is called simply the Mellin
transform of a function $\varphi\in\script{S}$. Owing to (8.3) it can be
represented in the following form
$$M^\a[\varphi ]=(1-p^{-\a})^{-1}\int |x|_p^{\a-1}[\varphi (x)-\varphi ({x/p})]d_px, \quad \a\neq\a_k, k\in Z.$$
According to (8.2) and (8.5) the equality takes place 
$$M^\pi[\tilde\varphi ]=\G_p(\pi_{\a,\t})M^{\tilde\pi}[\varphi ] \quad \varphi\in\script{S}. \eqno (8.25)$$
For $\t=1$ formula (8.25) takes the form
$$M^\a[\tilde\varphi ]=\G_p(\a)M^{1-\a}[\varphi ]. \eqno (8.25')$$

{\bf The Mellin transform of $Z_p^\times$-invariant (generalized) functions and
its inversion.} A function $\varphi\in\script{S}(\QQ_p^\times )$ is called
$Z_p^\times$-{\it invariant} if $\varphi (x)=\varphi (t|x|_p),$ 
$t\in Z_p^\times$, $x\in\QQ_p^\times$ or, in other words,
$$\varphi (x)=(1-p^{-1})^{-1}\int_{Z_p}\varphi (t|x|_p)d_pt\equiv S[\varphi ](|x|_p).$$
Every $Z_p^\times$-invariant function $\varphi\in\script{S}(\QQ_p^\times )$
is represented uniquely in the form
$$\varphi (x)=\sum_\g\varphi_\g\delta (|x|_p-p^\g), \quad \varphi_\g=\varphi (p^\g)=S[\varphi ](p^\g).$$
Thus the subspace of space $\script{S}(\QQ_p^\times )$ consisting of
$Z_p^\times$-invariant functions is isomorphic to the space of finite sequences
$\{\varphi_\g,$ $\g\in N\}$ where $N$ is a bounded subset of $Z$.

A generalized function $f\in\script{S}(\QQ_p^\times )$ is called $Z_p^\times$-
invariant if
$$(f,\varphi )=(f(x),S[\varphi ](|x|_p)), \quad \varphi\in\script{S}(\QQ_p^\times ).$$
Any $Z_p^\times$-invariant generalized function 
$\varphi\in\script{S}'(\QQ_p^\times )$ is represented uniquely in the form
$$f(x)=\sum_{\g}f_{\g}\delta (|x|_p-p^\g), \quad f_\g=(1-p^{-1})^{-1}p^{-\g}(f(x),\delta (|x|_p-p^\g)),$$
so a subspace of the space $\script{S}'(\QQ_p^\times) $
consisting of $Z_p^\times$-invariant generalized functions is isomorphic to the
space sequences $\{f_\g, \g\in Z\}$.

If $\varphi\in\script{S}(\QQ_p^\times )$ is $Z_p^\times$-invariant function 
then its Mellin transform
$$M^\a[\varphi ]=\int |x|^{\a-1}S[\varphi ](|x|_p)d_px=(1-p^{-1})\sum_{\g\in M}\varphi_{\g}p^{\a\g}$$
is entire functin of $\a$, and the invertion formula is valid [16]
$$\varphi (x)={{\ln p}\over{2\pi i(1-p^{-1})}}\int_{\s-{{i\pi}/{\ln p}}}^{\s+{{i\pi}/{\ln p}}}M^\a[\varphi ]|x|_p^{-\a}d\a. \eqno(8.26)$$
The formula (8.26) is extended also on $Z_p^\times$-invariant generalized 
functions $f$ from $\script{S}'(\QQ_p^\times )$ satisfying the condition
$$\sum_{\g\in Z}|f_\g|p^{c\g}<\infty$$
for some $c$. Its Mellin transform
$$M^\a[f]=(f(x),|x|_p^{\a-1})=(1-p^{-1})\sum_{\g\in Z}f_{\g}p^{\g\a}$$
is a holomorphic function of $\a$ in half-plane $\Re\a<c$, and the invertion
formula (8.26) is valid for $f$, and also integral (8.26) does not depends on
$\s<c$.

{\bf Space $\QQ_p^n$.} We restrict ourself by the case of a principal    
quasi-character $|x|_p^\a.$ The generalized function $|x|_p^{\a-n}$ is 
homogeneous of degree $\a-n$, holomorphic on $\a$ everywhere exept simple
poles $\a_k={{2k\pi i}/{\ln p}}, k\in Z$ with residue 
${{1-p^{-n}}\over{\ln p}}\delta (x);$ the formula of the Fourier transform is 
valid [10]
$$\widetilde{|x|_p^{\a-n}}=\G_p^{(n)}(\a)|\xi |_p^{-\a}, \quad \a\neq\a_k,k\in Z \eqno (8.27)$$
where $\G_p^{(n)}$ is the gamma-function of vector space $\QQ_p^n$
$(\G_p^{(1)}=\G_p)$,
$$\G_p^{(n)}(\a)=\int |x|_p^{\a-n}\chi_p(x_1)d_p^nx={{1-p^{\a-n}}\over{1-p^{-\a}}}, \quad \a\neq\a_k, k\in Z, \eqno (8.28)$$
$$\G_p^{(n)}(\a)\G_p^{(n)}(n-\a)=1, \eqno (8.29)$$
$$\G_p^{(n)}(\a)=(-1)^{n-1}p^{(n-1)({n/2}-\a)}\prod_{k=1}^{n-1}\G_p(\a-k). \eqno (8.30)$$

Beta-function $B_p^{(n)}$ of space $\QQ_p^n$ is defined similar to (8.11)
($B_p^{(1)}=B_p$) by the equality 
$$|x|_p^{\a-n}\ast |x|_p^{\b-n}=B_p^{(n)}(\a,\b)|x|_p^{\a+\b-n}, \eqno (8.31)$$
$$B_p^{(n)}(\a,\b)=\G_p^{(n)}(\a)\G_p^{(n)}(\b)\G_p^{(n)}(\g),$$
$$\a+\b+\g=n, (\a,\b)\neq (\a_k,\b_j), (k,j)\in Z^2. \eqno (8.32)$$

{\bf Adelic formulae for gamma- and beta-functions.} For gamma-functions the
following adelic formula is valid [2e)]
$$\G_\infty(\a)\reg\prod_{p=2}^\infty\G_p(\a)=1, \quad \a\neq 0,1 \eqno (8.33)$$
where $\G_\infty$ is the gamma-function of field $\RR$,
$$\G_\infty (\a)=\int |x|_p^{\a-1}\exp (-2\pi ix)dx$$
$$=2(2\pi)^{-\a}\G(\a)\cos{{\pi\a}\over 2}={{\zeta (1-\a)}\over{\zeta (\a)}} \eqno (8.34)$$
where $\G$ is the Euler gamma-function and $\zeta$ is the Riemann 
zeta-function,
$$\zeta (\a)=\sum_{n=1}^\infty n^{-\a}=\prod_{p=2}^\infty (1-p^{-\a})^{-1}, \quad \Re\a>1.$$

Regularization of the divergent product in (8.33) is defined by means of the 
formula
$$\prod_{p=2}^P\G_p(\a)\AC\prod_{p=P_1}^\infty (1-p^{-\a})^{-1} $$
$$={{\zeta (\a)}\over{\zeta (1-\a)}}\AC\prod_{p=P_1}^\infty(1-p^{\a-1})^{-1} \quad P=\infty,2,3,5,\ldots \eqno (8.35)$$
which folows from Tate's formula. Here $P_1$  is the prime number following the
prime $P$; $\AC f(\a)$ is the analytic continuation on $\a$ of function $f(\a)$
which is holomorphic in some domain of the complex plane of the variable $\a$.

Passing on to the limit in (8.35) as $P\to\infty$ in half-plane $\Re\a<0$, 
denoting
$$\reg\prod_{p=2}^\infty\G_p(\a)=\lim_{P\to\infty}\prod_{p=2}^P\G_p(\a)\AC\prod_{p=P_1}^\infty (1-p^{-\a})^{-1},$$
and using equality (8.34) we get adelic formula (8.33). For $\Re\a\leq 0$ 
the $\reg\prod\G_p(\a)$ is defined from (8.33) as the analytic continuation on
$\a$.

The similar adelic formula is valid also for beta-functions:
$$B_\infty(\a,\b)\reg\prod_{p=2}^\infty B_p(\a,\b)=1 \eqno (8.36)$$
where
$$B_\infty (\a,\b)=\G_\infty (\a)\G_\infty (\b)\G_\infty (\g), \quad \a+\b+\g=1 \eqno (8.37)$$
is beta-function of the field $\RR$, and according to (8.13)
$$\reg\prod_{p=2}^\infty B_p(\a,\b)=\prod_{x=\a,\b,\g}\reg\prod_{p=2}^\infty\G_p(x). \eqno (8.38)$$
Note others symmetric expressions for $B_\infty$:
$$B_\infty (\a,\b)=B(\a,\b)+B(\a,\g)+B(\b,\g) $$
$$={{\G(\a)\G(\b)}\over{\G(\a+\b)}}+{{\G(\a)\G(\g)}\over{\G(\a+\g)}}+{{\G(\b)\G(\g)}\over{\G(\b+\g)}}$$
$$={4\over\pi}\prod_{x=\a,\b,\g}\G(x)\cos{{\pi x}\over 2}=\prod_{x=\a,\b,\g}{{\zeta (1-x)}\over{\zeta (x)}}. \eqno (8.39)$$

{\bf Adelic formula for the Riemann zeta-function,}
$$\zeta (\a)=\sum_{n=1}^\infty n^{-\a}=\prod_{p=2}^\infty (1-p^{-\a})^{-1}.$$
$\zeta (\a)$ satisfies relation
$$\pi^{-\a/2}\G(\a/2)\zeta (\a)=\pi^{-(1-\a)/2}\G((1-\a)/2)\zeta (1-\a). \eqno(8.40)$$
Denote
$$\zeta_\infty (\a)=\int\exp^{-\pi x^2}|x|^{\a-1}dx=\pi^{-\a/2}\G (\a/2), \eqno (8.41)$$
$$\zeta_p (\a)={1\over{1-p^{-1}}}\int_{Z_p} |x|_p^{\a-1}d_px=(1-p^{-\a})^{-1}, \eqno (8.42)$$ 
$$\zeta_A (\a)=\zeta_\infty (\a)\zeta (\a). \eqno (8.43)$$
 Then the following formulae are valid:
$$\zeta_A (\a)=\zeta_A (1-\a), (\hbox{ cf. (8.40) }),\eqno (8.44)$$
$$\G_\infty (\a)={{\zeta_\infty (\a)}\over{\zeta_\infty (1-\a)}}={{\zeta (\a)}\over{\zeta (1-\a)}} \eqno (8.45)$$ 
(cf. (8.34)),
$$\zeta_\infty (\a)\prod_{p=2}^\infty\zeta_p (\a)=\zeta_A (\a). \eqno (8.46)$$
Formula (8.45) is the adelic formula for the Riemann zeta-function.

\bigskip
\centerline{\bf \Par 9. Quadratic Extensions of the Field $\QQ_p$}
\medskip

Let $d\not\in\QQ_p^{\times 2}$ be a $p$- dic number.{\it Quadratic extension}
of the field $\QQ_p$ is the field $\QQ_p(\sqrt d)=\QQ_p+\sqrt d\QQ_p.$ Let us
describe all non-isomorphic fields $\QQ_p(\sqrt d)$. According to what has been
said in $\Par 1$ it is sufficient to consider integer rational numbers $d$, 
free of squares, i.e.
$d=\pm p_1p_2\ldots p_n, \quad d\neq 1,$
where $p_1,p_2,\ldots,p_n$ are different prime numbers.

The following cases are possible:
$$p\neq 2,p_1,\ldots,p_n, \quad \Bigl ({d\over p}\Bigr )=1, \quad \QQ_p(\sqrt d)\sim\QQ_p;$$
$$p\neq 2,p_1,\ldots,p_n, \quad \Bigl ({d\over p}\Bigr )=-1, \quad \QQ_p(\sqrt d)\sim\QQ_p(\sqrt\epsilon), \epsilon\not\in\QQ_p^{\times 2}, |\epsilon |_p=1;$$
$$p\neq 2, p=p_i, \quad \Bigl ({{d/{p_i}}\over p}\Bigr )=1, \quad \QQ_p(\sqrt d)\sim\QQ_p(\sqrt p);$$
$$p\neq 2, p=p_i, \quad \Bigl ({{d/{p_i}}\over p}\Bigr )=-1, \quad \QQ_p(\sqrt d)\sim\QQ_p(\sqrt {p\epsilon}), \epsilon\not\in\QQ_p^{\times 2}, |\epsilon |_p=1;$$
$$p=2, \quad d\equiv 3,5,7(\mod 8), \quad \QQ_2(\sqrt d)\sim\QQ_2(\sqrt\epsilon ), \epsilon =3,5,7 \hbox{ resp. };$$
$$p=2, \quad {d/2}\equiv 1,3,5,7(\mod 8), \quad \QQ_2(\sqrt d)\sim\QQ_2(\sqrt {2\epsilon} ), \epsilon =1,3,5,7 \hbox{ resp. }.$$
Note that $\QQ_p(\sqrt d)$ is the closure of the field 
$\QQ(\sqrt d)=\QQ+\sqrt d\QQ$ on metric $\sqrt {|z\bar z|_p}$ where 
$z=x+\sqrt dy, \bar z=x-\sqrt dy,$ 
$z\bar z=x^2-dy^2, \quad x,y\in\QQ.$
 
The Haar mesure $d_pz$ of field $\QQ_p(\sqrt d)$ we choose in the form
$$d_pz={1/\delta}d_pxd_py, \quad z=x+\sqrt dy, x,y\in\QQ_p \eqno (9.1)$$
where $\delta =\delta_{p,d}=2$ if $p=2, d\equiv 5(\mod 8)$ and $\delta=1$ 
otherwise. The mesure $d_pz$ is normalized by the condition (see [2d)])
$$\int_{B_0^2}d_pz=1, \quad B_0^2=[z\in\QQ_p(\sqrt d): |z\bar z|_p\leq 1]. \eqno (9.2)$$
The following equality is valid
$$d_p(az)=|a\bar a|_pd_pz, \quad a\in\QQ_p^\times (\sqrt d). \eqno (9.3)$$
The quantity $|a\bar a|_p$ is called {\it the module of automorphism} $z\to az$
of field $\QQ_p(\sqrt d)$.

The maximal compact subring $Z_p(\sqrt d)$ of field $\QQ_p(\sqrt d)$ is
$$Z_p(\sqrt d)=[z\in\QQ_p(\sqrt d): |z\bar z |_p\leq 1], \quad Z_p=B_0^2;$$
its multiplicative subgroup is
$$Z_p^\times(\sqrt d)=[z\in\QQ_p(\sqrt d): |z\bar z |_p=1];$$
its maximal ideal is
$$I_p(\sqrt d)=[z\in\QQ_p(\sqrt d): |z\bar z |_p<1].$$  
Residue classes ${Z_p(\sqrt d)}/{I_p(\sqrt d)}$ form the finite field of 
characteristic $p$ called {\it residue field}; a number of its elements 
$q=q_{p,d}$ (is equal to $p$ or $p^2$) is called {\it the module of field} 
$\QQ_p(\sqrt d)$. For special cases we have: for $p=2, d\equiv 5(\mod 8)$, 
$q=4$ the residue field is $\{0,1,{1/2}\pm{{\sqrt 5}/2}\};$ for 
$d\not\equiv 5(\mod 8)$, $q=2$ the residue field is $\{0,1\}$; 
for $p\neq 2, |d|_p=1,$ $q=p^2$ the residue field is 
$\{k+\sqrt d j, k,j=0,1,\ldots,p-1\}$; for $|d|_p=1/p$, $q=p$ the residue field
is $\{0,1,\ldots,p-1.\}$
  
The Fourier transform $\tilde{\varphi}(\zeta ), \zeta=\xi+\sqrt d\eta$ of a 
test function $\varphi (z)$ $\equiv\varphi (x,y)$ in
$\script{S}(\QQ_p(\sqrt d))\sim\script{S}(\QQ_p^2)$ we define by the following
formula
$$\tilde{\varphi}(\zeta )=\delta\sqrt{|4d|_p}\int\varphi (z)\chi_p(z\zeta +\tilde z\tilde\zeta )d_pz$$
$$=\sqrt{|4d|_p}\int\varphi (x,y)\chi_p(2x\xi +2dy\eta )d_pxd_py.$$

The invers Fourier transform is expressed by the equality
$$\varphi (z)=\delta\sqrt{|4d|_p}\int\tilde{\varphi}(\zeta )\xi_p(-z\zeta -\tilde z\tilde\zeta)d_p\zeta.$$
Thus the mesure $\delta\sqrt{|4d|_p}d_pz$ is self-dual with respect to the
charater $\chi_p(z+\bar z).$

The generalized function
$$|z\bar z |_p^{\a-1}=|x^2-dy^2|_p^{\a-1}$$
is defined by the equality (see \Par 8)
$$(|z\bar z|_p^{\a-1},\varphi )=\int_{|z\bar z|_p\leq 1}|z\bar z|_p^{\a-1}[\varphi (z)-\varphi (0)]d_pz$$
$$+\int_{|z\bar z|_p>1}|z\bar z|_p^{\a-1}d_pz+\varphi (0){{1-q^{-1}}\over{1-q^{-\a}}}, \quad \varphi\in\script{S}(\QQ_p(\sqrt d))$$
or, equivalently,
$$(|z\bar z|_p^{\a-1},\varphi )=\int[|z\bar z|_p^{\a-1}[\varphi (z)-\varphi (0)]d_pz, \quad \varphi\in\script{S}(\QQ_p(\sqrt d)).$$
Here we used formulas:
$$\int_{B_0^2}|z\bar z|_p^{\a-1}d_pz={{1-q^{-1}}\over{1-q^{-\a}}}, \quad \a\neq\a_k, k\in Z, \eqno (9.4)$$
$$\int |z\bar z|_p^{\a-1}d_pz=0, \quad \a\neq\a_k, k\in Z \eqno (9.5)$$
where
$$\a_k={{2k\pi i}/{\ln q}}, \quad k\in Z. \eqno (9.6)$$   

The generalized function $|z\bar z|_p^{\a-1}$ (degree of homogeneity $2\a-2$) 
is holomorphic on $\a$ everywhere exept simple poles $\a=\a_k, k\in Z$ (see
(9.5)) with the residue ${{q-1}\over{q\ln q}}\delta (x,y)$.

The Fourier transform formula is valid [2d)]
$$F[|z\bar z|_p^{\a-1}]=\G_{p,d}(\a)|\zeta\bar\zeta|_p^{-\a}, \quad \a\neq\a_k, k\in Z \eqno (9.7)$$
where
$$\G_{p,d}(\a)=\delta\sqrt {|4d|_p}\int |z\bar z|_p^{\a-1}\xi_p(z+\bar z)d_pz=\rho_{p,d}(\a)\G_q(\a) \eqno (9.8)$$
is gamma-function of the field $\QQ_p(\sqrt d)$;
$$\G_q(\a)={{1-q^{\a-1}}\over{1-q^{-\a}}} \eqno (9.9)$$
is {\it reduced gamma-function} of the field $\QQ_p(\sqrt d)$; and
$$ \eqalignno{
\rho_{p,d}(\a) &=1, \hbox{ if } |d|_p=1, p\neq 2 \hbox{ or } d\equiv 5(\mod 8), p=2,\cr
               &=p^{\a-{1/2}}, \hbox{ if } |d|_p={1/p}, p\neq 2,\cr
               &=p^{2\a-1}, \hbox{ if } d\equiv 3(\mod 4), p=2,\cr
               &=p^{3\a-{3/2}}, \hbox{ if } |d|_2={1/2}, p=2. & (9.10) \cr
}$$
From (9.8)--(9.10) it follows the following relation for gamma-function of the 
field $\QQ_p(\sqrt d)$:
$$\G_{p,d}(\a)\G_{p,d}(1-\a)=1. \eqno (9.11)$$
Beta-function of the field $\QQ_p(\sqrt d)$ is introduced similar to \Par 8. 
The convolution $|z\bar z|_p^{\a-1}\-\ast |z\bar z|_p^{\b-1}$ exists for all
complex $(\a,\b)$ from the tube domain $\Re\a>0, \Re\b>0, \Re (\a+\b)<1$, and
it is expressed by the integral
$$|z\bar z|_p^{\a-1}\ast |z\bar z|_p^{\b-1}=\int |\zeta\bar\zeta |_p^{\a-1}|(z-\zeta )(\bar z-\bar\zeta )|_p^{\b-1}d_p\zeta $$
$$=B_{p,d}(\a,\b)|z\bar z|_p^{\a+\b-1} \eqno (9.12)$$
where $B_{p,d}$ is beta-function of the field $\QQ_p(\sqrt d)$ [4]:
$$B_{p,d}(\a,\b)=\int |\zeta\bar\zeta |_p^{\a-1}|(1-\zeta)(1-\bar\zeta )|_p^{\b-1}d_p\zeta $$
$$={{\G_{p,d}(\a)\G_{p,d}(\b)}\over{\delta\sqrt {|4d|_p}\G_{p,d}(\a+\b)}}. \eqno (9.13)$$
From equalities (9.8)--(9.13) it follows such symmetric expessions for 
beta-function:
$$B_{p,d}(\a,\b)={1\over{\delta\sqrt{|4d|_p}}}\G_{p,d}(\a)\G_{p,d}(\b)\G_{p,d}(\g)=B_q(\a,\b)$$
$$=\G_q(\a)\G_q(\b)\G_q(\g), \quad \a+\b+\g=1, (\a,\b)\neq (\a_k,\b_j), (k,j)\in Z^2. \eqno (9.14)$$
Note that equalities (9.12)--(9.14) are valid for all $(\a,\b)$ such that
$(\a,\b)\neq (\a_k,\a_j), (k,j)\in Z^2.$

We call {\it upper (lower) half-plain} of the field $\QQ_p(\sqrt d)$ a set of 
points $z=x+\sqrt dy$ for which ${\textstyle\sgn_{p,d}}y=1$ (resp. 
${\textstyle\sgn_{p,d}}y=-1$).

Generalized functions $(x\pm\sqrt d0)^{-1}$ are defined as the 
Fourier transform of functions
$$\t_d^{\pm}(\xi )={1\over 2}(1\pm{\textstyle\sgn_{p,d}}\xi ), \quad (x\pm\sqrt d0)^{-1}=\tilde{\t}_d^{\pm}(x). \eqno (9.15)$$
The following equalities are valid [4]
$$F[\t_d^{\pm}](x)=(x\pm\sqrt d0)^{-1}={1\over 2}\delta (x)+C_{p,d}{{{\textstyle\sgn_{p,d}}x}\over {|x|_p}}, \quad p\neq 2 \eqno (9.16)$$
which are similar to the Sochozki formulae (for the field $\RR$). Here a
generalized function ${{{\textstyle\sgn_{p,d}}x}\over {|x|_p}}$ is defined by 
the equality
$$\Bigl ({{{\textstyle\sgn_{p,d}}x}\over{|x|_p}},\varphi\Bigr )=\int{{{\textstyle\sgn_{p,d}}x}\over{|x|_p}}\varphi (x)d_px, \quad \varphi\in\script{S}, \eqno (9.17)$$
$$C_{p,d}=\cases{\sqrt{p\over{p+1}}, \hbox{ if } |d|_p=1,} \\ {\pm{1\over 2}\sqrt{p{\textstyle\sgn_{p,d}}(-1)}, \quad\hbox{ if } |d|_p={1/p}.} \endcases$$  

For $\G_q$-function the following adelic formulae are valid [2d)]
$$\G_\infty^2(\a)\reg\prod_{p=2}^\infty\G_q^\nu(\a)=D^{{1/2}-\a}, \quad d>0, \eqno (9.18)$$
$$\G_{\omega}(\a)\reg\prod_{p=2}^\infty\G_q^\nu(\a)=|D|^{{1/2}-\a}, \quad d<0 \eqno (9.18')$$
where $\G_\infty$ and $\G_{\omega}$ are gamma-functions of fields $\RR$ and 
$\CC$ resp.;
$$\G_{\omega}(\a)=2\int |z\bar z|^{\a-1}\exp (-4\pi ix)dxdy=(2\pi )^{1-2\a}{{\G(\a)}\over{\G(1-\a)}};$$
$$=2(2\pi)^{-2\a}\G^2(\a)\sin\pi\a=i\G_\infty (\a)\tilde{\G}(\a),$$
where
$$\tilde{\G}(\a)=\int\sgn x |x|^{\a-1}\exp(-2\pi ix)dx=-2i(2\pi)^{-\a}\G(\a)\sin{{\pi\a}\over 2};$$
$\nu=2$ if $d\in\QQ_p^{\times 2}$ and $\nu=1$ if $d\not\in\QQ_p^{\times 2}$; 
$D$ is the discriminant of the field $Q(\sqrt d)$,
$$D=\cases{d, \hbox{ if } d\equiv 1(\mod 4)}\\{4d, \hbox{ if } d\equiv 2,3(\mod 4)}\endcases.$$
(We took the Haar mesure of field $\CC$ in the form 
$|dz \wedge \bar z|=2dxdy$, $
z=x+iy.$)

Regularization of the divergent infinite products in (9.18) is defined by the 
formula (cf. (8.35))
$$\prod_{p=2}^P\G_q^\nu (\a)\AC\prod_{p=P_1}^\infty (1-q^{-\a})^{-\nu}={{\zeta_d(\a)}\over{\zeta_d(1-\a)}}\AC\prod_{p=P_1}^\infty (1-q^{-\a})^{-\nu},$$
$$\quad P=\infty,2,3,5,\ldots \eqno (9.19)$$
which follows from general Tate's formula. Here $\zeta_d$ is Dedekind's 
zeta-function of the field $\QQ(\sqrt d)$,
$$\zeta_d(\a)=\prod_{p=2}^\infty(1-q^{-\a})^{-\nu}, \quad \zeta_1(\a)=\zeta^2(\a).$$
 The Dedekind zeta-function satisfies relation (cf. (8.40))
$$(2\pi)^{1-\a}\G (\a)\zeta_d (\a)=(2\pi)^{\a}\G (1-\a)\zeta_d (1-\a)|D|^{1/2-\a},$$ 
which is equivalent to relation (cf. (8.46))
$$\zeta_{A_d} (\a)=\zeta_{A_d} (1-\a)|D|^{{1/2}-\a},$$
where it is denoted
$$\zeta_{A_d} (\a)=(2\pi)^{1-\a}\G (\a)\zeta_d (\a).$$
 Passing on in (9.19) to the limit $P\to\infty$, denoting 
$$\reg\prod_{p=2}^\infty\G_q^\nu (\a)=\lim_{P\to\infty}\prod_{p=2}^P\G_q^\nu (\a)\AC\prod_{p=P_1}^\infty (1-q^{-\a})^{-\nu}$$
and using equalities
$$\G_\infty^2(\a)=D^{{1/2}-\a}{{\zeta_d (1-\a)}\over{\zeta_d (\a)}}, \quad d>0, \eqno (9.20)$$
$$\G_{\omega}(\a)=|D|^{{1/2}-\a}{{\zeta_d (1-\a)}\over{\zeta_d (\a)}}, \quad d<0 \eqno (9.20')$$
in the halp-plain $\Re\a<0$ we obtain the adelic formulae (9.18). For remaining
$\a$ $\reg\prod\G_q^{-\nu}(\a)$ is defined from formulae (9.18)  s analytic
continuation on $\a$.

Similar adelic formulae are valid also for beta-functions
$$B_\infty^2(\a,\b)\reg\prod_{p=2}^\infty B_q^\nu(\a,\b)=\sqrt D, \quad d>0, \eqno (9.21)$$
$$B_{\omega}(\a,\b)\reg\prod_{p=2}^\infty B_q^\nu(\a,\b)=\sqrt {|D|}, \quad d<0 \eqno (9.21')$$
where $B_\infty$ and $B_{\omega}$ are beta-functions of fields $\RR$ and $\CC$
resp.,
$$B_{\omega}(\a,\b)=\G_\omega(\a)\G_\omega(\b)\G_\omega(\g), \quad \a+\b+\g=1 \eqno (9.22)$$
and in accordance with the formula (9.14) (cf. (8.36)) 
$$\reg\prod_{p=2}^\infty B_q^\nu (\a,\b)=\prod_{x=\a,\b.\g}\reg\prod_{p=2}^\infty\G_q^\nu (x).$$
Note another symmetric expressions for $B_\omega$
$$B_\omega (\a,\b)=2\pi\prod_{x=\a,\b,\g}{{\G (x)}\over{\G (1-x)}}={2\over{\pi^2}}\prod_{x=\a,\b,\g}\G^2(x)\sin\pi x. \eqno (9.23)$$
\bigskip
\centerline{\bf \Par 10. The operator $D^\a$}
\medskip
The generalized function
$$f_\a(x)={{|x|_p^{\a-1}}\over{\G_p(\a)}}$$
is holomorphic on $\a$ everywhere exept simples poles 
$1+\a_k, \quad \a_k={{2k\pi i}/{\ln p}},$ $ k\in Z$ with the residue
${{1-p}\over{p\ln p}}$, and also $f_{\a_k}=\delta$ and
$$f_\a\ast f_\b=f_{\a+\b}, \quad \a\neq 1+\a_k, \b\neq 1+\a_j, \a+\b\neq 1+\a_i, (k,j,i)\in Z^3.$$
Let $\a\in\RR, \a\neq{-1}$ and $f\in\script{S}'$ be such that the convolution
$f_{-\a}\ast f$ exists in $\script{S}'$. Operator $D^\a f=f_{-\a}\ast f$ is 
called for $\a>0$ the operator (fractional) {\it differentiation} of order 
$\a$,  nd for $\a<0$ the operator (fractional) {\it integration} of order 
$-\a$; for $\a=0$ $D^0f=\delta\ast f=f$ is the identical operator [2a)].

{\bf Example.} If $\a=1$ ¨ $\varphi\in\script{S}$ then
$$(D\varphi )(x)={{p^2}\over{p+1}}\int{{\varphi (x)-\varphi (y)}\over{|x-y|_p^2}}d_py=\int |\xi |_p\tilde{\varphi}(\xi )\chi_p(-\xi x)d_p\xi. \eqno (10.1)$$
Thus the operator $D$ is hyper-singular pseudo-differential operator (PDO) with
the symbol $|\xi |_p$.

Let $\a=1$ be. Consider a locally-integrable in $\QQ_p$ function
$$f_1(x)=-{{1-p^{-1}}\over{\ln p}}\ln |x|_p. \eqno (10.2)$$
It possesses the following properties:
$$\int f_\a(x)\varphi (x)d_px\to\int f_1(x)\varphi (x)d_px, \quad \a\to 1, \eqno (10.3)$$
if $\varphi\in\script{S}$ satisfies the condition
$$\int\varphi (x)d_px=0; \eqno (10.4)$$
$$\tilde{f_1}(\xi )=\reg |\xi |_p^{-1}+{1\over p}\delta(\xi ) \eqno (10.5)$$
where a generalized function $\reg |\xi |_p^{-1}$ is defined in \Par 6;
$$f_1\ast f_\a=f_{1-\a}, \quad \a\geq 1. \eqno (10.6)$$

The operator of integration of order 1 corresponding to the value of $\a=-1$ is
equal
$$D^{-1}f=f_1\ast f, \quad f\in\script{S}' \eqno (10.7)$$
if the convolution $f_1\ast f$ exists. Then
$$D^{-\a}f\to D^{-1}f, \a\to 1 \hbox{ in } \script{S}' \eqno (10.8)$$
if $f\in\script{E}'$ and
$$G\!\!\!\!\int f(x)d_px=0. \eqno (10.9)$$

Summarizing we get the following properties of the operator $D^\a, \a\in\RR$:
$$D^\a{D}^\b f=D^{\a+\b}f=D^\b{D}^\a f, \quad f\in\script{S}' \eqno (10.10)$$
if $(\a,\b,\a+\b)\neq (-1,-1,-1)$ or $\a\leq 0, \b=-1$ or
$\a=-1, \b\leq 0$; if $f$ satisfies the codition (10.9) then the equalities 
(10.10) are valid for all real $\a$ and $\b$, and $D^\a f$ continuously depends
on $\a$ in $\script{S}'$.

{\bf Example.}
$$D^\a\chi_p(ax)=|a|_p^\a\chi_p(ax), \quad \a\in\RR, \quad a\in\QQ_p^\times. \eqno (10.11)$$
The equation
$$D^\a\psi =g, \quad g\in\script{E}' \eqno (10.12)$$
is solvable for all $\a\in\RR$ and also for $\a>0$ its general solution is
expressed by the formula
$$\psi=D^{-\a}g+C \eqno (10.13)$$
where $C$ is arbitrary constant; for $\a\leq 0$ its solution is unique and it
expressed by the formula (10.13) for $C=0$.

The fundamental solution $\script{E}(x)$ of the operator $D^\a$,
$$D^\a\script{E}(x)=\delta (x), \quad \script{E}\in\script{S}' \eqno (10.14)$$
has been calculated in [2a)]. It is equal to
$$\script{E}(x)=\cases{\G_p^{-1}(\a)|x|_p^{\a-1}, \quad \a\neq 1,} \\
{-{{1-p^{-1}}\over{\ln p}}\ln |x|_p, \quad \a=1.} \endcases \tag 10.15 $$

Note that a fundamental solution does not exist in $\script{S}'$ for any PDO. 
For example, for the operator $D_t^\a-D_x^\a$ it is the case. Indeed, if a
solution $\script{E}$ of the equation
$$(D_t^\a-D_x^\a)\script{E}(t,x)=\delta (t,x)$$
would exist in $\script{S}'$ so we would have the contradictory equation
$$(|\eta |_p^\a-|\xi |_p^\a)F[\script{E}](\eta,\xi )=1, \quad (\eta,\xi )\in\QQ_p^2$$
in which left-hand side vanishes in the open set $|\eta |_p$ $=|\xi |_p$ of 
space $\QQ_p^2.$
 
The operator $D^\a$ for $\a>0$ in a clopen set $G$ is defined on those 
$\psi\in\script{L}^2(G)$ (see \Par 4) for which
$|\xi |_p^\a\tilde\psi\in\script{L}^2.$ This set of functions is called
{\it domain of definition} of the operator $D^\a$ in the clopen set $G$ and it
is denoted 
$\script{D}(D^\a,G); \quad \script{D}(D^\a,\QQ_p)=\script{D}(D^\a)$.
The following equality is valid
$$(D^\a\psi,\varphi )=\int |\xi |_p^\a\tilde{\psi}(\xi )\bar{\tilde{\varphi}}(\xi )d_p\xi, \quad \psi, \varphi\in\script{D}(D^\a,G). \eqno (10.16)$$

The operator $D^\a$ in $G$ is self-adjoint positive-definite, and also owing to
(10.16) for all $\psi\in\script{D}(D^\a,G)$ we have
$$(D^\a\psi,\psi )=(D^{\a/2}\psi,D^{\a/2}\psi )=\int |\xi |_p^\a|\psi (\xi )|^2d_p\xi\geq 0, \eqno (10.17)$$
so its spectrum is situated on semi-axis $\lambda\geq 0.$

For the operator $D^\a, \a>0$ we consider the eigen-value problem
$$D^\a\psi =\lambda\psi, \quad \psi\in\script{D}(D^\a,G). \eqno (10.18)$$

{\bf Theorem [1],[1b)].} {\it The spctrum of the operator $D^\a$ in $\QQ_p$ 
consists of countable number of eigen-values} 
${\textstyle\lambda_N=p^{\a N}}, N\in Z$
{\it every of which is infinite multiplicity, and the point $0$. There exists
an ortho-normalized bases of eigen-functions in $\script{L}^2(\QQ_p)$ of the
operator $D^\a$, and it have the following form: for} $p\neq 2$
$$\psi_{N,j,\epsilon}^\ell (x)=p^{{N+1-\ell}\over 2}\delta (|x|_p-p^{\ell-N})\delta (x_0-j)\chi_p(\epsilon_\ell p^{\ell -2N}x^2), \eqno (10.19)$$
$$\quad \ell=2,3,\ldots, j=1,2,\ldots,p-1, \epsilon_\ell=\varepsilon_0+\varepsilon_1p+\ldots +\varepsilon_{\ell -2}p^{\ell-2},$$
$$ \varepsilon_s=0,1,\ldots ,p-1, \varepsilon_0\neq 0, s=0,1,\ldots,\ell-2, \varepsilon_0\neq 0, s=0,1,\ldots, \ell-2;$$
$$\psi_{N,j,0}^1(x)=p^{{N-1}\over 2}\Omega (p^{N-1}|x|_p)\chi_p(jp^{-N}x), \quad \ell =1, $$
$$\quad j=1,2,\ldots,p-1, \epsilon_\ell=0; \eqno(10.19')$$
{\it for} $p=2$
$$\psi_{N,j,\epsilon_\ell}^\ell (x)=2^{{N-\ell}\over 2}\delta (|x|_2-2^{\ell +1-N})\chi_2(\epsilon_\ell 2^{\ell -2N}x^2+2^{\ell -N+j}x), \eqno (10.20)$$
$$\quad \ell =2,3,\ldots, j=0,1, \epsilon_\ell =1+\varepsilon_12+\ldots +\varepsilon_{\ell -2}2^{\ell -2}, \varepsilon_s=0,1, s=1,2,\ldots ,\ell -2;$$
$$\psi_{N,j,0}^1(x)=2^{{N-1}\over 2}[\Omega (2^N|x-j2^{N-2}|_2)-\delta (|x-j2^{N-2}|_2-2^{1-N})], $$
$$\quad \ell=1, j=0,1, \epsilon_\ell =0. \eqno(10.20')$$	

{\bf Theorem [2b)],[2c)].} {\it If $G$ is a clopen compact then eigen-values
$\lambda_k, k=0,1,\ldots$ of the operator $D^\a, \a>0$ in $G$ are of finite
multplicity and eigen-functions $\psi_k(x)$ form an ortho-normalized bases in} 
$\script{L}^2(G)$.

{\bf Example.} Eigen-values and ortho-normalized bases of eigen-functions of 
the operator $D^\a$ in $B_\g, \g\in Z$ [2b)]. For $p\neq 2$:
$$\lambda_0={{p-1}\over{p^{\a+1}-1}}p^{\a (1-\g )}, \quad \psi_0(x)=p^{-{\g/2}}, \hbox{ multipl. } 1;$$
$$\lambda_k=p^{\a (k-\g)}, \quad \psi_k(x)=\psi_{k-\g,j,\epsilon_\ell}^\ell (x), \quad \ell=1,2,\ldots ,k, j=1,2,\ldots ,p-1, \epsilon_\ell,$$
$$\quad \hbox{ multipl. } (p-1)p^{k-1}, \quad k=1,2,\ldots.$$
For $p=2$:
$$\lambda_0={{2^{\a (1-\g )}}\over{2^{\a+1}-1}}, \quad \psi_0(x)=2^{-{\g/2}}, \hbox{ multipl. } 1;$$
$$\lambda_1=2^{\a (1-\g )}, \quad \psi_1(x)=\psi_{1-\g,0,0}^1(x), \hbox{ multipl. } 1;$$
$$\lambda_k=2^{\a (k-\g )}, \quad \psi_k(x)=\psi_{k-\g,j,\epsilon_\ell}^\ell (x), \quad \ell=1,2,\ldots,k-1, j=0,1,$$
$$\hbox{ multipl. } 2^{k-1}, \quad k=2,3,\ldots.$$

{\bf Example.} Eigen-values and normalized bases of eigen-functions of the
operator $D^\a$ in $S_\g, \g\in Z$ [2b)]. For $p\neq 2$:
$$\lambda_0={{p^\a+p-2}\over{p^{\a+1}-1}}p^{\a(1-\g)}, \quad \psi_0(x)=p^{{1-\g}\over 2}(p-1)^{1/2}, \hbox{ multipl. } 1;$$
$$\lambda_1=p^{\a(1-\g)}, \quad \psi_1(x)=2^{-{1/2}}[\psi_{1-\g,j,0}^1(x)-\psi_{1-\g,j+1,0}^1(x)], \hbox{ multipl. } p-2;$$
$$\lambda_k=p^{\a(k-\g)}, \quad \psi_k(x)=\psi_{k-\g,j,\epsilon_k}^k(x), \quad j=1,2,\ldots,p-1, \quad \epsilon_k,$$
$$\hbox{ multipl. } (p-1)^2p^{k-2}, \quad k=2,3,\ldots.$$
For $p=2$:
$$\lambda_0={{2^{\a(2-\g)}}\over{2^{\a+1}-1}}, \quad \psi_0(x)=2^{{1-\g}\over 2}, \hbox{ multipl. } 1;$$
$$\lambda_1=2^{\a(2-\g)}, \quad \psi_1(x)=\psi_{1-\g,1,0}^1(x), \hbox{ multipl. } 1;$$
$$\lambda_k=2^{\a(k+1-\g)}, \quad \psi_k(x)=\psi_{k+1-\g,j,\epsilon_k}^k(x), \quad j=0,1, \quad \epsilon_k,$$
$$\hbox{ multipl. } 2^{k-1}, \quad k=2,3,\ldots.$$

It should be pointed out that multiplicative characters of rank $k$ of the
group $Z_p^\times$ are eigen-functions of the operator $D^\a$ in $S_0$
correspondig to the eigen-value $\lambda_k$ [11a)]. On the other hand, a number
of linearly idependent multiplicative characters of rank $k$ of the group 
$Z_p^\times$ was calculated (see [16]) and it coincides to the multiplicity 
$n_k$ of the eigen-value $\lambda_k$ ®f the operator $D^\a, \a>0$ in $S_0$ 
[2b)]. From here it follows such result: 

{\it There is exist an ortho-nomalized bases of eigen-functions of the operator
$D^\a, \a>0$ in $S_0$ consisting of all multiplicative characters of the group
$Z_p^\times$.} 

On the other hand, any multiplicative character of the group $Z_p^\times$ of
rank $k$ is expanded on eigen-functions $\psi_{a_k+j}(x), j=1,2,\ldots,n_k$ 
(by a suitable choose of $a_k$ [2b)], that is it is expanded on additive
characters of the field $\QQ_p$. 

Indicate concrete values for $\lambda_k$ and $n_k$. Assuming $\g=0$ we get 
[2b)]: for $p\neq 2$
$$\lambda_0={{p^\a+p-2}\over{p^{\a+1}-1}}p^\a, \quad n_0=1;$$
$$\lambda_1=p^\a, \quad n_1=p-2; \quad\lambda_k=p^{\a{k}}, \quad n_k=(p-1)^2p^{k-2}, \quad k=2,3,\ldots;$$
for $p=2$ 
$$\lambda_0={{2^{2\a}}\over{2^{\a+1}-1}}, \quad n_0=1; \quad \lambda_k=2^{\a(k+1)}, \quad n_k=2^{k-1}, \quad k=1,2,\ldots.$$

\newpage

\centerline{\BF Part II}
\bigskip
\centerline{\BF Tables of integrals}
\bigskip
\centerline{\bf \Par 11. Primary integrals, one variable}
\medskip
$$\int_{B_0}d_px=1. \eqno(11.1)$$
$$\int_{B_\g}d_px=p^\g. \eqno(11.2)$$
$$\int_{S_\g}d_px=(1-{1/p})p^\g. \eqno(11.3)$$
$$\int f(x)d_px=\sum_{\g=-\infty}^\infty\int_{S_\g}f(x)d_px. \eqno(11.4)$$
$$\int_{B_\g}f(|x|_p)d_px=(1-{1/p})\sum_{k=-\infty}^\g p^kf(p^k). \eqno(11.5)$$
$$\int f(|x|_p)d_px=(1-{1/p})\sum_{k=-\infty}^\infty p^kf(p^k). \eqno(11.6)$$
$$\int_Df(x)d_px=|a|_p\int_{{D-b}\over a}f(ay+b)d_py, \quad a\neq 0. \eqno (11.7)$$
$$\int_{S_\g}f(x)d_px=p^{2\g}\int_{S_{-\g}}f(1/y)d_py. \eqno (11.8)$$
$$\int_{B_\g}f(x)d_px=\int_{\QQ_p\backslash B_{1-\g}}f(1/y)|y|_p^{-2}d_py. \eqno(11.9)$$
$$\int f(x)d_px=\int f(1/y)|y|_p^{-2}d_py. \eqno(11.10)$$
$$\int f(|x|_p)d_px=\int f({1/{|y|_p}})|y|_p^{-2}d_py. \eqno(11.11)$$
$$\int_{G_p}f(x)d_px=\int_{G_p}f(\sin y)d_py. \eqno(11.12)$$
$$\int_{G_p}f(x)d_px=\int_{G_p}f(\arcsin y)d_py. \eqno(11.13)$$
$$\int_{G_p}f(x)d_px=\int_{G_p}f(\tg y)d_py. \eqno(11.14)$$
$$\int_{G_p}f(x)d_px=\int_{G_p}f(\arctg y)d_py. \eqno(11.15)$$
$$\int_{G_p}f(x)d_px=\int_{J_p}f(\ln y)d_py. \eqno(11.16)$$
$$\int_{J_p}f(x)d_px=\int_{G_p}f(\exp y)d_py. \eqno(11.17)$$
$$\int_{B_\g}|x|_p^{\a-1}d_px={{1-p^{-1}}\over{1-p^{-\a}}}p^{\a\g}, \quad \Re\a>0. \eqno(11.18)$$
$$\int_{S_0}|x-1|_p^{\a-1}d_px={{p-2+p^{-\a}}\over{p(1-p^{-\a})}}, \quad \Re\a>0 \quad [2a)]. \eqno(11.19)$$.
$$\int_{S_\g}|x-a|_p^{\a-1}d_px={{p-2+p^{-\a}}\over{p(1-p^{-\a})}}|a|_p^\a, \quad |a|_p=p^\g, \Re\a>0. \eqno(11.20)$$
$$\int_{B_\g}\ln |x|_pd_px=\Bigl (\g-{1\over{p-1}}\Bigr )p^\g\ln p. \eqno(11.21)$$
$$\int_{S_0}\ln |x-1|_pd_px=-{{\ln p}\over{p-1}} \quad [2a)]. \eqno(11.22)$$
$$\int_{S_\g}\ln |x-a|_pd_px=\Bigl [(1-{1/p})\ln |a|_p-{{\ln p}\over{p-1}}\Bigr ]|a|_p, \quad |a|_p=p^\g. \eqno(11.23)$$
$$\int_{S_\g}\ln |x|_pd_px=\g(1-{1/p})p^\g\ln p. \eqno(11.24)$$
$$\int |x|_p^{\a-1}|1-x|_p^{\b-1}d_px=B_p(\a,\b), $$
$$\Re\a>0, \Re\b>0, \Re (\a+\b)<1 \quad [4]. \eqno(11.25)$$
$$\int |x|_p^{\a-1}|y-x|_p^{\b-1}d_px=B_p(\a,\b)|y|_p^{\a+\b-1},$$
$$\Re\a>0, \Re\b>0, \Re (\a+\b)<1. \eqno(11.26)$$
$$\int_{B_\g}|x^2+a^2|_p^{(\a-1)/2}d_px=p^\g|a|_p^{\a-1}, \quad p^\g<|a|_p. \eqno(11.27)$$
$$={{1-p^{\a-1}}\over{1-p^\a}}|a|_p^\a+{{1-p^{-1}}\over{1-p^{-\a}}}p^{\a\g}, $$
$$p^\g\geq |a|_p\neq 0, \Re\a>0, p\equiv 3(\mod 4) \quad [3a)]. \eqno (11.28)$$
$$=\biggl [1-{2/p}+\Bigl (1-{1/p}\Bigr )\Bigl ({2\over{p^{(\a+1)/2}-1}}-{1\over{1-p^{-\a}}}\Bigr )\biggr ]|a|_p^\a-{{1-p^{-1}}\over{1-p^{-\a}}}p^{\a\g},$$
$$p^\g\geq |a|_p\neq 0, \Re\a>0, p\equiv 1(\mod 4) \quad [3a)]. \eqno (11.29)$$
$$\eqalignno{
&\int |x^2+a^2|_p^{{(\a-1)}/2}d_px, \quad a\neq 0 \cr
&={{1-p^{\a-1}}\over{1-p^\a}}|a|_p^\a, \quad \Re\a<0, p\equiv 3(\mod 4). & (11.30) \cr
&=\biggl [1-{2/p}+\Bigl (1-{1/p}\Bigr )\Bigl ({2\over{p^{{(\a+1)}/2}-1}}-{1\over{1-p^{-\a}}}\Bigr )\biggr ]|a|_p^\a, \cr
&\Re\a<0, p\equiv 1(\mod 4). & (11.31) \cr
}$$
$$\int_{S_0}| 1+x^2|_p^{\a-1}d_px=1-{3/p}-2{{1-p^{-1}}\over{1-p^\a}}, \quad \Re\a>0, p\equiv 1(\mod 4). \eqno(11.32)$$
$$\int_{S_{\g,k_0}}d_px=p^{\g-1}, \quad k_0=1,2,\ldots,p-1 \quad [2a)]. \eqno(11.33)$$
$$\int_{S_\g^{k_0}}d_px=(1-{2/p})p^\g, \quad k_0=1,2,\ldots,p-1 \quad [2a)]. \eqno(11.34)$$
$$\int_{S_{\g,k_n}}d_px=(1-{1/p})p^{\g-1}, \quad k_n=0,1,\ldots,p-1, n\in Z_+ \quad [2a)]. \eqno(11.35)$$
$$\int_{S_\g^{k_n}}d_px=(1-{1/p})^2p^\g, \quad k_n=0,1,\ldots,p-1, n\in Z_+ \quad [2a)]. \eqno(11.36)$$
$$\int_{S_{\g,k_0k_1\ldots k_n}}d_px=p^{\g-n-1},$$
$$k_j=0,1,\ldots,p-1, k_0\neq 0, n\in Z_+ \quad [2a)]. \eqno(11.37)$$
$$\int_{S_\g^{k_0k_1\ldots k_n}}d_px=(1-p^{-1}-p^{-n-1})p^\g,$$
$$k_j=0.,1,\ldots,p-1, k_0\neq 0, n\in Z_+ \quad [2a)]. \eqno(11.38)$$
$$\int_{\cap_{1\leq i\leq k}[|x-x_i|_p=1]}d_px=1-{k/p}, $$
$$1\leq k\leq p, |x_j-x_j|_p=1, i,j=1,2,\ldots,k, i\neq j \quad [9a)]. \eqno(11.39)$$
Let $\pi$ be a multiplicative character of the field $\QQ_p$ of rank $k\geq 1$.
$$\int_{S_\g}\pi (x)d_px=0 \quad [4]. \eqno(11.40)$$
Denote: $V_0=S_0, V_j=[x\in S_0: |1-x|_p\leq p^{-j}], j\in Z_+$.
$$\eqalignno{
\int_{V_j\backslash V_{j+1}}\pi (x)d_px&=0, \quad 0\leq j<k-1. & (11.41) \cr
                                       &=-p^{-k}, \quad j=k-1. & (11.42) \cr
                                       &=(1-{1/p})p^{-j}, \quad j\geq k \quad [11a)]. & (11.43) \cr
}$$
$$\int_{S_0}|1-x|_p^{\a-1}\pi (x)d_px=\G_p(\a)p^{-k\a}, \quad \Re\a>0 \quad [11a)]. \eqno (11.44)$$
$$\int_{S_\g}{\textstyle\sgn_{p,\epsilon}}xd_px=(1-{1/p})(-p)^\g,\quad \epsilon\notin\QQ_p^{\times 2}, |\epsilon |_p=1, p\neq 2 \quad [4]. \eqno(11.45)$$
$$\int_{B_\g}{\textstyle\sgn_{p,\epsilon}}xd_px={{p-1}\over{p+1}}(-p)^\g, \quad \epsilon\notin\QQ_p^{\times 2}, |\epsilon |_p=1, p\neq 2 \quad [4]. \eqno(11.46)$$
$$\eqalignno{
\int_{B_0}&{\textstyle\sgn_{p,\epsilon}}xd_px={{p-1}\over{p+1}}, \quad \epsilon\notin\QQ_p^{\times 2}, |\epsilon |_p=1, p\neq 2 \quad [4]. & (11.47) \cr
          &={1/3}, \quad \epsilon\equiv 5(\mod 8), p=2, & (11.48) \cr
          &=0, \quad |\epsilon |_p={1/p}, p\neq 2 \hbox{ or } \epsilon\not\equiv 1,5(\mod 8), p=2 \quad [4]. & (11.49) \cr
}$$
$$\eqalignno{
&\int_{B_0}\t_\epsilon^+(x)d_px={p\over{p+1}}, \quad \epsilon\notin\QQ_p^{\times 2}, |\epsilon |_p=1, p\neq 2. & (11.50) \cr
&={2/3}, \quad \epsilon\equiv 5(\mod 8), p=2. & (11.51) \cr
&={1/2}, \quad |\epsilon |_p={1/p}, p\neq 2 \hbox{ or } \epsilon\not\equiv 1,5(\mod 8), p=2 \quad [4]. & (11.52) \cr
}$$
$$\eqalignno{
&\int_{B_0}\t_\epsilon^-(x)d_px={1\over{p+1}}, \quad \epsilon\notin\QQ_p^{\times 2}, |\epsilon |_p=1, p\neq 2. & (11.53) \cr
&={1/3}, \quad \epsilon\equiv 5(\mod 8), p=2. & (11.54) \cr
&={1/2}, \quad |\epsilon |_p={1/p}, p\neq 2 \hbox{ or } \epsilon\not\equiv 1,5(\mod 8), p=2 \quad [4]. & (11.55) \cr
}$$
$$\eqalignno{
\int_{(B_0)^2}d_px & ={p\over{2(p+1)}}, \quad p\neq 2. & (11.56) \cr
                   & ={1/6}, \quad p=2 & (11.57) \cr
}$$		 
where $(B_0)^2$ is the set of squares of integers $p$-adic numbers $Z_p$.
$$\int_{\g (x)=2k\leq 0}d_px={p\over{p+1}}. \eqno(11.58)$$
$$\int_{\g (x)=2k\leq 0}f(|x|_p)d_px=(1-{1/p})\sum_{\g=0}^\infty p^{-2\g}f(p^{-2\g}). \eqno(11.59)$$
$$\int_{\g (x)-1=2k\leq 0}d_px={1\over{p+1}}. \eqno(11.60)$$
$$\int_{\g (x)-1=2k\leq 0}f(|x|_p)d_px=(1-{1/p})\sum_{\g=0}^\infty p^{-2\g-1}f(p^{-2\g-1}). \eqno(11.61)$$
$$\eqalignno{
\int_{B_0}\lambda_p(x)|x|_p^{-{1/2}}d_px&=1, \quad p\neq 2 \quad [1b)]. & (11.62) \cr
                                        &=2^{-{3/2}}, \quad p=2 \quad [1b)]. & (11.63) \cr
}$$
Let a function $f$ has the property
$$\int_{B_0}f(x+k)d_px=f(k), \quad k\in I_p$$
where $I_p$ is the set of indexes,
$$I_p=[k\in\QQ_p: k=p^{-\g}(k_0+k_1+\ldots +k_{\g-1}p^{\g-1}),$$
$$k_j=0,1,\ldots ,p-1, k_0\neq 0, j=0,1,\ldots ,\g-1, \g\in Z_+].$$
$$\int_{\QQ_p\backslash B_0}f(x)d_px=\sum_{k\in I_p}f(k) \quad [14]. \eqno(11.64)$$
$$\eqalignno{
&\int_{B_{-1}\backslash B_{-2n}}\lambda_p^2(x)|x|_p^{-1}d_px, \quad n\in Z_+ \cr
&=1-{1/p}, \quad p\equiv 3(\mod 4) \quad [1b)] & (11.65) \cr
&=(1-{1/p})(2n-1), \quad p\equiv 1(\mod 4) \quad [1b)] & (11.66) \cr
}$$
$$\int_{B_{-2}\backslash B_{-2n}}\lambda_2^2(x)|x|_2^{-1}d_2x=0, \quad p=2, n\geq 2 \quad [1b)]. \eqno (11.67)$$
Denote $|(x,m)|_p=\max (|x|_p,|m|_p).$
$$\int |(y,m)|_p^{\a-1}|(x-y,m)|_p^{\b-1}d_py=B_p(\a,\b)|(x,m)|_p^{\a+\b-1}$$
$$-\G_p(\a)|pm|_p^\a |(x,m)|_p^{\b-1}-\G_p(\b)|pm|_p^\b |(x,m)|_p^{\a-1}, $$
$$m\neq 0, \Re (\a+\b)<1 \quad [9a)]. \eqno(11.68)$$
Denote:
$$\script{K}_t(x,y)=\lambda_p(t)\sqrt {|2/t|_p}\chi_p\Bigl ({{2xy}\over{\sin t}}-{{x^2+y^2}\over{\tg t}}\Bigr ), \quad t\in G_p, x, y\in\QQ_p,$$
$$\script{K}_t(x)=\lambda_p(t)\sqrt {|2/t|_p}\chi_p\bigl (-{{x^2}/t}\bigr ), \quad t\in\QQ_p^\times, x\in\QQ_p.$$
$$\int\script{K}_t(x,y')\script{K}_\u(y',x)d_py'=\script{K}_{t+\u}(x,y), $$
$$t, \u\in G_p, x, y\in\QQ_p \quad [7]. \eqno(11.69)$$
$$\int_{B_0}\script{K}_t(x,y)d_py=\Omega (|x|_p), \quad t\in G_p, x\in\QQ_p \quad [7]. \eqno(11.70)$$
$$\script{K}_t(x,y)\to\delta (x-y), t\to 0 \hbox{ ¢ } \script{S}'(\QQ_p^2) \quad [7]. \eqno(11.71)$$
$$\int\script{K}_t(x-y)\script{K}_{\u}(y)d_py=\script{K}_{t+\u}(x), \quad t, \u\in\QQ_p^\times, x\in\QQ_p \quad [7]. \eqno(11.72)$$
$$\script{K}_t(x)\to\delta (x), t\to 0 \hbox{ ¢ } \script{S}' \quad [7]. \eqno(11.73)$$
$$\int_{|x|_p\neq 1}f(|x|_p)|1-x|_p^{-1}d_px=(1-p^{-1})\sum_{\g\neq 0}f(p^\g)\min (1,p^\g ). \eqno(11.74)$$
\bigskip
\centerline{\bf \Par 12. The Fourier integrals}
\medskip
{\it The Fourier integral} is called an integral of the form
$$\int f(x)\chi_p(\xi x)d_px, \quad \xi\in\QQ_p^\times.$$

$$\int_{B_\g}\chi_p(\xi x)d_px=p^\g\Omega (p^\g|\xi |_p) \quad [2a)]. \eqno(12.1)$$
$$\int_{S_\g}\chi_p(\xi x)d_px=(1-{1/p})p^\g\Omega (p^\g|\xi |_p)-p^{\g-1}\delta (|\xi |_p-p^{1-\g}) \quad [2a)]. \eqno(12.2)$$
$$\int\chi_p(\xi x)d_px=0, \quad \xi\neq 0 \quad [2a)]. \eqno(12.3)$$
$$\eqalignno{
&\int_{B_\g}f(|x|_p)\chi_p(\xi x)d_px \cr
&=(1-{1/p})\sum_{k=-\g}^\infty p^{-k}f(p^{-k}), \quad |\xi |_p\leq p^{-\g}. & (12.4) \cr
&=(1-{1/p})|\xi |_p^{-1}\sum_{k=0}^\infty p^{-k}f(p^{-k}|\xi |_p^{-1})-|\xi |_p^{-1}f(p|\xi |_p^{-1}), \cr
&|\xi |_p>p^{-\g} \quad [2a)]. & (12.5) \cr
&=\int f(|x|_p)\chi_p(\xi x)d_px, \quad |\xi |_p>p^{-\g}. & (12.6) \cr
}$$
$$\int f(|x|_p)\chi_p(\xi x)d_px, \quad \xi\neq 0$$
$$=(1-{1/p})|\xi |_p^{-1}\sum_{k=0}^\infty p^{-k}f(p^{-k}|\xi |_p^{-1})-|\xi |_p^{-1}f(p|\xi |_p^{-1}) \quad [2a)]. \eqno(12.7)$$
$$\int |x|_p^{\a-1}\chi_p(x)d_px={{1-p^{\a-1}}\over{1-p^{-\a}}}=\G_p(\a), \quad \Re\a>0 \quad [4]. \eqno(12.8)$$
$$\int |x|_p^{\a-1}\chi_p(\xi x)d_px=\G_p(\a)|\xi |_p^{-\a}, \quad \xi\neq 0, \Re\a>0 \quad [4]. \eqno(12.9)$$
$$\int\ln |x|_p\chi_p(x)d_px=-(1-{1/p})^{-1}\ln p \quad [2a)]. \eqno(12.10)$$
$$\int\ln |x|_p\chi_p(\xi x)d_px=-(1-{1/p})^{-1}\ln p|\xi |_p^{-1}, \quad \xi\neq 0 \quad [2a)]. \eqno(12.11)$$
$$\eqalignno{
&\int{{\chi_p(\xi x)}\over{|x|_p^2+m^2}}d_px, \quad m\neq 0 \cr
&=(1-{1/p})\sum_{k=-\infty}^\infty{{p^k}\over{p^{2k}+m^2}}, \quad \xi =0. & (12.12) \cr
&=(1-{1/p}){{|\xi |_p}\over{p^2+m^2|\xi |_p^2}}\sum_{k=0}^\infty p^{-k}{{p^2-p^{-2k}}\over{p^{-2k}+m^2|\xi |_p^2}}, \quad \xi\neq 0 \quad [2a)]. & (12.13) \cr
&\sim{{p^4+p^3}\over{p^2+p+1}}m^{-4}|\xi |_p^{-3}+O(|\xi |_p^{-5}), \quad |\xi |_p\to\infty \quad [2a)]. & (12.14) \cr
}$$
$$
\allowdisplaybreaks
\eqalignno{
&\mu_t^\a(x)= \int\exp (-t|\xi |_p^\a)\chi_p(\xi x)d_p\xi, \quad t>0, \a>0 \cr
&=(1-{1/p})|x|_p^{-1}\sum_{\g=0}^\infty p^{-\g}\exp (-t|px|_p^{-\a}) \cr
&\times\Bigr (\exp [t|px|_p^{-\a}(1-p^{-\a\g-\a})]-1\Bigl )>0. & (12.15) \cr
&=\sum_{n=1}^\infty{{(-t)^n}\over{n!}}\G_p (\a n+1)|\xi |_p^{-\a n-1} \quad [1a)]. & (12.16) \cr
&\int\mu_t^\a(x)d_px=1, \quad t>0. & (12.17) \cr
&\mu_t^\a(x)\to\delta (x), \quad t\to 0 \hbox{ in } \script{S}' \quad [1a)]. & (12.18) \cr
&\mu_t^\a\ast\mu_\u^\a=\mu_{t+\u }^\a, \quad t,\u >0 \quad [1a)]. & (12.19) \cr
&\int_0^\infty\mu_t^\a(x)dt=\G_p^{-1}(\a )|x|_p^{\a-1}=f_\a(x), \quad \a\neq\a_k, k\in Z. & (12.20) \cr
&{\partial\over{\partial t}}\mu_p^\a(x)|_{t=0}=\G_p(\a+1)|x|_p^{-\a-1}, \quad \a>0. & (12.21) \cr
&|x|_p^\a=-\G_p^{-1}(-\a)\int [1-\Re\chi_p(x\xi )]|\xi |_p^{-\a-1}d_p\xi & (12.22) \cr
}$$
and also
$$-\G_p^{-1}(-\a )|\xi |_p^{-\a-1}d_p\xi >0, \a> 0.$$
$$
\allowdisplaybreaks
\eqalignno{
&\int_{B_{-1}}\chi_p(a^2\tg\xi-x\xi )d_p\xi, \quad p\neq 2 \quad [1b)] \cr
&={1/2}\Omega (|px|_p), \quad |a|_p\leq 1. & (12.23) \cr
&={1/2}\delta (|x|_p-p^2)\delta (x_0-(a^2)_0), \quad |a|_p=p. & (12.24)  \cr
&={1/2}\delta (|x|_p-|a|_p^2)\delta (x_0-(a^2)_0)\delta (x_1-(a^2)_1)\varphi_a(x), \cr
&|a|_p\geq p^2 & (12.25) \cr
}$$
where $\varphi_a(x)$ is a continuous function.
$$\eqalignno{
&\int |x|_p^{\a-1}|x-a|_p^{\b-1}\chi_p(x)d_px, \quad \Re\a>0, \Re\b>0, \Re (\a+\b)<1 \cr
&=B_p(\a,\b)|a|_p^{\a+\b-1}+\G_p(\a+\b-1), \quad |a|_p\leq 1. & (12.26) \cr
&=\G_p(\a)|a|_p^{\b-1}+\G_p(\b)|a|_p^{\a-1}\chi_p(a), \quad |a|_p\geq p. & (12.27) \cr
}$$
$$\int_{S_\g}|x-a|_p^{\a-1}\chi_p(x-a)d_px=\G_p(\a),$$
$$|a|_p=p^\g, \g\geq 2, \Re\a>0. \eqno (12.28)$$

Let $n\in Z_+$ be not divisible by $p$ and $P$ be a polynom of degree $n$,
$$P(x)=\a_1x+\a_2x^2+\ldots+\a_nx^n, |\a_k|_p\leq 1, k=1,2,\ldots,n-1, |\a_n|_p=1.$$
$$\eqalignno{
&\int_{S_\g}\chi_p[P(x)]d_px=(1-{1/p})p^\g, \quad \g\leq 0, n\in Z_+ \quad [2a)]. & (12.29) \cr
&=0, \quad \g=2,3,\ldots, n\in Z_+ \hbox{ ¨«¨ } \g=1, n=2,3,\ldots \quad [2a)]. & (12.30) \cr
&=-1, \quad \g=1,n=1 \quad [2a)]. & (12.31) \cr
}$$
$$\eqalignno{
&\int_{B_\g}\chi_p[P(x)]d_px=p^\g, \quad \g\leq 0, n\in Z_+. & (12.32) \cr
&=1, \quad \g=2,3,\ldots, n\in Z_+ \hbox{ ¨«¨ } \g=1, n=2,3,\ldots. & (12.33) \cr
&=0, \quad \g\in Z_+, n=1. & (12.34) \cr
}$$
$$\eqalignno{
\int\chi_p[P(x)]d_px&=1, \quad n=2,3,\ldots. & (12.35) \cr
                    &=0, \quad n=1. & (12.36) \cr
}$$
Let (complex) numbers $\eta_1,\eta_2,\ldots,\eta_{p-1}$ be such that
$$\sum_{k=1}^{p-1}\eta_k=0, \quad p\neq 2,$$
and numbers $\eta_1',\eta_2',\ldots,\eta_{p-1}'$ are mutual to 
$\{\eta_k\}, k=1,2,\ldots,p-1$,
$$\eta_j'=\sum_{k=1}^{p-1}\eta_k\exp (2\pi i{{kj}/p}), \quad \sum_{j=1}^{p-1}\eta_j'=0.$$
$$\int_{S_\g}\eta_{x_0}\chi_p(\xi x)d_px=p^{\g-1}{\eta}'_{\xi_0}\delta (|\xi |_p-p^{1-\g}) \quad [2b)]. \eqno (12.37)$$
$$\eqalignno{
&\int_{B_\g}|x|_p^{\a-1}\chi_p(\xi x)d_px, \quad \Re\a>0 \cr
&={{1-p{-1}}\over{1-p^{-\a}}}p^{\a\g}, \quad |\xi |_p\leq p^{-\g}. & (12.38) \cr
&=\G_p(\a)|\xi |_p^{-\a}, \quad |\xi |_p>p^{-\g} \quad [1a)]. & (12.39) \cr
&=\G_p(\a), \quad \xi =1, \g\geq 1 \quad [2e)]. & (12.40) \cr
}$$
$$\int_{S_0}\delta (x_0-k)\chi_p(\xi x)d_px$$
$$=p^{-1}\chi_p(k\xi )\Omega (|p\xi |_p), \quad k=1,2,\ldots,p-1. \eqno(12.41)$$
$$\int\delta (x_0-k)\chi_p(\xi x)d_px $$
$$=|\xi |_p^{-1}\Bigl ({1\over{p-1}}+\chi_p({{k\xi_0}/ p})\Bigr ), \quad \xi\neq 0, \quad k=1,2,\ldots,p-1. \eqno(12.42)$$
$$\int_{B_1}\chi_p[(k-\xi )x]d_px=p\delta (|\xi |_p-1)\delta (\xi_0-k), \quad k=1,2,\ldots,p-1. \eqno(12.43)$$
$$\int_{S_0}\delta (x_1-k)\chi_p(\xi x)d_px={1/p}(1-{1/p})\Omega (|\xi |_p)-p^{-2}\delta (|\xi |_p-p)$$
$$+p^{-2}{{\chi_p(\xi )-\chi_p(p\xi )}\over{1-\chi_p(\xi )}}\chi_p(kp\xi )\delta (|\xi |_p-p^2), \quad k=0,1,\ldots,p-1. \eqno(12.44)$$
$$\int\delta (x_1-k)\chi_p(\xi x)d_px $$
$$=|\xi |_p^{-1}{{\chi_p(p^{-2}|\xi |_p\xi )-\chi_p(p^{-1}\xi_0 )}\over{1-\chi_p(p^{-1}\xi_0 )}}\chi_p(kp^{-1}\xi_0),$$
$$\xi\neq 0, p=0,1,\ldots,p-1. \eqno(12.45)$$
$$\int_{S_0}\delta (x_2-k)\chi_p(\xi x)d_px={1/p}(1-{1/p})\Omega (|\xi |_p)-p^{-2}\delta (|\xi |_p-p)$$
$$+p^{-3}{{\chi_p(\xi )-\chi_p(p\xi )}\over{1-\chi_p(\xi )}}{{\chi_p(kp^2\xi )-\chi_p((k+1)p^2\xi )}\over{1-\chi_p(p\xi )}}\delta (|\xi |_p-p^3),$$
$$k=0,1,\ldots,p-1. \eqno(12.46)$$
$$\int\delta (x_2-k)\chi_p(\xi x)d_px $$
$$=|\xi |_p^{-1}{{\chi_p(p^{-3}|\xi |_p\xi )-\chi_p(p^{-2}|\xi |_p\xi )}\over{1-\chi_p(p^{-3}|\xi |_p\xi )}}{{\chi_p(kp^{-1}\xi_0)-\chi_p((k+1)p^{-1}\xi_0)}\over{1-\chi_p(p^{-2}|\xi |_p\xi )}},$$
$$\xi\neq 0, k=0,1,\ldots,p-1. \eqno(12.47)$$
$$\int |x,m|_p^{\a-1}\chi_p(\xi x)d_px $$
$$=\G_p(\a)\bigl (|\xi |_p^{-\a}-|pm|_p^\a\bigr )\Omega (|m\xi |_p), \quad m\neq 0, \Re\a<0 \quad [9a)]. \eqno(12.48)$$
$$\int |x,1|_p^{-\a}\chi_p(\xi x)d_px $$
$$=\G_p(1-\a)(|\xi |_p^{\a-1}-p^{\a-1})\Omega (|\xi |_p)\equiv J_p^\a(\xi ), \quad \Re\a>0. \eqno(12.49)$$
$$J_p^1(\xi )=(1-{1/p})\biggl (1-{{\ln |\xi |_p}\over{\ln p}}\biggr )\Omega (|\xi |_p), \quad \a=1. \eqno(12.50)$$
$$\int J_p^\a(\xi )J_p^\b(x-\xi )d_p\xi =J_p^\a\ast J_p^\b=J_p^{\a+\b}, \quad \a, \b\in\CC. \eqno(12.51)$$
$$\ln |x,1|_p=\int\bigl (1-\Re\chi_p(x\xi )\bigr )d\s (\xi ) $$
$$=\ln p\sum_{\g=0}^\infty p^\g\Omega (p^\g |\xi |_p), \quad d\s (\xi )\geq 0. \eqno(12.52)$$
$$\eqalignno{
&\int_{B_{-1}\backslash B_{-2n}}\lambda_p^2(x)|x|_p^{-1}\chi_p(\xi x)d_px, \quad n\in Z_+ \cr
&=(1-{1/p})(2n-1), \quad \xi =0 \hbox{ ¨«¨ } \g (\xi )\leq 1, p\equiv 1(\mod 4) \quad [1b)]. & (12.53) \cr
&=(1-{1/p})(2n-\g (\xi ))-{1/p}, \cr
& 2\leq\g (\xi )\leq 2n, p\equiv 1(\mod 4) \quad [1b)]. & (12.54) \cr
&=1-{1/p}, \quad \xi =0 \hbox{ ¨«¨ } \g (\xi )\leq 1, p\equiv 3(\mod 4) \quad [1b)]. & (12.55) \cr
&={1/2}(-1)^{\g(\xi )}(1+{1/p})-{1/2}(1-{1/p}), \cr
& 2\leq\g (\xi )\leq 2n, p\equiv 3(\mod 4) \quad [1b)]. & (12.56) \cr
}$$
$$\int_{B_{-1}}\lambda_p^2(x)|x|_p^{-1}\chi_p(\xi x)d_px $$
$$=1-{1/p}, \quad \xi =0 \hbox{ ¨«¨ } \g (\xi )\leq 1, p\equiv 3(\mod 4) \quad [1b)]. \eqno (12.57) $$
$$={1/2}(-1)^{\g(\xi )}(1+{1/p})-{1/2}(1-{1/p}),$$
$$\g (\xi )\geq 2, p\equiv 3(\mod 4) \quad [1b)]. \eqno(12.58)$$
$$\eqalignno{
&\int_{B_{-2}\backslash B_{-2n}}\lambda_2^2(x)|x|_2^{-1}\chi_2(\xi x)d_2x, \quad n=2,3,\ldots \cr
&=0, \quad \g(\xi )\leq 3 \quad [1b)]. & (12.59) \cr
&={1/4}(-1)^{\xi_1+1}, \quad \g(\xi )\geq 4 \quad [1b)]. & (12.60) \cr
}$$
$$\eqalignno{
&\int_{B_{-2}}\lambda_2^2(x)|x|_2^{-1}\chi_2(\xi x)d_2x=0, \quad \xi =0 \hbox{ or } \g (\xi )\leq 3 \quad [1b)]. & (12.61) \cr
&={1/2}(-1)^{\xi_1+1}, \quad \g(\xi )\geq 4 \quad [1b)]. & (12.62) \cr
}$$
$$\eqalignno{
&\int {\textstyle\sgn_{p,d}}x|x|_p^{\a-1}\chi_p(\xi x)d_px, \quad d\not\in\QQ_p^{\times 2} \cr
&=\tilde{\G}_p(\a ){\textstyle\sgn_{p,d}}\xi |\xi |_p^{-\a}, \quad |d|_p=1, \Re\a>0 \quad [4]. & (12.63) \cr
&=\pm p^{\a-{1/2}}\sqrt{{\textstyle\sgn_{p,d}}(-1)}{\textstyle\sgn_{p,d}}\xi |\xi |_p^{-\a}, \quad |d|_p={1/p}, \a\in\CC \quad [4]. & (12.64) \cr
}$$
Let $\varepsilon =\pm be.$
$$
\allowdisplaybreaks
\eqalignno{
&\int_{S_\g}\lambda_p(x)\chi_p(\varepsilon\xi x)d_px \quad [1b)],[15] \cr
&=p^\g(1-{1/p}), \quad |\xi |_p\leq p^{-\g}, \g=2k. & (12.65) \cr
&=0, \quad |\xi |_p\leq p^{-\g}, \g=2k+1. & (12.66) \cr
&=-p^{\g-1}, \quad |\xi |_p=p^{-\g+1}, \g=2k. & (12.67) \cr
&=\Bigl ({{\xi_0}\over p}\Bigr )p^{\g-{1/2}}, \quad |\xi |_p\leq p^{-\g+1}, \g=2k+1, p\equiv 1(\mod 4). & (12.68) \cr
&=-\varepsilon\Bigl ({{\xi_0}\over p}\Bigr )p^{\g-{1/2}}, \quad |\xi |_p\leq p^{-\g+1}, \g=2k+1, p\equiv 3(\mod 4). & (12.69) \cr
&=0, \quad |\xi |_p\geq p^{-\g+2}. & (12.70) \cr
}$$
$$
\allowdisplaybreaks
\eqalignno{
&\int_{S_\g}\lambda_2(x)\chi_2(\varepsilon\xi x)d_2x \quad [10b)],[15] \cr
&=2^{\g-{3/2}}, \quad |\xi |_2\leq 2^{-\g}, \g=2k. & (12.71) \cr
&=0, \quad |\xi |_2\leq 2^{-\g}, \g=2k+1. & (12.72) \cr
&=-2^{\g-{3/2}}, \quad |\xi |_2=2^{-\g+1}, \g=2k. & (12.73) \cr
&=0, \quad |\xi |_2=2^{-\g+1}, \g=2k+1. & (12.74) \cr
&=-\varepsilon (-1)^{\xi_1}2^{\g-{3/2}}, \quad |\xi |_2=2^{-\g+2}, \g=2k. & (12.75) \cr
&=0, \quad |\xi |_2=2^{-\g+2}, \g=2k+1. & (12.76) \cr
&=0, \quad |\xi |_2\geq 2^{-\g+3}, \g=2k. & (12.77) \cr
&=i^{\xi_1}(-1)^{\xi_2}2^{\g-3}(1+i)(1+i\varepsilon )[1-\varepsilon (-1)^{\xi_1}], \cr
&|\xi |_2=2^{-\g+3}, \g=2k+1. & (12.78) \cr
&=0, \quad |\xi |_2\geq 2^{-\g+4}, \g=2k+1. & (12.79) \cr
}$$
$$\eqalignno{
&\int_{|x|_p\geq 1}\lambda_p(x)|x|_p^{\a-1}\chi_p(\varepsilon\xi^2 x)d_px=0, \quad |\xi |_p\geq p, p\neq 2. & (12.80) \cr
&=(1-{1/p}){{1-p^{2\a}|\xi |_p^{-2\a}}\over{1-p^{2\a}}}+p^{\a-{1/2}}|\xi |_p^{-2\a}, \cr
&|\xi |_p\leq 1, p\equiv 1(\mod 4). & (12.81) \cr
&=(1-{1/p}){{1-p^{2\a}|\xi |_p^{-2\a}}\over{1-p^{2\a}}}-\varepsilon p^{\a-{1/2}}|\xi |_p^{-2\a}, \cr
&|\xi |_p\leq 1, p\equiv 3(\mod 4). & (12.82) \cr
}$$
$$\int_{|x|_p\geq 1}\lambda_p(x)|x|_p^{-{3/2}}\chi_p(-\xi^2 x)d_px=\Omega (|\xi |_p), \quad p\neq 2 \quad [15]. \eqno(12.83)$$
$$\int_{|x|_2\geq 4}\lambda_2(x)|x|_p^{-{3/2}}\chi_p(-\xi^2x)d_2x=\sqrt 2\Omega (|\xi |_p), \quad p=2 \quad [15]. \eqno(12.84)$$
\bigskip
\centerline{\bf \Par 13. The Gaussian integrals}
\medskip
{\it The Gaussian integral} is called an integral of the form
$$\int f(x)\chi_p(ax^2+bx)d_px, \quad a\in\QQ_p^\times, b\in\QQ_p.$$
Various formulae for the Gaussian integrals are contained in [2a)],[6]--[8],
[10b)]. The most full lists of them are collected in [1a)],[1b)]. Here
$$\epsilon =\varepsilon_0+\varepsilon_1p+\varepsilon_2p^2+\ldots.$$
$$\eqalignno{
&\int_{S_\g}\chi_p[\epsilon (x-y)^2]d_py \cr
&=p^\g\chi_p(\epsilon x^2)\bigl [(1-{1/p})\Omega (p^\g|x|_p)-{1/p}\delta (|x|_p-p^{1-\g})\bigr ], \cr
&\g\leq 0, p\neq 2. & (13.1) \cr
&=\delta (|x|_p-p^\g), \quad \g\geq 1, p\neq 2. & (13.2) \cr
&=2^{\g-1}\chi_2(\epsilon x^2)\bigl [\Omega (2^{\g-1}|x|_2)-\delta (|x|_2-2^{2-\g})\bigr ], \quad \g\leq 0, p=2. & (13.3) \cr
&=[\sqrt 2\lambda_2(\epsilon )-1]\Omega (|x|_2)+\delta (|x|_2-2), \quad \g=1, p=2. & (13.4) \cr
&=\sqrt 2\lambda_2(\epsilon )\delta (|x|_2-2^\g), \quad \g\geq 2, p=2. & (13.5) \cr
}$$
$$\eqalignno{
&\int_{S_\g}\chi_p[\epsilon p(x-y)^2]d_py \cr
&=p^\g\chi_p(\epsilon px^2)\bigl [(1-{1/p})\Omega (p^{1-\g}|x|_p) \cr
&-{1/p}\delta (|x|_p-p^{2-\g})\bigr ], \quad\g\leq 0, p\neq 2. & (13.6) \cr
&=[\sqrt p\lambda_p(\epsilon p)-\chi_p(\epsilon px^2)]\Omega (|px|_p), \quad \g=1, p\neq 2. & (13.7) \cr
&=\sqrt p\lambda_p(\epsilon p)\delta (|x|_p-p^\g), \quad \g\geq 2, p\neq 2. & (13.8) \cr
&=2^{\g-1}\chi_2(2\epsilon x^2)[\Omega (2^{\g-2}|x|_2)-\delta (|x|_2-2^{3-\g})], \quad\g\leq 0, p=2. & (13.9) \cr
&=-\Omega (|x|_2)+\delta (|x|_2-2)+\lambda_2(2\epsilon )\delta (|x|_2-4), \quad \g=1, p=2. & (13.10) \cr
&=2\lambda_2(2\epsilon )\Omega (|2x|_2), \quad \g=2, p=2. & (13.11) \cr
&=2\lambda_2(2\epsilon )\delta (|x|_2-2^\g), \quad \g\geq 3, p=2. & (13.12) \cr
}$$
$$\eqalignno{
&\int_{S_\g}\chi_p(ax^2+\xi x)d_px \cr
&=\lambda_p(a)|2a|_p^{-{1/2}}\chi_p\bigl (-{{\xi^2}/{4a}}\bigr )\delta\bigl (|{\xi/{2a}}|_p-p^\g\bigr ), \cr
&|4a|_p\geq p^{2-2\g}. & (13.13) \cr
&=|2a|_p^{-{1/2}}\Bigl [\lambda_p(a)\chi_p\bigl (-{{\xi^2}/{4a}}\bigr )-{1\over{\sqrt p}}\Bigr ]\Omega (p^{1-\g}|\xi |_p), \cr
&|a|_p=p^{1-2\g}. & (13.14) \cr
}$$
$$\eqalignno{
&\int_{B_\g}\chi_p(ax^2+\xi x)d_px=p^\g\Omega (p^\g|\xi |_p), \quad |a|_pp^{2\g}\leq 1. & (13.15) \cr
&=\lambda_p(a)|2a|_p^{-{1/2}}\chi_p\bigl (-{{\xi^2}/{4a}}\bigr )\Omega (\bigl (p^{-\g}|{\xi/{2a}}|_p\bigr ), \quad |4a|_pp^{2\g}>1. & (13.16) \cr
&=2^\g\lambda_2(a)\chi_2\bigl (-{{\xi^2}/{4a}}\bigr )\delta (|\xi |_2-2^{1-\g}), \quad |a|_22^{2\g}=2, p=2. & (13.17) \cr
&=2^{\g-{1/2}}\lambda_2(a)\chi_2\bigl (-{{\xi^2}/{4a}}\bigr )\Omega (2^\g|\xi |_2), \quad |a|_22^{2\g}=4, p=2. & (13.18) \cr
}$$
$$\eqalignno{
&\int\chi_p(ax^2+\xi x)d_px, \quad a\neq 0 \cr
&=\lambda_p(a)|2a|_p^{-{1/2}}\chi_p\bigl (-{{\xi^2}/{4a}}\bigr ). & (13.19) \cr
&=\chi_p(-{{\xi^2}/2}), \quad a={1/2}, p\neq 2. & (13.20) \cr
&=\exp (i{\pi}/4)\chi_p(-{{\xi^2}/2}), \quad a={1/2}, p=2. & (13.21) \cr
}$$
$$
\allowdisplaybreaks
\eqalignno{
&\int\exp{(-|y|_p^2)}\chi_p[a(x-y)^2]d_py, \quad a\neq 0, \g=\g(a) \cr
&=|a|_p^{-{1/2}}S(|a|_p^{-1},{1/p}), \quad |x|_p\sqrt {|a|_p}\leq 1, \g=2k, p\neq 2. & (13.22) \cr
&={1/{\sqrt p}}|a|_p^{-{1/2}}S({1/p}|a|_p^{-1},{1/p})+[\lambda_p(a)-{1/{\sqrt p}}]|a|_p^{-{1/2}}\exp (-|pa|_p^{-1}), \cr
&|x|_p\sqrt{p|a|_p}\leq 1, \g=2k+1, p\neq 2. & (13.23) \cr
&=\lambda_p(a)|a|_p^{-{1/2}}\exp (-|x|_p^2)+|ax|_p^{-1}\chi_p(ax^2)[S(|ax|_p^{-2},{1/p}) \cr
&-\exp (-|pax|_p^{-2})], \quad |x|_p\sqrt{|a|_p}\geq\sqrt p, p\neq 2. & (13.24) \cr
&=[\sqrt 2\lambda_2(a)-1]|a|_2^{-{1/2}}\exp (-|4a|_2^{-1})+|a|_2^{-{1/2}}S(|a|_2^{-1},{1/2}), \cr
&|x|_2\sqrt{|a|_2}\leq 1, \g=2k, p=2. & (13.25) \cr
&=|a|_2^{-{1/2}}\exp (-|4a|_2^{-1})+[\sqrt 2\lambda_2(a)-1]|a|_2^{-{1/2}}S(|a|_2^{-1},{1/2}), \cr
&|x|_2\sqrt{|a|_2}=2, \g=2k, p=2. & (13.26) \cr
&=(2|a|_2)^{-{1/2}}S((2|a|_2)^{-1},{1/2})-(2|a|_2)^{-{1/2}}\exp (-|2a|_2^{-1}) \cr
&+\lambda_2(a)|2a|_2^{-{1/2}}\exp (-|8a|_2^{-1}), \cr
&|x|_2\sqrt{2|a|_2}\leq 1, \g=2k+1, p=2. & (13.27) \cr
&=|2a|_2^{-{1/2}}S(|2a|_2^{-1},{1/2})+\lambda_2(a)|2a|_2^{-{1/2}}\exp (-|8a|_2^{-1}), \cr
&|x|_2\sqrt{|a|_2}=\sqrt 2, \g=2k+1, p=2. & (13.28) \cr
&=\lambda_2(a)(2|a|_2)^{-{1/2}}S(|2a|_2^{-1},{1/2}), \cr
&|x|_2\sqrt{|a|_2}=2\sqrt 2, \g=2k+1, p=2. & (13.29) \cr
&=\lambda_2(a)|2a|_2^{-{1/2}}\exp (-|x|_2^2)+|2ax|_2^{-1}\chi_2(ax^2)[S(|2ax|_2^{-2},{1/2}) \cr
&-2\exp (-|4ax|_2^{-2})], \quad |x|_2\sqrt{|a|_2}>2, p=2. & (13.30) \cr
&\sim{{p^4+p^3}\over{p^2+p+1}}|2ax|_p^{-3}\chi_p(ax^2)+O(|x|_p^{-5}), |x|_p\to\infty. & (13.31) \cr
&\sim |a|_p^{-{1/2}}S(|a|_p^{-1},p^{-1})+O[|a|_p^{-{1/2}}\exp (-|p^2a|_p^{-1})], \cr
&|a|_p\to 0, \g=2k. & (13.32) \cr
&\sim (p|a|_p)^{-{1/2}}S((p|a|_p)^{-1},p^{-1})+O[|a|_p^{-{1/2}}\exp (-|pa|_p^{-1})], \cr
&|a|_p\to 0, \g=2k+1. & (13.33) \cr
}$$
Here
$$S(\a,q)=(1-q)\sum_{k=0}^\infty{{(-\a)^k}\over{k!(1-q^{2k+1})}}, \quad |q|<1, \a\in\CC.$$
This function satisfies the relation
$$S(\a q^2,q)={1/q}S(\a,q)+(1-{1/q})e^{-\a}. \eqno(13.34)$$
\bigskip
\centerline{\bf\Par 14. Two variables}
\medskip
$$\int_{B_0^2}d_p^2x=1. \eqno (14.1)$$
$$\int_{B_{\g}^2}d_p^2x=p^{2\g}. \eqno (14.2)$$
$$\int_{S_{\g}^2}d_p^2x=(1-p^{-2})p^{2\g}. \eqno(14.3)$$
$$\int_{B_{\g}^2}f(|x|_p)d_p^2x=(1-p^{-2})\sum_{k=-\infty}^{\g}p^{2k}f(p^k). \eqno(14.4)$$
$$\int f(|x|_p)d_p^2x=(1-p^{-2})\sum_{k=-\infty}^{\infty}p^{2k}f(p^k). \eqno(14.5)$$
$$\int_{B_{\g}^2}|x|_p^{\a-2}d_p^2x={{1-p^{-2}}\over{1-p^{-\a}}}p^{\a\g}, \quad \Re\a>0. \eqno(14.6)$$
$$\int_{S_{\g}^2}|x|_p^{\a-2}d_p^2x=(1-p^{-2})p^{\a\g}. \eqno(14.7)$$
$$\eqalignno{
&\int_{B_{\g}^2}|(x,x)|_p^{\a-1}\chi_p((\xi,x))d_p^2x, \quad \Re\a>0, |(\xi,\xi )|_p>p^{-\g} \cr
&=\G_p^2(\a)|(\xi,\xi )|_p^{-\a}, \quad p\equiv 1(\mod 4) \quad [3a)]. & (14.8) \cr
&=\G_p(\a)\tilde{\G}_p(\a)|(\xi,\xi )|_p^{-\a}, \quad p\equiv 3(\mod 4) \quad [3a)]. & (14.9) \cr
}$$
$$\eqalignno{
&\int |(x,x)|_p^{\a-1}\chi_p((\xi ,x))d_p^2x, \quad \Re\a>0, (\xi,\xi )\neq 0 \cr
&=\G_p^2(\a)|(\xi,\xi )|_p^{-\a}, \quad p\equiv 1(\mod 4) \quad [3a)]. & (14.10) \cr
&=\G_p(\a)\tilde{\G}_p(\a)|(\xi,\xi )|_p^{-\a}, \quad p\equiv 3(\mod 4) \quad [3a)]. & (14.11) \cr
}$$
$$\eqalignno{
&\int f((x,x))\chi_p((\xi,x))d_p^2x, \quad (\xi,\xi )\neq 0 \cr
&=|(\xi,\xi )|_p^{-1}\bigl [(1-p^{-2})\sum_{\g=0}^\infty p^{-2\g}f(p^{-2\g}|(\xi,\xi )|_p^{-1})-f(p^2|(\xi,\xi )|_p^{-1})\bigr ], \cr
&p\equiv 3(\mod 4) [1a)]. & (14.12) \cr
&=|(\xi,\xi )|_p^{-1}\Bigl [(1-{1/p})^{-2}\sum_{\g=0}^\infty \Bigl (\g+{{p-3}\over{p-1}}\Bigr )p^{-\g}f(p^{-\g}|(\xi,\xi )|_p^{-1}) \cr
&-2(1-{1/p})f(p|(\xi,\xi )|_p^{-1})+f(p^2|(\xi,\xi )|_p^{-1})\Bigr ],\cr
&p\equiv 1(\mod 4) \quad [1a)]. & (14.13) \cr
}$$
$$\int{{\chi_p((\xi,x))}\over{|(x,x)|_p+m^2}}d_p^2x, \quad m\neq 0, (\xi,\xi )\neq 0$$
$$={{1-p^{-2}}\over{p^2+m^2|(\xi,\xi )|_p}}\sum_{\g=0}^\infty{{p^2-p^{-2\g}}\over{1+p^\g m^2 |(\xi,\xi )|_p}}$$
$$=\sum_{\g=0}^\infty{{1}\over{1+p^{2\g}m^2|(\xi,\xi )|_p}}-{1\over{p^2+p^{2\g}m^2|(\xi,\xi )|_p}},$$
$$p\equiv 3(\mod 4) \quad [1a)],[3a)]. \eqno(14.14)$$
$$=(1-{1/p})^2\sum_{\g=0}^\infty \Bigl (\g+{{p-3}\over{p-1}}\Bigr ){{1}\over{1+p^\g{m^2}|(\xi,\xi )|_p}}$$
$$-2(1-{1/p}){{1}\over{p+m^2|(\xi,\xi )|_p}}+{{1}\over{p^2+m^2|(\xi,\xi )|_p}}$$
$$=\sum_{\g=0}^\infty (\g+1)\Bigl ({{1}\over{1+p^\g{m^2}|(\xi,\xi )|_p}}-{{2}\over{p+p^\g{m^2}|(\xi,\xi )|_p}}$$
$$+{{1}\over{p^2+p^\g{m^2}|(\xi,\xi )|_p}}\Bigr ), \quad p\equiv 1(\mod 4) \quad [3a)]. \eqno(14.15)$$
$$\sim{{p^4}\over{p^2+1}}m^{-4}|(\xi,\xi )|_p^{-2}, |(\xi,\xi )|_p\to\infty, \quad \equiv 3(\mod 4) \quad [1a)],[3a)]. \eqno(14.16)$$
$$\sim -{{p^4}\over{(p+1)^2}}m^{-4}|(\xi,\xi )|_p^{-2}, |(\xi,\xi )|_p\to\infty, \quad p\equiv 1(\mod 4) \quad [3a)]. \eqno(14.17)$$
$$\int |x|_p^{\a-1}|1-x|_p^{\b-1}|x-y|_p^{\g}|y|_p^{\a'-1}|1-y|_p^{\b'-1}d_pxd_py $$
$$=\G_p(\g)\int |t|_p^{2-\a-\b-\a'-\b'}B_p(t;\a,\b)B_p(-t;\a',\b')d_pt $$
$$=B_p(\a,\b)B_p(\a',\b')+B_p(\a,\b)B_p(\g,\a'+\b'-1)+B_p(\a',\b')B_p(\g,\a+\b-1) $$
$$+B_p(\a+\b-1,\a'+\b'-1)B_p(\g,3-\a-\b-\a'-\b')-B_p(\a,\a')B_p(\g,\a+\a') $$
$$-B_p(\b,\b')B_p(\g,\b+\b')+\G_p(\g)p^{-\g}\Bigl\{[\G_p(\a+\b-1)p^{1-\a-\b}+B_p(\a,\b)] $$
$$\times[\G_p(\a'+\b'-1)p^{1-\a'-\b'}+B_p(\a',\b')]-[\G_p(\a)p^{-\a}+\G_p(\b)p^{-\b}] $$
$$\times[\G_p(\a')p^{-\a'}+\G_p(\b')p^{-\b'}]\Bigr\} , $$
$$\Re\a>0, \Re\b>0, \Re\g>0, \Re\a'>0, \Re\b'>0. \eqno(14.18) $$
Here
$$B_p(t;\a,\b)=\int |x|_p^{\a-1}|t-x|_p^{\b-1}\chi_p(x)d_px, \quad B_p(1;\a,\b)=B_p(\a,\b).$$

Below in formulas (14.19)--(14.29) we use the notations for the field
$\QQ_p(\sqrt d)$, $d\notin\QQ_p^{\times 2}$ (see \Par 9). In particular, (see
(9.2) and (9.6))
$$B_\g^2=[z\in\QQ_p(\sqrt d): |z\bar z|_p\leq q^\g]; \quad \a_k={{2k\pi i}/{\ln q}}, k\in Z.$$
$$\int_{B_0^2}d_pz=1. \eqno (14.19)$$
$$\int_{B_0^2}|z\bar z|_p^{\a-1}d_pz={{1-q^{-1}}\over{1-q^{-\a}}}, \quad \Re\a>0. \eqno(14.20)$$
$$\int_{B_\g^2}|z\bar z|_p^{\a-1}d_pz={{1-q^{-1}}\over{1-q^{-\a}}}q^{\a\g}, \quad \Re\a>0. \eqno(14.21)$$
$$\int_{B_0^2}f(z\bar z)d_pz={1\over {C_{p,d}}}\int_{B_0}f(x)\t_d^+(x)d_px, \quad p\neq 2 \quad [4] \eqno(14.22)$$
where quantity $C_{p,d}$ is defined in (9.16) and (9.17).
$$\eqalignno{
&\delta\sqrt{|4d|_p}\int |z\bar z|_p^{\a-1}\chi_p(z\zeta +\bar z\bar\zeta)d_pz, \quad \Re\a>0 \cr
&={\textstyle \G_{p,d}}(\a)|\zeta\bar\zeta |_p^{-\a}, \quad \zeta\neq 0 \quad [2a)]. & (14.23) \cr
&={\textstyle \G_{p,d}}(\a), \quad \zeta=1. & (14.24) \cr
}$$
$$\delta\sqrt{|4d|_p}\int_{B_\g^2}|z\bar z|_p^{\a-1}\chi_p(z+\bar z)d_pz=\G_{p,d}(\a ), $$
$$\Re\a>0, \g\geq 1 \quad [2a)]. \eqno (14.25)$$
$$\eqalignno{
&\int |\zeta\bar\zeta |_p^{\a-1}|(z-\zeta )(\bar z-\bar\zeta )|_p^{\b-1}d_p\zeta, \quad \Re\a>0, \Re\b>0, \Re (\a+\b )<1 \cr
&=B_q(\a,\b)|z\bar z|_p^{\a+\b-1}, \quad z\neq 0 \quad [2a)]. & (14.26) \cr
&=B_q(\a,\b), \quad z=1. & (14.27) \cr
}$$
$$\eqalignno{
&\int\chi_p(\xi z\bar z)d_pz, \quad \xi\neq 0, p\neq 2 \cr
&={{\sgn_{p,d}\xi}\over{|\xi |_p}}, \quad |d|_p=1, d\notin\QQ_p^{\times 2} \quad [4]. & (14.28) \cr
&=\pm\sqrt{p\textstyle{\sgn_{p,d}}(-1)}{{\textstyle{\sgn_{p,d}}\xi}\over{|\xi |_p}}, \quad |d|_p={1/p} \quad [4]. & (14.29) \cr
}$$
\bigskip
\centerline{\bf \Par 15. n-Variables}
\medskip
$$\int_{B_0^n}d_p^nx=1. \eqno (15.1)$$
$$\int_{S_0^n}d_p^nx=1-p^{-n}. \eqno (15.2)$$
$$\int_{B_\g^n}d_p^nx=p^{n\g}. \eqno (15.3)$$
$$\int_{S_\g^n}d_p^nx=(1-p^{-n})p^{n\g}. \eqno (15.4)$$
$$\int_{B_{\g}^n}f(|x|_p)d_p^nx=(1-p^{-n})\sum_{k=-\infty}^{\g}p^{nk}f(p^k). \eqno(15.5)$$
$$\int f(|x|_p)d_p^nx=(1-p^{-n})\sum_{k=-\infty}^\infty p^{nk}f(p^k). \eqno(15.6)$$
$$\int_{B_{\g}^n}|x|_p^{\a-n}d_p^nx={{1-p^{-n}}\over{1-p^{-\a}}}p^{\a\g}, \quad \Re\a>0. \eqno(15.7)$$
$$\int_{S_{\g}^n}|x|_p^{\a-n}d_p^nx=(1-p^{-n})p^{\a\g}. \eqno (15.8)$$
$$\int_{|x|_p>p^\g}|x|_p^{\a-n}d_p^nx=-{{1-p^{-n}}\over{1-p^{-\a}}}p^{\g\a}, \Re\a<0. \eqno (15.9)$$
$$\int_{S_\g^n}\chi_p((\xi,x))d_p^nx $$
$$=(1-p^{-n})p^{\g n}\Omega (p^\g|\xi |_p)-p^{(\g-1)n}\delta (|\xi |_p-p^{1-\g}) \quad [12],[2a)]. \eqno(15.10)$$
$$\int_{B_\g^n}\chi_p((\xi,x))d_p^nx=p^{\g n}\Omega (p^\g|\xi |_p) \quad [12],[2a)]. \eqno(15.11)$$
$$\eqalignno{
&\int_{B_\g^n}|(x,x)|_p^{\a-{n/2}}\chi_p((\xi ,x))d_p^nx, \quad |(\xi ,\xi )|_p>p^{-\g}, \Re\a>0 \cr
&=\G_p(\a-{n/2}+1)\G_p(\a )|(\xi,\xi )|_p^{-\a}, \quad n\equiv 0(\mod 4), \cr
&p\neq 2 \hbox{ or } n\equiv 2(\mod 4), p\equiv 1(\mod 4) \quad [3b)]. & (15.12) \cr
&=(-1)^{\g((\xi,\xi ))}\G_p(\a-{n/2}+1)\tilde{\G}_p(\a )|(\xi ,\xi )|_p^{-\a}, \cr
&n\equiv 2(\mod 4), p\equiv 3(\mod 4) \quad [3b)]. & (15.13) \cr
}$$
$$\eqalignno{
&\int |(x,x)|_p^{\a-{n/2}}\chi_p((\xi ,x))d_p^n, \quad (\xi,\xi )\neq 0, \Re\a>0 \cr
&=\G_p(\a-{n/2}+1)\G_p(\a )|(\xi,\xi )|_p^{-\a}, \cr
&n\equiv 0(\mod 4), p\neq 2 \hbox{ ¨«¨ } n\equiv 2(\mod 4), p\equiv 1(\mod 4) \quad [3b)]. & (15.14) \cr
&=(-1)^{\g((\xi,\xi ))}\G_p(\a-{n/2}+1)\tilde{\G}_p(\a )|(\xi,\xi )|_p^{-\a}, \cr
&n\equiv 2(\mod 4), p\equiv 3(\mod 4) \quad [3b)]. & (15.15) \cr
}$$
$$\int_{B_\g^n}|x|_p^{\a-n}\chi_p(x_1)d_p^nx=\G_p^{(n)}(\a), \quad \Re\a>0, \g\geq 1. \eqno(15.16)$$
$$\int |x|_p^{\a-n}\chi_p(x_1)d_p^nx=\G_p^{(n)}(\a), \quad \Re\a>0. \eqno(15.17)$$
$$\int |x|_p^{\a-n}\chi_p((\xi,x))d_p^nx=\G_p^{(n)}(\a)|\xi |_p^{-\a}, \quad \Re\a>0, \xi\neq 0. \eqno(15.18)$$
$$\int |x,m|_p^{\a-n}\chi_p((\xi,x))d_p^nx $$
$$=\G_p^{(n)}(\a)\bigl (|\xi |_p^{-\a}-|pm|_p^\a\bigr )\Omega (|m\xi |_p), \quad m\neq 0 \quad [1a)],[9a)]. \eqno(15.19)$$
$$\int |x,1|_p^{-\a}\chi_p((\xi,x))d_p^nx$$
$$=\G_p^{(n)}(n-\a)(|\xi |_p^{\a-n}-p^{\a-n})\Omega (|\xi |_p)\equiv J_p^\a(\xi ), \quad \Re\a>n. \eqno(15.20)$$
$$J_p^n(\xi )=(1-p^{-n})\bigl (1-{{\ln |\xi |_p}/{\ln p}}\bigr )\Omega (|\xi |_p), \quad \a=n. \eqno(15.21)$$
$$\int J_p^\a(\xi )J_p^\b(x-\xi )d_p^n\xi =J_p^\a\ast J_p^\b=J_p^{\a+\b}, \quad \a, \b\in\CC. \eqno(15.22)$$
$$\int |x|_p^{\a-n}|\varepsilon -x|_p^{\b-n}d_p^nx=B_p^{(n)}(\a,\b),$$
$$\Re\a>0, \Re\b>0, \Re (\a+\b)<n, |\varepsilon |_p=1. \eqno(15.23)$$
$$\int |y|_p^{\a-n}|x-y|_p^{\b-n}d_p^ny=B_p^{(n)}(\a,\b)|x|_p^{\a+\b-n},$$
$$\Re\a>0, \Re\b>0, \Re (\a+\b)<n \quad [1a)],[9a)]. \eqno(15.24)$$
$$\int |y,m|_p^{\a-n}|x-y,m|_p^{\b-n}d_p^ny=B_p^{(n)}(\a,\b)|x,m|_p^{\a+\b-n}$$
$$-\G_p^{(n)}(\a )|pm |_p^\a|x,m|_p^{\b-n}-\G_p^{(n)}(\a )|pm|_p^\b|x,m|_p^{\a-n},$$
$$\Re (\a+\b)<n, m\neq 0 \quad [9a)]. \eqno(15.25)$$
$$\int_{\QQ_p^{n-1}}\chi_p\Bigl \{\sum_{k=0}^{n-1}\Bigl ({{2x_kx_{k+1}}\over {\sin t_k}}-{{x_k^2+x_{k+1}^2}\over{\tg t_k}}\Bigr )\Bigr \}d_px_1d_px_2\ldots d_px_{n-1} $$
$$={{\lambda_p(T_n)}\over{\sqrt{|T_n|_p}}}\prod_{k=0}^{n-1}{{\sqrt{|t_k |_p}}\over{\lambda_p(t_k)}}\chi_p\Bigl ({{2x_0x_n}\over{\sin T_n}}-{{x^2+x_n^2}\over{\tg T_n}}\Bigr ),$$
$$n=2,3,\ldots, p\neq 2, |t_k|_p\leq{1/p}, k=0,1,\ldots,n-1, T_n=\sum_{k=0}^{n-1}t_k \quad [9a)]. \eqno(15.26)$$

Let $x_i\in\QQ_p^n, |x_i|_p=1, i=1,2,\ldots,k<p^n$ ¨ $|x_i-x_j|_p=1,$
$i,j=1,2,\ldots, i\neq j.$ Denote
$$D_k^n=[x\in\QQ_p^n: |x-x_i|_p=1, i=1,2,\ldots,k].$$
$$\int_{D_k^n}d_p^nx=1-kp^{-n}, \quad k\leq p^n, p\neq 2 \quad [9b)]. \eqno(15.27)$$
Let $G_k^{n}=[(x_1,x_2,\ldots,x_k)\in\QQ_p^{kn}: |x_i|_p=1,$
$|x_i-x_j|_p=1, i,j=1,2,\ldots,k, i\neq j].$
$$\int_{G_k^n}d_p^nx_1d_p^nx_2\ldots d_p^nx_k=\prod_{\ell =1}^k(1-\ell p^{-n})=c_{p,k}^n,$$
$$k\leq p^n, p\neq 2 \quad [9b)]. \eqno(15.28)$$
Let $x_0\in\QQ_p^n, |x_0|_p=1$ ¨ $G_k^n(x_0)=[(x_1,\ldots,x_k)\in\QQ_p^{kn}:$
$|x_i|_p=1$, $i=0,1,\ldots,k, \quad |x_i-x_j|_p=1, i,j=0,1,\ldots,k, i\neq j],$
$$\int_{G_k^n(x_0)}d_p^nx_1d_p^nx_2\ldots d_p^nx_k={{1-(k+1)p^{-n}}\over{1-p^{-n}}}c_{k,p}^n,$$
$$k+1\leq p^n, p\neq 2 \quad [9b)]. \eqno(15.29)$$

{\rm The Missarov--Lerner integral [12],[17].} Let $G$ be a connected finite 
graph, $V=V(G)$ and $L=L(G)$ are sets of its vertices and edges respectively. 
To every line $l\in L$ we associate a complex number $a_l$, and denote the set
$a=\{a_l, l\in L\}$. To every vertex $v\in V$ we associate $n$-dimensional
$p$-adic vector $x_v=(x_{v1},x_{v2},\ldots,x_{vn})\in \QQ_p^n$. On the set of
vertices $V$ we introduce a hierarchy $A$ by the following way. The hierarchy
$€$ is a family of subsets of the set $V$ such that: 1)$V\in A$, 2)$v\in A$ for
all $v\in V$ and 3)for any pairs $V'\in A, V''\in A$ either $V'\cap V''=\O$ or
$V'\in V''$ or $V''\in V'$. For any $V'\in A, V'\neq V$ we denote by $\t(V')$ 
a minimal subset in $A$ containing $V'$ but not coinciding with it. Let 
$K(V')=[V''\in A: \t(V'')=V']$. We consider only such hierarchies $A$ for which
$$1<|K(V')|\leq p^n, V'\in A', \hbox { where } A'=[V'\in A: |V'|>1].$$
Denote
$$a(V')=\sum_{l\in L(G(V'))}a_l, \quad \b(V')=a(V')+n(|V'|-1)$$
where $L(G(V'))$ is the set of edges $\{l\}$ of the graph $G$ beginning $i(l)$
and end $f(l)$ of which lay in $V'\subset V=V(G)$. 
By the condition $\b(V')>0, V'\in A'$ the following equality is valid
$$F_G(a)\equiv\int_{Z_p^{n|V|}}\prod_{l\in L}|x_{i(l)}-x_{f(l)}|_p^{a_l}\prod_{v\in V}d_p^n x_v $$
$$=p^{a(V)}\sum_A\prod_{V'\in A'}{1\over{p^{\b(V')}-1}}{{(p^n-1)!}\over{(p^n-|K(V')|!}} \eqno(15.30)$$ 
where the summing is taken over all hierarchies $A$. (Simbol $|V|$ denotes a
number of elements of the set $V$.) Evaluation of various Feynman integrals
is reduced to the integral $F_G(a)$ [12].
\bigskip
\centerline{\bf \Par 16. Integrals and convolutions of generalized functions}
\medskip
Integral (see \Par 6) of a generalized function $f\in\script{S}'(\script{O})$
on a clopen set $D\in\script{O}\in\QQ_p^n$ is called the limit (if it exists!)
$$G\!\!\!\!\int_Df(x)d_p^nx=\lim_{k\to\infty}(f\t_D,\Omega_k).$$
Integrals of generalized functions are contained also in \Par\Par 12--15 and in
\Par 17.
\medskip
$$G\!\!\!\!\int_{B_0^n}d_p^nx=1. \eqno(16.1)$$
$$G\!\!\!\!\int_{B_\g^n}d_p^nx=p^{\g n}. \eqno(16.2)$$
$$G\!\!\!\!\int_{S_\g^n}d_p^nx=(1-p^{-n})p^{\g n}. \eqno(16.3)$$
$$G\!\!\!\!\int f(x)d_p^nx=\int f(x)d_p^nx, \quad f\in\script{L}^1. \eqno(16.4)$$
$$G\!\!\!\!\int f(x)d_p^nx=\lim_{\g\to\infty}\int_{B_\g^n}f(x)d_p^nx, \quad f\in\script{L}_{\loc}^1. \eqno(16.5)$$
$$G\!\!\!\!\int_Df(x)d_p^nx=\int_Df(x)d_p^nx, \quad f\in\script{L}^1(D). \eqno(16.6)$$
$$G\!\!\!\!\int f(x)d_p^nx=\lim_{\g\to\infty}\int_{B_\g^n}f(x)d_p^nx, \quad f\in\script{L}^p. \eqno(16.7)$$
$$G\!\!\!\!\int f(x)d_p^nx=(f,\Omega_N), \quad f\in\script{S}', \spt\in B_N^n. \eqno(16.8)$$
$$G\!\!\!\!\int_Df(x)d_p^nx=(f,\t_D), \quad f\in\script{S}'(\script{O}) \eqno(16.9)$$
where $D$ is an open compact in $\script{O}$.
$$G\!\!\!\!\int f(x)d_p^nx=\lim_{\g\to\infty}(f,\Omega_\g), \quad f\in\script{S}'. \eqno(16.10)$$
$$G\!\!\!\!\int\delta (x)d_p^nx=1. \eqno(16.11)$$
$$G\!\!\!\!\int_{S_\g}\pi (x)d_px=0, \quad \pi\not \equiv 1, \a\in\CC \quad (\hbox{ cf. } (11.40)). \eqno(16.12)$$
$$G\!\!\!\!\int_{B_\g}|x|_p^{\a-1}\pi (x)d_px=0, \quad \pi\not \equiv 1, \a\in\CC. \eqno(16.13)$$
$$G\!\!\!\!\int |x|_p^{\a-1}\pi (x)d_px=0, \quad \pi\not \equiv 1, \a\in\CC. \eqno(16.14)$$
$$G\!\!\!\!\int_{B_\g}|x|_p^{\a-1}d_px={{1-p^{-1}}\over{1-p^{-\a}}}p^{\a\g},$$
$$\a\neq\a_k, k\in Z \quad (\hbox{ cf. } (11.18)). \eqno(16.15)$$
$$G\!\!\!\!\int |x|_p^{\a-1}d_px=0, \quad \a\neq\a_k, k\in Z \quad (\hbox{ cf. } (11.18)). \eqno(16.16)$$
$$G\!\!\!\!\int_{S_\g}|x-a|_p^{\a-1}d_px={{p-2+p^{-\a}}\over{p(1-p^{-\a})}}|a|_p^\a,$$
$$|a|_p=p^\g, \a\neq\a_k, k\in Z \quad (\hbox{ cf. } (11.20)). \eqno(16.17)$$
$$G\!\!\!\!\int_{B_\g}|x^2+a^2|_p^{(\a-1)/2}d_px={{1-p^{\a-1}}\over{1-p^{\a}}}|a|_p^\a+{{1-p^{-1}}\over{1-p^{-\a}}}p^{\a\g},$$
$$0\neq |a|_p\leq p^\g, \a\neq\a_k, k\in Z, p\equiv 3(\mod 4) \quad (\hbox{ cf. } (11.28)). \eqno(16.18)$$
$$G\!\!\!\!\int |x^2+a^2|_p^{(\a-1)/2}d_px={{1-p^{\a-1}}\over{1-p^{\a}}}|a|_p^\a,$$
$$a\neq 0, \a\neq\a_k, k\in Z, p\equiv 3(\mod 4) \quad (\hbox{ cf. } (11.30)). \eqno(16.19)$$
$$G\!\!\!\!\int_{B_\g}|x^2+a^2|_p^{\a-1}d_px$$
$$=\Bigl [1-{2/p}+(1-{1/p})\Bigl ({2\over{p^\a-1}}+{1\over{p^{1-2\a}-1}}\Bigr )\Bigr ]|a|_p^{2\a-1}-{{(1-{1/p})p^{(2\a-1)\g}}\over{1-p^{2\a-1}}},$$
$$0\neq |a|_p\leq p^\g, \a\neq\{\a_k, (1-\a_k)/2, k\in Z\}, p\equiv 1(\mod 4). \eqno(16.20)$$
$$G\!\!\!\!\int |x^2+a^2|_p^{\a-1}d_px $$
$$=\Bigl [1-{2/p}+(1-{1/p})\Bigl ({2\over{p^\a-1}}+{1\over{p^{1-2\a}-1}}\Bigr )\Bigr ]|a|_p^{2\a-1},$$
$$\a\neq\{\a_k, (1-a_k)/2, k\in Z\}, a\neq 0, p\equiv 1(\mod 4) \quad (\hbox{ cf. } (11.31)). \eqno(16.21)$$
$$G\!\!\!\!\int_{S_0}|x^2+1|_p^{\a-1}d_px=1-{3/p}-2{{1-p^{-1}}\over{1-p^\a}},$$
$$\a\neq\a_k, k\in Z, p\equiv 1(\mod 4) \quad (\hbox{ cf. } (11.32)). \eqno(16.22)$$
$$|x|_p^{\a-1}\ast |x|_p^{\b-1}=B_p(\a,\b )|x|_p^{\a+\b-1},$$
$$(\a,\b)\neq (\a_k,\a_j), (k,j)\in Z^2. \eqno(16.23)$$
$$G\!\!\!\!\int |x|_p^{\a-1}|1-x|_p^{\b-1}d_px=B_p(\a,\b), $$
$$(\a,\b)\neq (\a_k,\a_j), (k,j)\in Z^2. \eqno(16.24)$$
$$|x,m|_p^{\a-1}\ast |x,m|_p^{\b-1}=B_p(\a,\b)|x,m|_p^{\a+\b-1} $$
$$-\G_p(\a)|pm|_p^\a|x,m|_p^{\b-1}-\G_p(\b)|pm|_p^\b|x,m|_p^{\a-1},$$
$$m\neq 0, (\a,\b)\neq (\a_k,\a_j), (k,j)\in Z^2 \quad (\hbox{ cf. } (11.68)). \eqno(16.25)$$
$$G\!\!\!\!\int |x,m|_p^{\a-1}\chi_p(\xi x)d_px $$
$$=\G_p(\a)(|\xi |_p^{-\a}-|pm|_p^\a)\Omega (|m\xi |_p),$$
$$m\neq 0, \a\in\CC \quad (\hbox{ see } (12.48)). \eqno(16.26)$$
$$G\!\!\!\!\int_{S_\g}\delta (x_0-k)d_px=p^{\g-1}, \quad k=1,2,\ldots ,p-1 \quad (\hbox{ see } (11.33)). \eqno(16.27)$$
$$G\!\!\!\!\int_{S_\g}[1-\delta (x_0-k)]d_px=(1-{2/p})p^\g, $$
$$k=1,2,\ldots ,p-1 \quad (\hbox{ see } (11.34)). \eqno(16.28)$$
$$G\!\!\!\!\int_{S_\g}\delta (x_n-k)d_px=(1-{1/p})p^{\g-1}, $$
$$k=0,1,\ldots ,p-1, n\in Z_+ \quad (\hbox{ see } (11.35)). \eqno(16.29)$$
$$G\!\!\!\!\int_{S_\g}[1-\delta (x_n-k)]d_px=(1-{1/p})^2p^\g, $$
$$k=0,1,\ldots ,p-1, n\in Z_+ \quad (\hbox{ see }(11.36)). \eqno(16.30)$$
$$G\!\!\!\!\int_{S_\g}\delta (x_0-k_0)\prod_{l=1}^n\delta (x_l-k_l)=p^{\g-n-1},$$
$$k_l=0,1,\ldots ,p-1, k_0\neq 0, n=0,1,\ldots \quad (\hbox{ see } (11.37)). \eqno(16.31)$$
$$G\!\!\!\!\int_{S_\g}\Bigl [1-\delta (x_0-k_0)\prod_{l=1}^n\delta (x_l-k_l)\Bigr ]=(1-p^{-1}-p^{-n-1})p^\g, $$
$$k_l=0,1,\ldots ,p-1, k_0\neq 0, n=0,1,\ldots \quad (\hbox{ see } (11.38)). \eqno(16.32)$$
$$G\!\!\!\!\int_{S_\g}\Bigl (\prod_{l=1}^n\delta (x_l-k_l)\Bigr )d_px=(1-{1/p})p^{\g-n},$$
$$k_l=0,1,\ldots ,p-1, n\in Z_+. \eqno(16.33)$$
$$G\!\!\!\!\int_{S_\g}\Bigl [1-\prod_{l=1}^n\delta (x_{i_l}-k_{i_l})\Bigr ]d_px=(1-{1/p})(1-p^{-n})p^\g,$$
$$k_l=0,1,\ldots ,p-1, n\in Z_+. \eqno(16.34)$$
$$G\!\!\!\!\int_{S_\g^n}|x|_p^{\a-n}d_p^nx=(1-p^{-n})p^{\a\g}, \quad \a\in\CC \quad (\hbox{ see } (15.8)). \eqno(16.35)$$
$$G\!\!\!\!\int_{B_\g^n}|x|_p^{\a-n}d_p^nx={{1-p^{-n}}\over{1-p^{-\a}}}p^{\a\g}, \quad \a\neq\a_k, k\in Z \quad (\hbox{ cf. } (15.7)). \eqno(16.36)$$
$$G\!\!\!\!\int |x|_p^{\a-n}d_p^nx=0, \quad \a\neq\a_k, k\in Z, n\in Z_+. \eqno(16.37)$$
$$\eqalignno{
&G\!\!\!\!\int_{B_\g^n}|(x,x) |_p^{\a-{n/2}}\chi_p((\xi ,x))d_p^nx, \quad |(\xi ,\xi )|_p>p^\g \cr
&=\G_p(\a-{n/2}+1)\G_p(\a )|(\xi ,\xi )|_p^{-\a}, \quad \a\neq\{\a_k, \a_k+{n/2}-1, k\in Z\},\cr
&n\equiv 0(\mod 4), p\neq 2 \hbox{ or } n\equiv 2(\mod 4), p\equiv 1(\mod 4). & (16.38) \cr
&=(-1)^{\g((\xi, \xi ))}\G_p(\a-{n/2}+1)\tilde{\G}_p(\a )|(\xi, \xi )|_p^{-\a}, \cr
&\a\neq\{\a_k-{{\pi i}/{\ln p}}, \a_k+{n/2}-1, k\in Z\},\cr
&n\equiv 2(\mod 4), p\equiv 3(\mod 4) \quad (\hbox{ cf. } (15.13)). & (16.39) \cr
&=\G_p^2(\a )|(\xi ,\xi )|_p^{-\a}, \quad \a\neq\a_k, k\in Z, n=2, p\equiv 1(\mod 4). & (16.40) \cr
&=\G_p(\a )\tilde{\G}_p(\a )|(\xi ,\xi )|_p^{-\a}, \quad \a\neq\{\a_k, \a_k-{{\pi i}/{\ln p}}, k\in Z\},\cr
&n=2, p\equiv 3(\mod 4) \quad (\hbox{ see } (14.9)). & (16.41) \cr
}$$
$$\eqalignno{
&G\!\!\!\!\int |(x,x) |_p^{\a-{n/2}}\chi_p((\xi ,x))d_p^nx, \quad (\xi ,\xi )\neq 0 \cr
&=\G_p(\a-{n/2}+1)\G_p(\a )|(\xi,\xi )|_p^{-\a}, \cr
&\a\neq\{\a_k, \a_k+{n/2}-1, k\in Z\}, \cr
&n\equiv 0(\mod 4), p\neq 2 \quad \hbox{ or } n\equiv 2(\mod 4), p\equiv 1(\mod 4). & (16.42) \cr
&=(-1)^{\g((\xi ,\xi ))}\G_p(\a-{n/2}+1)\tilde{\G}_p(\a )|(\xi, \xi )|_p^{-\a}, \cr
&\a\neq\{\a_k-{{\pi i}/{\ln p}}, \a_k+{n/2}-1, k\in Z\},\cr
&n\equiv 2(\mod 4), p\equiv 3(\mod 4) \quad (\hbox{ see } (15.15)). & (16.43) \cr
&=\G_p^2(\a )|(\xi ,\xi )|_p^{-\a}, \quad \a\neq\a_k, k\in Z, \cr
&n=2, p\equiv 1(\mod 4) \quad (\hbox{ cf. } (14.10)). & (16.44) \cr
&=\G_p(\a )\tilde{\G}_p(\a )|(\xi ,\xi )|_p^{-\a}, \a\neq\{\a_k, \a_k-{{\pi i}/{\ln p}}, k\in Z\},\cr
&n=2, p\equiv 3(\mod 4) (\hbox{ cf. } (14.11)). & (16.45) \cr
}$$
$$G\!\!\!\!\int_{B_\g^n}|x|_p^{\a-n}\chi_p(x_1)d_p^nx=\G_p^{(n)}(\a), \quad \a\neq\a_k, k\in Z, \g\in Z_+. \eqno(16.46)$$
$$G\!\!\!\!\int |x|_p^{\a-n}\chi_p(x_1)d_p^nx=\G_p^{(n)}(\a), \quad \a\neq\a_k, k\in Z. \eqno(16.47)$$
$$G\!\!\!\!\int |x|_p^{\a-n}\chi_p\bigr ((\xi ,x)\bigr )d_p^nx=\G_p^{(n)}(\a)|\xi |_p^{-\a}, $$
$$\a\neq\a_k, k\in Z, \xi\neq 0 \quad (\hbox{ cf. } (15.18)). \eqno(16.48)$$
$$|x|_p^{\a-n}\ast |x|_p^{\b-n}=B_p^{(n)}(\a,\b)|x|_p^{\a+\b-n}, $$
$$(\a,\b)\neq (\a_k,\a_j), (k,j)\in Z^2. \eqno(16.49)$$
$$G\!\!\!\!\int |x,m|_p^{\a-n}\chi_p\bigl ((\xi, x)\bigr )d_p^nx=\G_p^{(n)}(\a)\bigl (|\xi |_p^{-\a}-|pm|_p^\a\bigr ) $$
$$\times\Omega (|m\xi |_p), \quad m\neq 0, \a\in\CC \quad (\hbox{ cf. } (15.19)). \eqno(16.50)$$
$$G\!\!\!\!\int |x,1|_p^{-\a}\chi_p\bigl ((\xi,x)\bigr )d_p^nx $$
$$=\G_p^{(n)}(n-\a)\bigl (|\xi |_p^{\a-n}-p^{\a-n}\bigr )\Omega (|\xi |_p), \quad \a\in\CC. \eqno(16.51)$$
$$|x,m|_p^{\a-n}\ast |x,m|_p^{\b-n}=B_p^{(n)}(\a,\b)|x,m|_p^{\a+\b-n}$$
$$=-\G_p^{(n)}(\a )|pm |_p^\a |x,m|_p^{\b-n}-\G_p^{(n)}(\b )|pm|_p^\b|x,m|_p^{\a-n}, $$
$$(\a,\b)\neq (\a_k,\a_j), (k,j)\in Z^2, m\neq 0 \quad (\hbox{ cf. } (15.25)). \eqno(16.52)$$
$$\eqalignno{
&(D^\a\varphi )(x)=(f_{-\a}\ast\varphi )(x), \quad \varphi\in\script{S} \cr
&=\G_p^{-1}(-\a )\int{{\varphi (y)-\varphi (x)}\over{|x-y|_p^{\a+1}}}d_py, \quad \Re\a>0. & (16.53) \cr
&=(1-p^{-\a-1})^{-1}\int [\varphi (x+y)-\varphi (x+{y/p})]|y|_p^{-\a-1}d_py, \cr
&\a\neq\a_k-1, k\in Z. & (16.54) \cr
&=-{{p-1}\over{p\ln p}}\int\varphi (y)\ln |x-y|_pd_py, \quad \int f(y)d_py=0 \cr
&\a=\a_k-1, k\in Z. & (16.55) \cr
&=\varphi (x), \quad \a=\a_k, k\in Z. & (16.56) \cr
&=\int |\xi |_p^\a\tilde{\varphi}(\xi )\chi_p(-\xi x)d_p\xi, \quad \Re\a>-1. & (16.57) \cr
&=\int |\xi |_p^\a[\tilde{\varphi}(\xi )\chi_p(-\xi x)-\tilde{\varphi}(0)]d_p\xi, \quad \Re\a<-1. & (16.58) \cr
&=\int_{Z_p}|\xi |_p^{-1}[\tilde{\varphi}(\xi )\chi_p(-\xi x)-\tilde{\varphi}(0)]d_p\xi+{1/p}\tilde{\varphi}(0) \cr
&+\int_{\QQ_p\backslash Z_p}|\xi |_p^{-1}\tilde{\varphi}(\xi )\chi_p(-\xi x)d_p\xi, \quad \a=\a_k-1, k\in Z. & (16.59) \cr
}$$
$$D^\a\chi_p(ax)=|a|_p^\a\chi_p(ax), \quad \a\in\CC, a\neq 0. \eqno(16.60)$$
$$D^\a\Phi (x)=p^{\g\a}\Phi (x), \quad \a\in\RR \quad [1a)]. \eqno(16.61)$$
¥á«¨ $\Phi (x)=F[\delta (|\xi |_p-p^\g)f(\xi )], f\in\script{S}'$.
$$D^\a[\delta (|x|_p-p^\g)\chi_p(ax^2)] $$
$$=p^{\g\a}|2a|_p^\a\delta (|x|_p-p^\g)\chi_p(ax^2), \quad \a\in\RR, |2a|_p\leq p^{2-2\g} \quad [1a)]. \eqno(16.62)$$
$$D^\a[\eta (x_0)\delta (|x|_p-p^\g )]=p^{\a (1-\g)}\eta (x_0)\delta (|x|_p-p^\g ), $$
$$\a\in\RR, p\neq 2, \sum_{k=1}^{p-1}\eta (k)=0 \quad [2b)]. \eqno(16.63)$$
$$\eqalignno{
&(D^\a{f})(x), \quad f\in\script{S}', \spt f\in B_N, |x|_p>p^N \cr
&=\G_p^{-1}(\a)|x|_p^{\a-1}(f,\Omega_N), \quad \a\neq -1 \quad [2a)]. & (16.64) \cr
&=-{{p-1}\over{p\ln p}}\ln |x|_p(f,\Omega_N), \quad \a=-1 \quad [2a)]. & (16.65) \cr
}$$
$$D^\a1=0, \quad \a>0. \eqno(16.66)$$
$$D^\a\delta (x-a)=f_{-\a}(x-a), \quad \a\in\CC, a\in\QQ_p. \eqno(16.67)$$
$$D^\a[\delta (|x|_p-p^{\ell-N})\delta (x_0-j)\chi_p(\epsilon_\ell p^{\ell -2N}x^2)] $$
$$=p^{\a{N}}\delta (|x|_p-p^{\ell-N})\delta (x_0-j)\chi_p(\epsilon_\ell p^{\ell -2N}x^2),$$
$$N\in Z, p\neq 2, \a>0, \ell=2,3,\ldots, j=1,2,\ldots,p-1, $$
$$\epsilon_\ell=\varepsilon_0+\varepsilon_1p+\ldots +\varepsilon_{\ell -2},$$
$$\varepsilon_s=0,1,\ldots ,p-1, \varepsilon_0\neq 0, s=0,1,\ldots,\ell-2 \quad [2c)]. \eqno(16.68)$$
$$D^\a[\Omega (p^{N-1}|x|_p)\chi_p(jp^{-N}x)]=p^{\a{N}}\Omega (p^{N-1}|x|_p)\chi_p(jp^{-N}x), $$
$$N\in Z, p\neq 2, \a>0, j=1,2,\ldots ,p-1 \quad [2b)]. \eqno (16.69)$$
$$D^\a[\delta (|x|_2-2^{\ell +1-N})\chi_2(\epsilon_\ell 2^{\ell -2N}x^2+2^{\ell -N-j}x)]$$
$$=2^{\a{N}}\delta (|x|_2-2^{\ell +1-N})\chi_2(\epsilon_\ell 2^{\ell -2N}x^2+2^{\ell -N-j}x),$$
$$N\in Z, p=2, \a>0, \ell =2,3,\ldots , j=0,1, \epsilon_\ell =1+\varepsilon_12+\ldots +\varepsilon_{\ell -2}2^{\ell -2},$$
$$\varepsilon_s=0,1, s=1,2,\ldots ,\ell -2 \quad [2b)]. \eqno(16.70)$$
$$D^\a[\Omega (2^N|x-j2^{N-2}|_2)-\delta (|x-j2^{N-2}|_2-2^{1-N})]$$
$$=2^{\a{N}}[\Omega (2^N|x-j2^{N-2}|_2)-\delta (|x-j2^{N-2}|_2-2^{1-N})], $$
$$N\in Z, p=2, \a>0, j=0,1 \quad [2b)]. \eqno(16.71)$$
$$D^\a\Omega (p^{-\g}|x|_p)={{p-1}\over{p^{\a+1}-1}}p^{\a(1-\g)}, \quad x\in B_\g, \a>0 \quad [2c)]. \eqno(16.72)$$
$$D^\a\delta (|x|_p-p^\g)={{p^\a+p-2}\over{p^{\a+1}-1}}p^{\a(1-\g)}, \quad x\in S_\g, \a>0 \quad [2c)]. \eqno(16.73)$$

Let $\script{K}(t,\u )$ be a real symmetric kernel
$$\script{K}(t,t)=0, \quad \script{K}(t,\u )=\rho (1-{1/p})^{-1}t^{-\a-1}, \u<t,$$
$$\s={{p^\a+p-2}\over{p^{\a+1}-1}}p^\a, \rho=-\G_p^{-1}(-\a )(1-{1/p}), \s+\rho=p^\a$$
and a function $f\in\script{L}_{\loc}^1$ such that 
$$\int_{|x|_p>1}|f(x)||x|_p^{-\a-1}d_px<\infty.$$
Then
$$(D^\a{f})(x)=-\int\script{K}(|x|_p,|y|_p)f(y)d_py+\s|x|_p^{-\a}f(x), \quad \a>0 \quad [2c)]. \eqno(16.74)$$

In the following formulas \Par 16 all integrals
$$G\!\!\!\!\int f(z,\bar z)d_pz$$
are understood on the normalized measure $d_pz=\delta^{-1}d_pxd_py,$\quad 
$z=x+\sqrt dy, \bar z=x-\sqrt dy$ of the field $\QQ_p(\sqrt d)$, 
$d\notin\QQ_p^{\times 2}$ (see (9.2)). In particular,
$B_\g^2=[z\in\QQ_p(\sqrt d): |z\bar z|_p\leq q^\g]$;
$\a_k={{2k\pi i}\over{\ln q}}, k\in Z$ (see (9.6)).

$$G\!\!\!\!\int_{B_0^2}d_pz=1. \eqno(16.75)$$
$$G\!\!\!\!\int_{B_0^2}|z\bar z|_p^{\a-1}d_pz={{1-q^{-1}}\over{1-q^{-\a}}}, \quad \a\neq\a_k, k\in Z. \eqno(16.76)$$
$$G\!\!\!\!\int |z\bar z|_p^{\a-1}d_pz=0, \quad \a\neq\a_k, k\in Z. \eqno(16.77)$$
$$G\!\!\!\!\int |z\bar z|_p^{\a-1}\chi_p(z+\bar z)d_pz={{\G_{p,d}(\a)}\over{\delta\sqrt{|4d|_p}}}, \quad \a\neq\a_k, k\in Z. \eqno(16.78)$$
$$G\!\!\!\!\int_{B_1^2}|z\bar z|_p^{\a-1}\chi_p(z+\bar z)d_pz={{\G_{p,d}(\a)}\over{\delta\sqrt{|4d|_p}}}, \quad \a\neq\a_k, k\in Z. \eqno(16.79)$$
$$|z\bar z|_p^{\a-1}\ast |z\bar z|_p^{\b-1}=B_q(\a,\b)|z\bar z|_p^{\a+\b-1}, $$
$$(\a,\b)\neq (\a_k,\a_j), (k,j)\in Z^2. \eqno(16.80)$$
$$G\!\!\!\!\int |z\bar z|_p^{\a-1}|(1-z)(1-\bar z)|_p^{\b-1}d_pz=B_q(\a,\b), $$
$$(\a,\b)\neq (\a_k,\a_j), (k,j)\in Z^2. \eqno(16.81)$$
$$G\!\!\!\!\int\chi_p(\xi z\bar z)d_pz={{\sgn_{p,d}\xi}\over{|\xi |_p}}+{{1+p}\over{2p}}\delta(\xi ),$$
$$p\neq 2, |d|_p=1, d\notin\QQ_p^{\times 2} \quad [4]. \eqno(16.82)$$
$$G\!\!\!\!\int\chi_p(\xi z\bar z)d_pz=\pm\sqrt{p\textstyle{\sgn_{p,d}(-1)}}{{\textstyle{\sgn_{p,d}}\xi}\over{|\xi |_p}}+\delta(\xi ),$$
$$p\neq 2, |d|_p={1/p} \quad [4]. \eqno(16.83)$$
\bigskip
\centerline{\bf \Par 17. Table of the Fourier transforms}
\medskip
For one-to-one correspondence between {\it preimage} $f\in\script{S}'$ and its
{\it image} $\tilde f\in\script{S}'$ -- the Fourier transform of $f$ -- we 
shall use the notation (see \Par 7) 
$$f(x)\iff\tilde f(\xi ).$$
\medskip
$$\omega_\g(x)\iff\delta_\g(\xi ). \eqno(17.1)$$
$$\delta (x)\iff 1(\xi ). \eqno(17.2)$$
$$f(Ax+b)\iff |\det A|_p^{-1}\chi_p\bigl (-(A^{-1}b,\xi )\bigr )\tilde f(\bar {A}'\xi ), $$
$$\det A\neq 0, b\in\QQ_p^n. \eqno(17.3)$$
$$f(x-b)\iff\chi_p((b,\xi ))\tilde f(\xi ), \quad b\in\QQ_p^n. \eqno(17.4)$$
$$\breve f(x)\iff\breve{\tilde f}(\xi ). \eqno(17.5)$$
$$f(x)\iff\int f(x)\chi_p((\xi ,x))d_p^nx, \quad f\in\script{L}^1. \eqno(17.6)$$ 
$$f(x)\iff\lim_{k\to\infty}\int_{B_k^n}f(x)\chi_p((\xi ,x))d_p^nx \hbox{ in } \script{S}', \quad f\in\script{L}_{\loc}^1. \eqno(17.7)$$ 
$$f(x)\iff\lim_{k\to\infty}\int_{B_k^n}f(x)\chi_p((\xi ,x))d_p^nx \hbox{ in } \script{L}^2, \quad f\in\script{L}^2. \eqno(17.8)$$
$$f(x)\iff\bigl (f(x),\Omega_N(x)\chi_p((\xi ,x))\bigr ), \quad\spt f\in B_N. \eqno(17.9)$$
$$f\ast g\iff\tilde f\cdot\tilde g. \eqno(17.10)$$
$$f\cdot g\iff\tilde f\ast\tilde g. \eqno(17.11)$$
$$\delta (|x|_p-p^\g )\iff (1-{1/p})p^\g\Omega (p^\g| \xi |_p)-p^{\g-1}\delta (|\xi |_p-p^{1-\g}). \eqno(17.12)$$
$$f(|x|_p)\Omega_\g( |x|_p)\iff (1-{1/p})\sum_{k=-\infty}^\g{p}^k{f}(p^k)\Omega (p^\g | \xi |_p)$$
$$+|\xi |_p^{-1}\Bigl [(1-{1/p})\sum_{k=0}^\infty p^{-\g}f(p^{-\g}|\xi |_p^{-1})-f(p|\xi |_p^{-1})\Bigr ][1-\Omega (p^\g |\xi |_p)]. \eqno(17.13)$$
$$f( |x|_p)\iff |\xi |_p^{-1}\Bigl [(1-{1/p})\sum_{k=0}^\infty p^{-\g}f(p^{-\g}|\xi |_p^{-1})-f(p|\xi |_p^{-1})\Bigr ]. \eqno(17.14)$$
$$|x|_p^{\a-1}\iff\G_p(\a)|\xi |_p^{-\a}, \quad \a\neq\a_k, k\in Z. \eqno(17.15)$$
$$\ln |x|_p\iff -(1-{1/p})^{-1}\ln p \bigl (\reg |\xi |_p^{-1}+{1/p}\delta (\xi )\bigr ). \eqno(17.16)$$
$${1\over {|x|_p^2+m^2}}\iff (1-{1/p}){|\xi |_p\over{p^2+m^2|\xi |_p^2}}$$
$$\times\sum_{\g=0}^\infty p^{-\g}{{p^2-p^{-2\g}}\over{p^{-2\g}+m^2|\xi |_p^2}}, \quad m\neq 0. \eqno(17.17)$$
$$|x|_p^{\a-1}|1-x|_p^{\b-1}\iff\bigl [\G_p(\a+\b-1)|\xi |_p^{1-\a-\b}+B_p(\a,\b)\bigr ]\Omega (|\xi |_p)$$
$$+\bigl [\G_p(\a )|\xi |_p^{-\a}+\G_p(\b )|\xi |_p^{-\b}\chi_p(\xi )\bigr ][1-\Omega (|\xi |_p)], $$
$$(\a,\b)\neq (\a_k,\a_j), (k,j)\in Z^2. \eqno(17.18)$$
$$\delta (|x|_p-1)|1-x|_p^{\a-1}\iff\G_p(\a )\chi_p(\xi )|\xi |_p^{-\a}, $$
$$\g (\xi )\geq 2, \a\neq\a_k, k\in Z. \eqno(17.19)$$
$$\eta_{x_0}\delta (|x|_p-p^\g)\iff p^{\g-1}\eta'_{\xi_0}\delta (|\xi |_p-p^{1-\g}), \quad p\neq 2 \eqno(17.20)$$
where
$$\sum_{k=1}^{p-1}\eta_k=0, \quad \eta'_j=\sum_{k=1}^{p-1}\eta_k\exp\bigl (2\pi i{{kj}\over p}\bigr ).$$
$$|x|_p^{\a-1}\Omega (p^{-\g}|x|_p)\iff{{1-p^{-1}}\over{1-p^{-\a}}}p^{\a\g}\Omega (p^\g|\xi |_p)$$
$$+\G_p(\a )|\xi |_p^{-\a}\bigl [1-\Omega(p^\g|\xi |_p)\bigr ], \quad \a\neq\a_k, k\in Z. \eqno(17.21)$$
$$\delta (|x|_p-1)\delta (x_0-p+1)\iff p^{-1}\chi_p(-\xi )\Omega (|p\xi |_p). \eqno(17.22)$$
$$\chi_p(x)\Omega (|px|_p)\iff p\delta (|\xi |_p-1)\delta (\xi_0-p+1). \eqno(17.23)$$
$$|x,m|_p^{\a-1}\iff\G_p(\a)\bigl (|\xi |_p^{-\a}-|pm |_p^\a\bigr )\Omega (|m\xi |_p), \quad m\neq 0, \a\in\CC. \eqno(17.24)$$
$$f((x,x))\iff |(\xi ,\xi )|_p^{-1}\Bigl [(1-p^{-2})\sum_{\g=0}^\infty p^{-2\g}f(p^{-2\g}|(\xi ,\xi )|_p^{-1})$$
$$-f(p^2|(\xi ,\xi )|_p^{-1})\Bigl ], \quad n=2, p\equiv 3(\mod 4). \eqno(17.25)$$
$$f((x,x))\iff |(\xi ,\xi )|_p^{-1}\Bigl [(1-{1/p})^2\sum_{\g=0}^\infty\Bigl (\g+{{p-3}\over{p-1}}\Bigr )p^{-\g}f(p^{-\g}|(\xi, \xi )|_p^{-1})$$
$$-2(1-{1/p})f(p|(\xi ,\xi )|_p^{-1})+f(p^2|(\xi ,\xi )|_p^{-1})\Bigr ], \quad n=2, p\equiv 1(\mod 4). \eqno(17.26)$$
$$|(x,x)|_p^{\a-1}\iff\G_p^2(\a )|(\xi ,\xi )|_p^{-\a}, $$
$$n=2, \a\neq\a_k, k\in Z, p\equiv 1(\mod 4). \eqno(17.27)$$
$$|(x,x)|_p^{\a-1}\iff\G_p(\a)\tilde{\G}_p(\a)|(\xi ,\xi )|_p^{-\a}, $$
$$\a\neq\{\a_k, \a_k-{{\pi i}/{\ln p}}, k\in Z\}, n=2, p\equiv 3(\mod 4). \eqno(17.28)$$
$${1\over{|(x,x)|_p+m^2}}\iff{{1-p^{-2}}\over{p^2+m^2|(\xi,\xi )|_p}}\sum_{\g=0}^\infty{{p^2-p^{-2\g}}\over{1+p^\g{m}^2|(\xi,\xi )|_p}}, $$
$$n=2, m\neq 0, p\equiv 3(\mod 4). \eqno (17.29)$$
$${1\over{|(x,x)|_p+m^2}}\iff (1-{1/p})^2\sum_{\g=0}^\infty \Bigr (\g+{{p-3}\over{p-1}}\Bigl ){1\over{1+p^\g{m}^2|(\xi,\xi )|_p}}$$
$$-2{{1-{1/p}}\over{p+m^2|(\xi,\xi )|_p}}+{1\over{p^2+m^2|(\xi,\xi )|_p}}, $$
$$n=2, m\neq 0, p\equiv 1(\mod 4). \eqno(17.30)$$
$$|(x,x)|_p^{\a-{n/2}}\iff\G_p(\a-{n/2}+1)\G_p(\a)|(\xi ,\xi )|_p^{-\a}, $$
$$\a\neq\{\a_k, a_k+{n/2}-1, k\in Z\}, n\equiv 0(\mod 4), p\neq 2$$
$$\hbox{ or } n\equiv 2(\mod 4), p\equiv 1(\mod 4). \eqno(17.31)$$
$$|(x,x)|_p^{\a-{n/2}}\iff (-1)^{\g ((\xi, \xi ))}\G_p(\a-{n/2}+1)\tilde{\G}_p(\a)|(\xi ,\xi )|_p^{-\a},$$
$$\a\neq\{\a_k-{{\pi i}/{\ln p}}, \a_k+{n/2}-1, k\in Z\}, $$
$$n\equiv 2(\mod 4), n\geq 6, p\equiv 3(\mod 4). \eqno(17.32)$$
$$|x|_p^{\a-n}\iff\G_p^{(n)}(\a )|\xi |_p^{-\a}, \quad \a\neq\a_k, k\in Z. \eqno(17.33)$$
$$|x,m|_p^{\a-n}\iff\G_p^{(n)}(\a )\bigl (|\xi |_p^{-\a}-|pm|_p^\a\bigr )\Omega (|m\xi |_p), \quad m\neq 0, \a\in\CC. \eqno(17.34)$$
$$\sqrt{|2a|_p}\chi_p(ax^2)\delta (|x|_p-p^\g)\iff\lambda_p(a)\chi_p(-{{\xi^2}/{4a}})$$
$$\times\delta (|\xi |_p-|2a|_pp^\g), \quad |4a|_p\geq p^{2-2\g}. \eqno(17.35)$$
$$\sqrt{|2a|_p}\chi_p(ax^2)\delta (|x|_p-p^\g)\iff\bigl [\lambda_p(a)\chi_p(-{{\xi^2}/{4a}})-{1/{\sqrt p}}\bigr ] $$ 
$$\times\Omega (p^{1-\g}|\xi |_p), \quad p\neq 2, |a|_p=p^{1-2\g}. \eqno(17.36)$$
$$\chi_p(ax^2)\Omega (p^{-\g}|x|_p)\iff p^\g\Omega (p^\g|\xi |_p), \quad |a|_pp^{2\g}\leq 1. \eqno(17.37)$$
$$\sqrt{|2a|_p}\chi_p(ax^2)\Omega (p^{-\g}|x|_p)\iff\lambda_p(a)\chi_p\bigl (-{{\xi^2}/{4a}}\bigr )$$
$$\times\Omega\bigl (p^{-\g}|2a|_p^{-1}|\xi |_p\bigr ), \quad |4a|_pp^{2\g}\geq p. \eqno(17.38)$$
$$\sqrt{|2a|_2}\chi_2(ax^2)\Omega (2^{-\g}|x|_2)\iff\lambda_2(a)\chi_2\bigl (-{{\xi^2}/{4a}}\bigr )$$
$$\times\delta (|\xi |_2-2^{1-\g}), \quad p=2, |a|_22^{2\g}=2. \eqno(17.39)$$
$$\sqrt{|2a|_2}\chi_2(ax^2)\Omega (2^{-\g}|x|_2)\iff\lambda_2(a)\chi_2\bigl (-{{\xi^2}/{4a}}\bigr )\Omega (2^\g|\xi |_2),$$
$$p=2, |a|_22^{2\g}=4. \eqno(17.40)$$
$$\chi_p(ax^2)\iff\lambda_p (a)|2a|_p^{-1/2}\chi_p\bigl (-{{\xi^2}/{4a}}\bigr ), \quad a\neq 0. \eqno(17.41)$$
$$\chi_p({x^2}/2)\iff\chi_p(-{\xi^2}/2), \quad p\neq 2. \eqno(17.42)$$
$$\chi_2({x^2}/2)\iff \exp (i{{\pi}/4})\chi_2(-{\xi^2}/2), \quad p=2. \eqno(17.43)$$
$$\sqrt{|a|_p}\exp (-|x|_p^2)\chi_p(ax^2)\iff S(|a|_p^{-1},{1/p})\chi_p\bigl (-{{\xi^2}/{4a}}\bigr )\Omega (|a|_p^{-1/2}|\xi |_p)$$
$$+\Bigl \{\lambda_p (a)\exp (-|{\xi}/a|_p^2)\chi_p\bigl (-{{{\xi}^2}/{4a}}\bigr )+|a|_p^{1/2}|\xi |_p^{-1}\bigl [S(|\xi |_p^{-2},{1/p})$$
$$-\exp (-|p\xi |_p^{-2})\bigr ]\Bigr \}[1-\Omega (|a|_p^{-1/2}|\xi |_p)], \quad p\neq 2, \g (a)=2k. \eqno(17.44)$$
$$\sqrt{|a|_p}\exp (-|x|_p^2)\chi_p(ax^2)\iff\Bigl\{ {1/{\sqrt p}}S(p^{-1}|a|_p^{-1},{1/p})+[\lambda_p(a)-{1/{\sqrt p}}]$$
$$\times\exp (-|pa|_p^{-1})\Bigr \}\chi_p\bigl (-{{\xi^2}/{4a}}\bigr )\Omega (\sqrt p|a|_p^{-1/2}|\xi |_p)$$
$$+\Bigl \{\lambda_p(a)\exp (-|{{\xi} /a}|_p^2)\chi_p\bigl (-{{\xi^2}/{4a}}\bigr )+|a|_p^{1/2}|\xi |_p^{-1}\bigl [S(|\xi |_p^{-2},{1/p})$$
$$-\exp (-|p\xi |_p^{-2})\bigr ]\Bigr \}[1-\Omega (\sqrt p|a|_p^{-1/2}|\xi |_p)], \quad p\neq 2, \g (a)=2k+1. \eqno(17.45)$$
$$\sqrt{|a|_2}\exp (-|x|_2^2)\chi_2(ax^2)\iff\Bigl \{[\sqrt 2\lambda_2(a)-1]\exp (-|4a|_2^{-1})$$
$$+S(|a|_2^{-1},{1/2})\Bigr \}\chi_2\bigl (-{{\xi^2}/{4a}}\bigr )\Omega (|4a|_2^{-1/2}|\xi |_2)+\Bigl \{\exp (-|4a|_p^{-1})$$
$$+[\sqrt 2\lambda_2(a)-1]S(|a|_2^{-1},{1/2})\Bigr \}\delta (|\xi |_2-|a|_2^{1/2})\chi_2\bigl (-{{\xi^2}/{4a}}\bigr )$$
$$+\Bigl \{\sqrt 2\lambda_2(a)\exp(-|2a|_2^{-2}|\xi |_2^2)\chi_2\bigl (-{{\xi^2}/{4a}}\bigr )+|a|_2^{1/2}|\xi |_2^{-1}[S(|\xi |_2^{-2},{1/2})$$
$$-2\exp (-|2\xi |_2^{-2})\bigr ]\Bigr \}[1-\Omega (|a|_2^{-1/2}|\xi |_2)], \quad p=2, \g (a)=2k. \eqno(17.46)$$
$$\sqrt{|a|_2}\exp (-|x|_2^2)\chi_2(ax^2)\iff {1/\sqrt 2}[S(|a/2|_2^{-1},{1/2})-\exp (-|2a|_2^{-1})$$
$$+2\lambda_2(a)\exp (-|8a|_2^{-1})]\chi_2\bigl (-{{\xi^2}/{4a}}\bigr )\Omega (|8a|_2^{-1/2}|\xi |_2)$$
$$+\sqrt 2\Bigl [S(|2a|_2^{-1},{1/2})+\lambda_2(a)\exp (-|8a|_2^{-1})\Bigr ]\chi_2\bigl (-{{\xi^2}/{4a}}\bigr )\delta (|\xi |_2-|2a|_2^{1/2})$$
$$+\sqrt 2\lambda_2(a)S(|2a|_2^{-1},{1/2)}\chi_2\bigl (-{{\xi^2}/{4a}}\bigr )\delta (|\xi |_2-\sqrt{2|a|_2}$$
$$+\Bigl\{|a|_2^{1/2}|\xi |_2^{-1}[S(|\xi |_2^{-2},{1/2})-2\exp (-|2\xi |_2^{-2})]$$
$$+\sqrt 2\lambda_2(a)\exp (-|2a|_2^{-2}|\xi |_2^2)\chi_2\bigr (-{{\xi^2}\over{4a}})\Bigr \}$$
$$\times [1-\Omega (2^{-{1/2}}|a|_2^{-1/2}|\xi |_2)], \quad p=2, \g (a)=2k+1. \eqno(17.47)$$
$$|x|_p^{\a-1}\t (x)\iff\G_p(\pi_{\a,\t})|\xi |_p^{-\a}\t^{-1}(\xi ), \quad \t\not\equiv 1, \a\in\CC. \eqno(17.48)$$
$$\t(p^kx)\delta (|x|_p-p^k)\iff p^ka_{p,k}(\t )\t^{-1}(\xi )\delta (|\xi |_p-1), \quad k=\rho (\t ) \eqno (17.49)$$
where quantity $a_{p,k}$ is defined in (8.17).
$$|z\bar z|_p^{\a-1}\iff\G_{p,d}(\a )|\zeta\bar\zeta |_p^{-\a}, $$
$$\a\neq\a_k, k\in Z, d\not\in\QQ_p^{\times 2}, \quad (\hbox{ see } (9.7)). \eqno(17.50)$$
$$|x|_p^{\a-1}\textstyle{\sgn_{p,d}}x\iff\tilde{\G}_p(\a )|\xi |_p^{-\a}\textstyle{\sgn_{p,d}}\xi, $$
$$\a\neq\a_k-{{\pi i}/{\ln p}}, k\in Z, p\neq 2, |d|_p=1, d\not\in\QQ_p^{\times 2} \quad (\hbox{ see } (8.8)). \eqno(17.51)$$
$$|x|_p^{\a-1}\textstyle{\sgn_{p,d}}x\iff\pm p^{\a-{1/2}}\sqrt{\textstyle{\sgn_{p,d}}(-1)}|\xi |_p^{-\a}\textstyle{\sgn_{p,d}}\xi,$$
$$p\neq 2, |d|_p={1/p} \quad (\hbox{ see } (8.24)). \eqno(17.52)$$

\centerline {\bf Literature}
\bigskip

\item{1.} Vladimirov V.~S., Volovich I.~V., Zelenov E.~I., a) $p$-Adic Analysis

          and Mathematical Physics. -- Singapore: World Scientific, 1994;

          b) Spectral Theory in $p$-Adic Quantum Mechanics and Representation

	  Theory // Math. USSR Izv., 1991, v.~36, no.~2, p.~281--309.

\item{2.} Vladimirov V.~S., a) Generalized Functions over the Field of $p$-Adic

	  Numbers // Russ. Math. Surveys, 1988, v.~43, no.~5, p.~19--64;

	  b) On Spectrum of Some Pseudo-differential Operators over the 
	  
	  $p$-Adic Number Field // Leningrad Math. J., 1991 v.~2, no.~6, 
	  
	  p.~1261--1276; c) On Spectral Properties of $p$-Adic 
	  
	  Pseudo-differential Schrodinger-type Operators // Acad. Sci. Izv., 
	  
	  Math., 1993, v.~41, no.~1, p.~55--73; d) The Adelic Freund-Witten 
	  
	  Formulas for the Veneziano and Virasoro-Shapiro Amplitudes // Russ. 
	  
	  Math. Surveys, 1993, v.~48, no.~6, p.~1--39; e) On the Freund-Witten 
	  
	  Adelic Formula for Veneziano Amplitudes // Lett. Math. Phys., 1993, 
	  
	  v.~27, p.~123--131.

\item{3.} Bikulov A.~H., a) Investigation on the $p$-adic Green function //

          Theor. Math. Phys., 1991, v.~87, no.~3, p.~376--390 (in Russian);

	  b) Private communication.

\item{4.} Gelfand I.~M., Graev M.~I., Pjatetskii-Shapiro I.~I., Representation

          Theory and Automorphic Functions. -- Philadelphia: Saunders, 1969.

\item{5.} Borevich Z.~I., Shafarevich I.~R., The Number Theory. -- N.-Y.:

          Academic Press, 1966.

\item{6.} Ruelle Ph., Thiran E., Verstegen D., Weyers J., a) Quantum Mechanics

          on $p$-Adic Fields // J. Math. Phys., 1989, v.~30, no.~12, 
	  
	  p.~2854--2874; b) Adelic string and superstring amplitudes // Mod. 
	  
	  Phys. Lett. A, 1989, v.~4, no.~18, p.~1745--1752.

\item{7.} Vladimirov V.~S., Volovich I.~V., $p$-Adic Quantum Mechanics //

          Commun. Math. Phys., 1989, v.~123, p.~659--676.

\item{8.} Meurice Y., Quantum Mechanics with $p$-Adic Numbers // Int. J. 

          Modern Phys. A, 1989, v.~4, no.~19, p.~5133--5147.

\item{9.} Smirnov V.~A., a) Renormalization in $p$-Adic Quantum Mechanics //
       
          Modern Phys. Lett. A, 1991, v.~6, no.~15, p.~1421--1427;

	  b) Calculation of General $p$-Adic Feynman Amplitude // Commun.

	  Math. Phys., 1992, v.~149, p.~623--636.

\item{10.}Zelenov E.~I., a) $p$-Adic Path Integrals // J. Math. Phys., 1991,
        
	  v.~32, p.~147--152; b) $p$-Adic quantum mechanics for $p=2$ // Theor.
	  
	  Math. Phys., 1989, v.~80, no.~2, p.~253--264 (in Russian).

\item{11.}Kochubei A.~N., a) Additive and Multiplicative Fractional
        
	  Differentiations over the Field of $p$-Adic Numbers. -- In: $p$-Adic
	
	  Functional Analysis. Proceedings of the  Fourth International
	
	  Conference. Lecture Notes in Pure and Appl. Math., 1997, v.~192,
	
	  p.~275--280. -- N.-Y.: Marsel Dekker; b) A Schrodinger-type Equation 
	  
	  over the Field of $p$-Adic Numbers // J. Math. Phys., 1993, v.~34(8),
	  
	  p.~3420--3428; c) Parabolic equations over the field of $p$- dic 
	  
	  numbers // Math. USSR Izv., 1992, v.~39, p.~1263--1280; d) Gaussian 
	  
	  Integrals and Spectral Theory over a Local Field // Russ. Acad. Sci.,
	  
	  Izv. Math., 1995, v.~45; e) On asymptotic expansion of $p$-adic Green
	  
	  functions. -- In: Proceedings of the Steklov inst., 1994, v.~203, 
	  
	  p.~116--125. -- M.:~Nauka (in Russian).

\item{12.}Missarov M.~D., Renormalization Group and Renormalization Theory 
         
	  in $p$-Adic and Adelic Scalar Models. In: Dynamical systems and 
	
	  statistcal machanics // Adv. Soviet Math., 1991, v.~3, p.~143--164.

\item{13.}Frampton P.~H., Retrospective on $p$-Adic String Theory. -- In:
        
	  Proceedings of the Steklov inst., 1994, v.~203, p.~287--291. 
	  
	   -- M.:~Nauka.

\item{14.}Bikulov A.~H., Volovich I.~V., $à$-Adic Brownian motion // Izv. RAS,
        
	  ser. math., 1997, v.~61, no.~3, p.~75--90 (in Russian).

\item{15.}Dragovi\'c B.~G., Private communication.

\item{16.}Taibleson M.~H., Fourier Analysis on Local Fields. -- Princeton:
        
	  Princeton Univ.~Press and Univ. of Tokio Press, 1975. 

\item{17.}Lerner E.~Yu., Missarov M.~D., $p$-Adic Feynman and String Amplitudes
 
         //Commun. Math. Phys., 1989, v.~121, p.~35--48.  

\item{18}Brekke L., Freund P.~G.~O., $p$-Adic Numbers in Physics. PHYSICS

         REPORTS (Review Sect. of Physics Letters), 1993, v.~233, no.~1,
	 
	 p.~1--66.

\enddocument